\documentclass[journal,12pt,draftclsnofoot,onecolumn,english]{IEEEtran}

\usepackage{epsfig}
\usepackage{times}
\usepackage{float}
\usepackage{afterpage}
\usepackage{sansmath}
\usepackage{amsmath}
\usepackage{amstext,cite}
\usepackage{amssymb,bm}
\usepackage{latexsym}
\usepackage{color}
\usepackage{graphicx}
\usepackage{amsmath}
\usepackage{amsthm}
\usepackage{graphicx}
\usepackage[center]{caption}
\usepackage{pstricks}
\usepackage{caption}
\usepackage{subcaption}
\usepackage{booktabs}
\usepackage{multicol}
\usepackage{lipsum}
\usepackage[T1]{fontenc}
\usepackage{hyperref}
\usepackage{aecompl}
\usepackage{mathrsfs}
\usepackage{blkarray}

\usepackage{arydshln}

\usepackage{soul}
\usepackage{cancel}

\usepackage{changes}
\definechangesauthor[color=red,name={Mingyue Ji}]{MJ}

\allowdisplaybreaks

\long\def\comment#1{}

\newcommand{\av}{{\mathbf a}}
\newcommand{\bv}{{\mathbf b}}

\newcommand{\ev}{{\mathbf e}}

\newcommand{\sv}{{\mathbf s}}

\newcommand{\Ac}{{\mathcal A}}

\newcommand{\Cc}{{\mathcal C}}

\newcommand{\Gc}{{\mathcal G}}

\newcommand{\Mc}{{\mathcal M}}

\newcommand{\Oc}{{\mathcal O}}

\newcommand{\Rc}{{\mathcal R}}
\newcommand{\Sc}{{\mathcal S}}
\newcommand{\Tc}{{\mathcal T}}
\newcommand{\Uc}{{\mathcal U}}

\newcommand{\Vc}{{\mathcal V}}
\newcommand{\Xc}{{\mathcal X}}

\newcommand{\asf}{{\mathsf a}}

\newcommand{\qsf}{{\mathsf q}}

\newcommand{\Ksf}{{\mathsf K}}
\newcommand{\Lsf}{{\mathsf L}}

\newcommand{\Nsf}{{\mathsf N}}

\newcommand{\Rsf}{{\mathsf R}}
\newcommand{\Ssf}{{\mathsf S}}

\newcommand{\Usf}{{\mathsf U}}

\newtheorem{thm}{Theorem}

\newtheorem{lem}{Lemma}

\newtheorem{example}{Example}

\providecommand{\definitionname}{Definition}

\usepackage{amsmath}
\usepackage{tikz}
\usetikzlibrary{calc}

\usepackage[nodisplayskipstretch]{setspace} 

\begin{document}

\title{On the Information Theoretic Secure Aggregation with Uncoded Groupwise Keys}  
\author{
Kai~Wan,~\IEEEmembership{Member,~IEEE,}
Xin~Yao,  
Hua~Sun,~\IEEEmembership{Member,~IEEE,}
Mingyue~Ji,~\IEEEmembership{Member,~IEEE,}   
and~Giuseppe Caire,~\IEEEmembership{Fellow,~IEEE}
\thanks{
K.~Wan and G.~Caire are with the Electrical Engineering and Computer Science Department, Technische Universit\"at Berlin, 10587 Berlin, Germany (e-mail:  kai.wan@tu-berlin.de; caire@tu-berlin.de). The work of K.~Wan and G.~Caire was partially funded by the European Research Council under the ERC Advanced Grant N. 789190, CARENET.}
\thanks{
X.~Yao and M.~Ji are with the Electrical and Computer Engineering Department, University of Utah, Salt Lake City, UT 84112, USA (e-mail: Xin.Yao@utah.edu; mingyue.ji@utah.edu). The work of X.~Yao and M.~Ji was supported in part by NSF Awards 1817154 and 1824558.}
\thanks{
H.~Sun is with the Department of Electrical Engineering, University of North Texas, Denton, TX 76203, USA (email: hua.sun@unt.edu). The work of H. Sun was supported in part by NSF Awards 2007108 and 2045656.
}
}
\maketitle

\begin{abstract}
   Secure aggregation, which is a core component of federated learning, 
aggregates locally trained models from distributed users at a central server. The ``secure'' nature of such aggregation consists of the fact that no information about the local users' data must be leaked to the server except the aggregated local models.
  In order to guarantee   security, some keys may be shared among the users (this is referred to as the key sharing phase). After the key sharing phase, each user masks its trained model  which is then sent to the  server (this is referred to as the model aggregation phase). 
   This paper follows the information theoretic secure aggregation problem originally formulated by Zhao and Sun,  with the objective to characterize the minimum communication cost from the $\Ksf$ users in the  model aggregation phase. Due to   user dropouts, which are    common   in real systems,  
   the server may not receive all messages from the users. A secure aggregation scheme   should tolerate the dropouts  of  at most $\Ksf-\Usf$  users, where $\Usf$ is a system parameter. 
   The optimal communication cost is characterized by Zhao and Sun, but with the assumption that the keys stored by the users could be any random variables with  arbitrary dependency.  
   On the motivation that uncoded groupwise keys are more convenient to be shared and could be used in large range of applications besides federated learning, in this paper we add one constraint into the above problem,  
namely, that the key variables are mutually independent and each key is shared by a group of  $\Ssf$ users, where $\Ssf$ is another system parameter. 
 To the best of our knowledge, all existing secure aggregation schemes (with information theoretic security or computational security) assign coded keys to the users.
We show that if  $\Ssf>\Ksf-\Usf$, a new secure aggregation scheme with uncoded groupwise keys can achieve the same   
    optimal communication cost as the best scheme with coded keys; 
    if $\Ssf\leq \Ksf-\Usf$, uncoded groupwise key sharing is strictly sub-optimal. 
    Finally, we also implement our proposed secure aggregation scheme into  Amazon EC2, which are then compared with the 
existing secure aggregation schemes with offline key sharing. 
\end{abstract}


\section{Introduction}
\label{sec:intro}
Federated learning is essentially a distributed machine learning framework, where  a central server  aims to solve a machine learning problem by the help of distributed users with local data~\cite{mcmahan2017communication,yang2019federated,li2020federated,mcmahan2021advances}.  A notable advantage  of federated learning compared to other distributed learning scenarios, is the security protection on the users' raw local data against the server. Instead of asking the users to directly upload the raw data, 
federated learning lets each user compute the model updates using its local data and securely aggregates these updates at the  server (secure aggregation).
In this paper, we use information theoretic tools to focus on two core challenges  of the secure aggregation process in federated learning, namely the effect of user   dropouts  and the communication efficiency~\cite{li2020federated}. First, in a real environment some users may drop or reply   slowly  during the training process due to the network connectivity or computational capability.
It is non-trivial to let the server  recover the aggregated updated models of the surviving users securely 
 while mitigating the effect of potential user   dropouts. Second,  additional communication among the users and server 
may be needed to guarantee the perfect  security and   mitigate the effect of the user dropouts, for example,   additional communications on exchanging the keys among the users may be taken. Since a  federated learning system   usually contains of a massive number of devices, the minimization of the communication cost is crucial.

  The secure aggregation problem with   user dropouts   was originally considered in~\cite{bonawitz2017practical},   and generally contains two phases: {\it offline key   sharing}   and {\it   model aggregation}, where the user dropouts may happen in either phase or both phases. 
In the first phase, the users generate  random seeds, and secretly  share their private random seeds such that some keys are shared among the users. 
The offline key sharing phase   is independent of the users' local training data,  and thus can take place   during off-peak traffic times when the network is not busy. 
For example, the secure aggregation schemes in~\cite{bonawitz2017practical,bell2020secure,choi2020communication,ITsecureaggre2021,lightsec2021so} all make use of offline key sharing protocols.\footnote{\label{foot:online key sharing}Online key sharing protocols (for example the ones proposed  in~\cite{so2021turbo,kadhe2020fastsecagg,nezhad2022swiftagg}) which are beyond the scope of this paper,   allow
users to communicate some information about the updated models and keys among each other,  while in offline protocols  users can only share keys.} 
If there is no private link among users,    
the communication among users should go through the central server,  and some  key agreement protocol  such as~\cite{hellman1976newdirection}    is needed,    whereby two or more parties can agree on a key
 by communicating some local information through a public link, such that even if some eavesdropper  can observe the communication in the public link, it cannot   determine   the shared key.
Once the keys are shared among the users, the users mask the updated models by the keys and send  
  masked   models to the server, such that the server could recover   the aggregated   models of the  surviving users without getting any other information about the users' local data. 

Recently, the authors in~\cite{ITsecureaggre2021} proposed an information theoretic formulation of the secure aggregation problem with   user dropouts originally considered in~\cite{bonawitz2017practical},  whose objective  is to characterize  the fundamental limits of the communication cost while preserving the  information theoretic security of the users' local data.\footnote{\label{foot:infor security}Among the existing secure aggregation schemes with     user dropouts,  the ones in~\cite{ITsecureaggre2021,lightsec2021so,nezhad2022swiftagg} considered the   information theoretic security constraint~\cite{shannonsecurity}, while the others considered the computational security.}
Due to the difficulty to characterize  the fundamental limits of the communication rates in both two phases,  with the assumption    that  the  key sharing phase  has been already performed during network off-traffic times and any keys with arbitrary dependency could be used in the model aggregation phase (i.e.,  we only consider the model aggregation phase and ignore the cost of the key sharing phase), 
the authors   in~\cite{ITsecureaggre2021}   formulated a $(\Ksf,\Usf )$ two-round information theoretic  secure aggregation problem for the server-users communication model, where  $\Ksf$ represents the number of users, $\Usf$ represents the minimum number of surviving users.\footnote{\label{foot:single epoch}The problem  in~\cite{ITsecureaggre2021} only considers one epoch of the       learning process.} 
  Each user can communicate with the server while the communication  among users is not allowed. 
The server aims to compute the element-wise sum of the vector inputs  (i.e., updated models) of $\Ksf$ users, where the input vector of  user $k$ is denoted by $W_k$ and contains $\Lsf$ uniform and i.i.d. symbols over a    finite field $\mathbb{F}_{\qsf}$.
Each user $k$ has stored a key $Z_k$, which can be any random variable    independent of  $W_1,\ldots,W_{\Ksf}$.  The transmissions (in the   model aggregation phase) contains two rounds.\footnote{\label{foot:two round}It was shown in~\cite{ITsecureaggre2021} that for the sake of security under user dropouts, at least two rounds communications must be taken.}  
In the first round of transmission, each user $k \in \{1,\ldots,\Ksf\}$ sends a coded message $X_k$ as a function of $W_k$ and $Z_k$ to the server. Since some users may drop during its transmission, the server only receives the messages from the users in $\Uc_1$ where $|\Uc_1|\geq \Usf$.
 Then the server informs the users    in the subset $\Uc_1$ of non-dropped users. 
 In the second round of transmission, after knowing the set $\Uc_1$, each user $k\in \Uc_1$ transmits another coded message $Y^{\Uc_1}_k$ as a function of $(W_k,Z_k,\Uc_1)$ to the server. Due to the user dropouts in the second round, letting $\Uc_2$ denote the set of surviving users in the second round with $\Uc_2 \subseteq \Uc_1$ and  $|\Uc_2| \geq  \Usf$,   the server receives $Y^{\Uc_1}_k$ where $k\in \Uc_2$. 
By receiving $(X_k:k\in \Uc_1)$ and $(Y^{\Uc_1}_k: k\in \Uc_2)$, the server should recover  the element-wise sum $\sum_{k\in \Uc_1}W_k$ without getting any other information about $W_1,\ldots,W_{\Ksf}$ even if the server can receive $(X_k:k\in  [\Ksf]\setminus \Uc_1)$, $ (Y^{\Uc_1}_k: k\in \Uc_1 \setminus\Uc_2) $ (e.g., the users are not really dropped but too slow in the transmission). 
 Since  the identity of the dropped users in each round is not known  a priori by the users unless they receive the list of surviving users from the server,
we should design $(X_k:k\in \{1,\ldots,\Ksf\})$ and $(Y^{\Uc_1}_{k}:k\in \Uc_1)$ for any sets $\Uc_1, \Uc_2$ where $\Uc_2 \subseteq \Uc_1\subseteq \{1,\ldots,\Ksf\}$ and $|\Uc_1|\geq |\Uc_2|\geq \Usf$, while minimizing the communication rates by the users in two rounds.
It was shown in~\cite{ITsecureaggre2021} that
 the minimum numbers of symbols that each user needs to send are $\Lsf$   over
the first round, and $\Lsf/\Usf$   over the second round,  which can be  achieved simultaneously by   a novel secure aggregation scheme. 
Another secure aggregation scheme was proposed in~\cite{lightsec2021so} for the above problem, which needs a less amount of generated keys  in the system than that of~\cite{ITsecureaggre2021}.

\begin{figure}
    \centering
    \begin{subfigure}[t]{1\textwidth}
        \centering
        \includegraphics[scale=0.25]{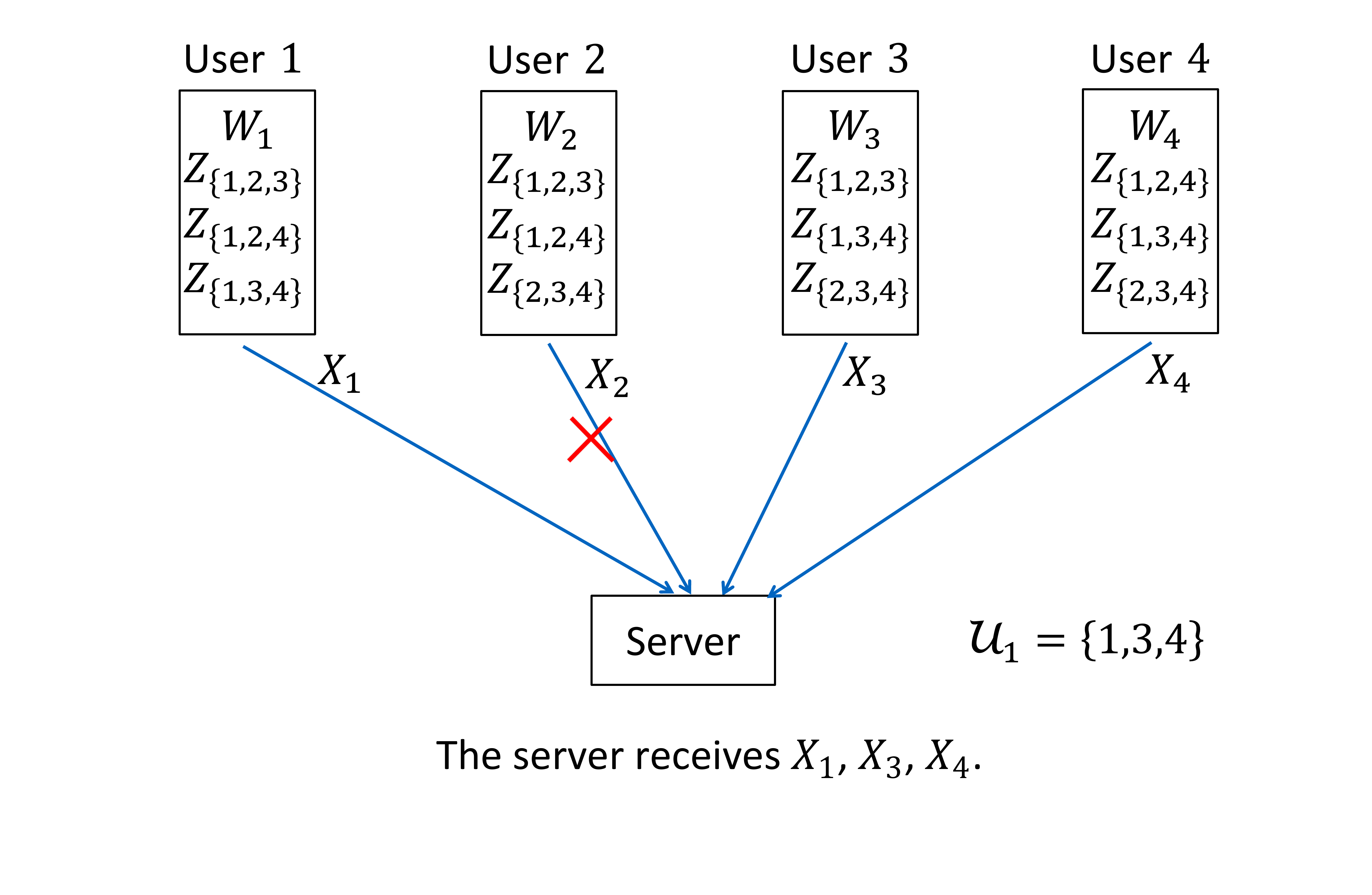}
        \caption{\small First round.}
        \label{fig:numerical 0a}
    \end{subfigure}%
   \\
    \begin{subfigure}[t]{1\textwidth}
        \centering
        \includegraphics[scale=0.25]{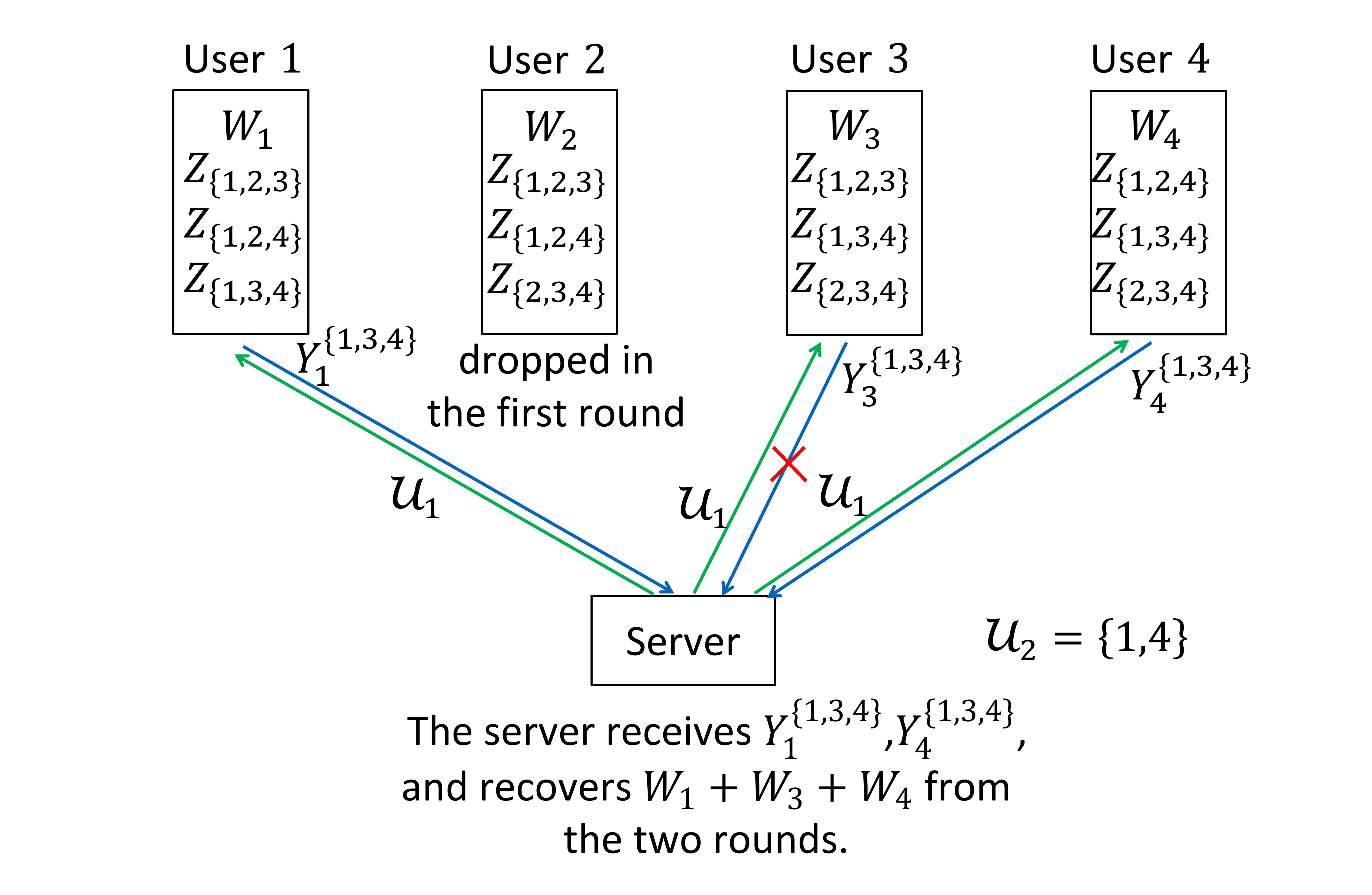}
        \caption{\small Second round.}
        \label{fig:numerical 0b}
    \end{subfigure}
    \caption{\small  $(\Ksf,\Usf,\Ssf)=(4,2,3)$ information theoretic secure aggregation problem with uncoded groupwise keys.}
    \label{fig:system model}
\end{figure}

To the best of our knowledge, all existing secure aggregation schemes with offline key sharing let the users share and store coded keys, through secret sharing    (such as~\cite{bonawitz2017practical,bell2020secure,choi2020communication}) or Minimum Distance Separable (MDS) codes (such as~\cite{ITsecureaggre2021,lightsec2021so}).\footnote{\label{foot:user private link}The key sharing protocols in~\cite{bonawitz2017practical,bell2020secure,choi2020communication} are designed for the network where  no private links exist among  users, under the constraint of computational security. The key sharing protocols in~\cite{ITsecureaggre2021,lightsec2021so} lead to information theoretic privacy, but under the constraint  that there are private links among users for the key sharing phase.}
In this paper, we  follow the information theoretic secure aggregation problem with   user dropouts in~\cite{ITsecureaggre2021}, while adding the additional constraint of uncoded groupwise keys as illustrated in Fig.~\ref{fig:system model}.\footnote{\label{foot:uncoded}The constraint of uncoded groupwise keys means that, the keys are independent among each other and each key is stored by a set of users.} 
By defining a system parameter $\Ssf\in \{1,\ldots,\Ksf\}$, for each $\Vc \subseteq \{1,\ldots,\Ksf\}$ where $|\Vc|= \Ssf$, there exists a   key   $Z_{\Vc}$ shared by the users in $\Vc$,  which is independent of other keys.\footnote{\label{foot:all schemes fail}Note that all existing secure aggregation schemes fail to satisfy this constraint  when $\Ssf<\Ksf$, due to the coded keys shared among users.}  
The uncoded groupwise keys could be directly generated and shared among users by some key agreement protocol  such as~\cite{hellman1976newdirection,maurer1993secretkey,ahlswede1993commonran,csiszar2004secrey,gohari2010itkeyaggre,sun2020securegroupcast,sun2020compound}, even if there do  not exist private links among users.\footnote{\label{foot:groupwise key generate}To generate an uncoded groupwise key shared among $\Ssf$ users, we need $\Ssf-1$ pairwise key agreement communications, each of which is between two users.} In addition, uncoded groupwise keys may be preferred   in   practice 
since  they can be generated    with low complexity and shared conveniently, and find a wide range of applications besides secure aggregation in federated learning.\footnote{\label{foot:groupwise key example}For example, the uncoded pairwise key shared among each two users are independent of the other keys and thus can guarantee the information theoretic secure communication between these two users, while the other users (who may collude) are eavesdropper  listening to the communication~\cite{shannonsecurity}.   However, the pairwise coded keys used in the scheme~\cite{lightsec2021so} cannot guarantee secure communication between any two users, because the coded key shared by these two users are correlated to other keys stored by the other users.} 
 Our objective is to characterize the capacity  region of the 
    numbers of transmissions  by the users in two rounds of the  model  aggregation phase (i.e., the rates region).

\subsection{Main Contributions}
In this paper, we first formulate  the new information theoretic   secure aggregation problem with uncoded groupwise keys. Then our main contributions on this new model   are as follows:
\begin{itemize}
\item When $\Ssf >\Ksf-\Usf$, we propose a new secure aggregation scheme which achieves exactly the same capacity region as in~\cite{ITsecureaggre2021}; this means that, when $\Ssf >\Ksf-\Usf$, secure aggregation with uncoded groupwise key sharing  has no loss on the communication efficiency. 
It is also interesting to see that by increasing $\Ssf$ above $\Ksf - \Usf+1$ yields no reduction in the transmission cost; i.e.,  $\Ssf = \Ksf - \Usf + 1$ is sufficient and no larger value of  $\Ssf$ provide improvements.  
  The main technical challenge  of the proposed scheme  based on linear coding is to   determine    the coefficients of the keys in the two round transmissions, satisfying the encodability (i.e., the keys cannot appear in the transmitted linear combinations by   the users who do not know them), decodability, and security constraints.  We overcome these challenges by designing new interference alignment strategies.\footnote{\label{foot:interference alignment}Interference alignment was originally proposed in~\cite{kuserinterference} for the wireless interference channel, which aligns the undesired packets (i.e., interference) by each user such that their linear space dimension is reduced.} 
 Note that, to achieve the optimal   rates region by our proposed scheme, not all the keys $Z_{\Vc}$ where $\Vc \subseteq \{1,\ldots,\Ksf\}$ and $|\Vc|= \Ssf$ are needed during the transmission.
The number of   needed keys is       either $\Oc(\Ksf)$ or  $\Oc(\Ksf^2)$, where each key has $(\Ksf-\Usf+1)\Lsf/\Usf$ symbols. 
\item When $\Ssf \leq \Ksf-\Usf$, we derive a new converse bound  to show that the optimal rates region of the considered problem is   a  strict subset of that in~\cite{ITsecureaggre2021} (which is without any constraint on the keys). This implies that in this regime using uncoded keys strictly hurts. 
\item Experimental results  over the Amazon EC2 cloud  show that the proposed secure aggregation scheme     
reduces the communication time in the model aggregation  by up to $53\%$ compared to the original secure aggregation scheme in~\cite{bonawitz2017practical}, and reduces    the key sharing time up to $31.7\%$ compared to the best existing information theoretic secure aggregation scheme with offline key sharing in~\cite{lightsec2021so}. 
\end{itemize}

\subsection{Paper Organization}
The rest of this paper is organized as follows. Section~\ref{sec:system} formulates the considered secure aggregation problem with uncoded groupwise keys. Section~\ref{sec:main} lists the main results of this paper. The proposed secure aggregation scheme is introduced in Section~\ref{sec:novel secure aggregation scheme}. Experimental results are provided in Section~\ref{sec:experiment}.  Section~\ref{sec:conclusion} concludes the paper, while some proofs can be found in the Appendices.

\subsection{Notation Convention}
\label{sub:notation}
Calligraphic symbols denote sets, 
bold symbols denote vectors and matrices,
and sans-serif symbols denote system parameters.
We use $|\cdot|$ to represent the cardinality of a set or the length of a vector;
$[a:b]:=\left\{ a,a+1,\ldots,b\right\}$ and $[n] := [1:n]$; 
$\mathbb{F}_{\qsf}$ represents a  finite field with order $\qsf$;  
 $\ev_{n,i}$ represents the vertical $n$-dimensional unit vector  whose entry in the  $i^{\text{th}} $ position is 1 and 0 elsewhere; 
 $\mathbf{1}_{n}$  and $\mathbf{0}_{n}$ represent  the vertical $n$-dimensional   vector  whose elements are all $1$ and all $0$, respectively;   
$\mathbf{A}^{\text{\rm T}}$  and $\mathbf{A}^{-1}$ represent the transpose  and the inverse of matrix $\mathbf{A}$, respectively;
$\text{rank}(\mathbf{A})$ represents the rank of matrix $\mathbf{A}$; 
$\mathbf{I}_n$ represents the identity matrix  of  dimension $n \times n$;
${\bf 0}_{m,n}$ represents all-zero matrix  of  dimension $m \times n$;
${\bf 1}_{m,n}$ represents all-one matrix  of  dimension $m \times n$;
$(\mathbf{A})_{m \times n}$ explicitly indicates that the matrix $\mathbf{A}$ is of dimension $m \times n$;
$<\cdot>_a$ represents the modulo operation with  integer quotient $a>0$ and in this paper we let $<\cdot>_a \in \{1,\ldots,a \}$ (i.e., we let $<b>_a=a$ if $a$ divides $b$);
  let $\binom{x}{y}=0$ if $x<0$ or $y<0$ or $x<y$;
let $\binom{\Xc}{y}=\{\Sc \subseteq \Xc: |\Sc|=y \}$ where $|\Xc|\geq y>0$.
  In the rest of the paper entropies will be in base $\qsf$, where $\qsf$ represents the field size.

 \section{System Model}
\label{sec:system}
 We formulate a $(\Ksf,\Usf, \Ssf)$ information theoretic secure aggregation problem with uncoded groupwise keys as illustrated in Fig~\ref{fig:system model},
which contains one   epoch of the learning process among 
 $\Ksf$ users and one server. For each $k\in [\Ksf]$,  user $k$ holds one input vector (i.e., updated model) $W_k$ composed of  $\Lsf$ uniform and i.i.d. symbols over a  finite field $\mathbb{F}_{\qsf}$. As in~\cite{ITsecureaggre2021}, we assume that $\Lsf$ is large enough. Ideally, the server aims to compute the element-wise sum of input vectors of all users. 
However, due to the user  dropouts,  the server may not be able to recover the sum of all input vectors. Hence, we let the server compute the sum of the input vectors from the surviving users, where the number of surviving users is at least $\Usf$. In this paper, we mainly deal with   the user dropouts and thus we assume that $\Usf\in [\Ksf-1]$.\footnote{\label{foot:U=K}When $\Usf=\Ksf$, it was shown in~\cite[Theorem 2]{Wan2022securecomp}  (by taking $\Nsf_{\rm r}=\Nsf$ in~\cite[Theorem 2]{Wan2022securecomp}) that one round transmission is enough and that the minimum number of transmitted symbols by each user is $\Lsf$.}
 In addition, by the secure aggregation constraint, the server must not retrieve any other information except the task from the received symbols. In order to guarantee the security, the users must share some secrets (i.e., keys) which are independent of the input vectors. Different from the  secure aggregation problem
  in~\cite{ITsecureaggre2021} which   assumes that the keys could be any random variables shared among   users, in this paper we consider uncoded groupwise keys, where  the keys are independent of each other and each key is shared among   $\Ssf$ users where $\Ssf\in [\Ksf]$, which is shared through private link between each two users or  by the key agreement  protocols such as~\cite{hellman1976newdirection,maurer1993secretkey,ahlswede1993commonran,csiszar2004secrey,gohari2010itkeyaggre,sun2020securegroupcast,sun2020compound}.
For each set $\Vc \in \binom{[\Ksf]}{\Ssf}$,  there exists a key $Z_{\Vc}$ independent of other keys.
  Thus 
\begin{align}
H\left( \big(Z_{\Vc}:  \Vc \in \binom{[\Ksf]}{\Ssf}), (W_1,\ldots,W_{\Ksf}\big)  \right) = \sum_{ \Vc \in \binom{[\Ksf]}{\Ssf} } H(Z_{\Vc}) +\sum_{k\in[\Ksf]} H(W_k). \label{eq:key constraint}
\end{align} 
We define  
$
Z_k := \left(Z_{\Vc}: \Vc \in \binom{[\Ksf]}{\Ssf}, k \in \Vc \right), 
$
as  the keys accessible by the user $k\in [\Ksf]$.
The whole secure aggregation procedure contains the following two rounds. 

{\it First round.}
In the first round, each user $k\in [\Ksf]$ generates a message $X_k$ as a function of $W_k$ and $Z_k$,  without knowing the identity of the dropped users.
 The communication rate of the first round $\Rsf_1$ is defined as the largest transmission load among all users normalized by $\Lsf$, i.e., 
\begin{align}
\Rsf_1:= \max_{k\in [\Ksf]} \frac{|X_k|}{\Lsf}. \label{eq:def of R1}
\end{align}
 User $k$ then sends $X_k$ to the server. 
 
Some users may drop in the first round transmission,  and the set of surviving users after the first round is denoted as $\Uc_1$,  where $\Uc_1 \subseteq [\Ksf]$ and $|\Uc_1|\geq \Usf$. Thus  the server receives  $X_k$ where $k\in \Uc_1$. 

{\it Second round.}
 In the second round, the server first sends the list of the surviving users (i.e., the set $\Uc_1$) to each user in $\Uc_1$. Then each user $k\in \Uc_1$ participates in the second round transmission by generating and sending a message $Y^{\Uc_1}_{k}$ as a function of $W_k$, $Z_k$, and $\Uc_1$.
 The communication rate of the second round $\Rsf_2$ is defined as the largest transmission load among all $\Uc_1$ and all users in $\Uc_1$ normalized by $\Lsf$, i.e.,
 \begin{align}
 \Rsf_2 := \max_{\Uc_1 \subseteq [\Ksf]:|\Uc_1|\geq \Usf} \ \max_{k\in \Uc_1}  \frac{|Y^{\Uc_1}_{k}|}{\Lsf}. \label{eq:def of R2}
 \end{align}
 
 Some users may also drop in the second round transmission,  and the set of surviving users after the second round is denoted as $\Uc_2$,  where $\Uc_2 \subseteq \Uc_1$ and $|\Uc_2|\geq \Usf$. Thus  the server receives  $Y^{\Uc_1}_{k}$  where $k\in \Uc_2$. 

 {\it Decoding.}
The server should recover $\sum_{k\in \Uc_1} W_k$ from $( X_{k_1} : k_1 \in \Uc_1 )$ and $( Y^{\Uc_1}_{k_2}: k_2 \in \Uc_2 )$, i.e.,
\begin{align}
H\left( \sum_{k\in \Uc_1} W_k \Big|  ( X_{k_1} : k_1 \in \Uc_1 ), ( Y^{\Uc_1}_{k_2}: k_2 \in \Uc_2 ) \right)=0, \ \forall \Uc_1 \subseteq [\Ksf], \Uc_{2} \subseteq \Uc_1 : |\Uc_1| \geq  |\Uc_2|\geq \Usf. \label{eq:decodability}
\end{align}
Meanwhile, the security constraint imposes that after receiving all messages sent by the users  including   the  dropped  users (e.g., the users are not really dropped but too slow in the transmission), the server cannot get any other information about the input vectors except  $\sum_{k\in \Uc_1} W_k$, i.e.,
\begin{align}
I\left(   W_1,\ldots,W_{\Ksf} ; X_1,\ldots, X_{\Ksf}, ( Y^{\Uc_1}_{k}: k \in \Uc_1 )   \Big|  \sum_{k\in \Uc_1} W_k \right)=0 , \ \forall \Uc_1 \subseteq [\Ksf] : |\Uc_1|\geq \Usf. \label{eq:security constraint}
\end{align}

{\it Objective.}
A rate tuple $(\Rsf_1,\Rsf_2)$ is achievable if there exist keys $\left(Z_{\Vc}:  \Vc \in \binom{[\Ksf]}{\Ssf} \right) $ satisfying~\eqref{eq:key constraint} and a secure aggregation scheme satisfying the decodability and security constraints in~\eqref{eq:decodability} and~\eqref{eq:security constraint}. Our objective is to determine the capacity region (i.e., the closure of all achievable rate tuples) of the considered problem, denoted by $\Rc^{\star}$.

{\it A converse bound from~\cite{ITsecureaggre2021}.}
By removing the uncoded groupwise constraint on the keys in our considered problem, we obtain the information theoretic aggregation problem in~\cite{ITsecureaggre2021}. Hence, the converse bound  on the capacity region in~\cite{ITsecureaggre2021}  is also a converse bound for our considered problem, which leads to the following lemma.
\begin{lem}[\cite{ITsecureaggre2021}]
\label{lem:converse}
For the $(\Ksf,\Usf, \Ssf)$ information theoretic secure aggregation problem with uncoded groupwise keys, any achievable rate tuple $(\Rsf_1,\Rsf_2)$ satisfies  
\begin{align}
\Rsf_1 \geq 1, \ \Rsf_2 \geq \frac{1}{\Usf}. \label{eq:converse from litterature} 
\end{align}
\hfill $\square$ 
\end{lem}

However,  the achievable secure aggregation schemes in~\cite{ITsecureaggre2021,lightsec2021so} cannot work in our considered problem with $\Ssf<\Ksf$, since 
the schemes in~\cite{ITsecureaggre2021,lightsec2021so} assign correlated coded keys to   users, while in our considered problem 
the keys are uncoded, groupwise-sharing and independent.

 
 Another observation is that the capacity region of the $(\Ksf,\Usf, \Ssf_1)$ information theoretic secure aggregation problem with uncoded groupwise keys covers that of the   $(\Ksf,\Usf, \Ssf_2)$ information theoretic secure aggregation problem with uncoded groupwise keys, where $\Ssf_1>\Ssf_2$. This is because, without collusion between the server and the users, having more users knowing the same key will not hurt. So any key $Z_{\Vc_2}$ could be generated by  extracting some symbols from 
  $Z_{\Vc_1}$  where $\Vc_2 \subseteq \Vc_1$. 
 

 \section{Main Results}
\label{sec:main}
We first present the main result of our paper.  
\begin{thm}
\label{thm:main result}
For the $(\Ksf,\Usf, \Ssf)$ information theoretic secure aggregation problem with uncoded groupwise keys, when $\Ssf > \Ksf-\Usf$, we have 
\begin{align}
\Rc^{\star} = \left\{(\Rsf_1,\Rsf_2): \Rsf_1 \geq 1, \Rsf_2\geq \frac{1}{\Usf} \right\}. \label{eq:main result thm}
\end{align}
\hfill $\square$ 
\end{thm}
The converse bound for Theorem~\ref{thm:main result} is directly from Lemma~\ref{lem:converse}. 
For the achievability, we propose a    new secure aggregation  scheme based on linear coding and interference alignment, which is described in
  Section~\ref{sec:novel secure aggregation scheme}.
  
When $\Ssf > \Ksf-\Usf$, the proposed scheme for Theorem~\ref{thm:main result} achieves the same capacity region as the optimal secure aggregation scheme without any constraint on the keys in~\cite{ITsecureaggre2021}. It is also interesting to see that  
 increasing $\Ssf$ above $\Ksf-\Usf+1$ will not reduce the communication cost.

 There are totally $ \binom{\Ksf}{\Ssf}$ subsets of $[\Ksf]$ with cardinality   $\Ssf$. 
By the problem setting,   we can   use at most $ \binom{\Ksf}{\Ssf}$ keys each of which is shared by  $\Ssf$ users.
However, we do not need to use  generate all these $ \binom{\Ksf}{\Ssf}$ keys in our proposed secure scheme for Theorem~\ref{thm:main result}.  
It will be clarified in Section~\ref{sec:novel secure aggregation scheme} that,  the number of needed keys by the proposed secure aggregation scheme for Theorem~\ref{thm:main result}  is  $\Ksf$ when $\Usf\leq \Ksf-\Usf+1$ and is $\Oc(\Ksf^2)$   when $\Usf> \Ksf-\Usf+1$, where each   key has $(\Ksf-\Usf+1)\Lsf/\Usf$ symbols.
  Note that if coded key assignment is allowed,  
the secure aggregation scheme in~\cite{ITsecureaggre2021} needs to generate  $\Usf$ coded keys with $\Lsf/\Usf$ symbols  for  each group of users $\Vc\subseteq [\Ksf]$ where $|\Vc|\in [\Usf:\Ksf]$, where  each user in the group stores a linear combination of these $\Usf$ coded keys; 
for each pair of users $\Vc\subseteq [\Ksf]$ where $|\Vc|=2$,  
 the secure aggregation scheme in~\cite{lightsec2021so}  lets each user in the pair generate  a coded    key  with $\Lsf/\Usf$ symbols and 
 share it to the other user in the pair.

For the case $\Ssf \leq \Ksf-\Usf$, the following theorem shows that the communication rate of the optimal  
 secure aggregation scheme without any constraint on the keys in~\cite{ITsecureaggre2021} cannot be achieved; i.e., the capacity region of the considered problem is a strict subset of the one in~\cite{ITsecureaggre2021}.
\begin{thm}
\label{thm:subregion}
For the $(\Ksf,\Usf, \Ssf)$ information theoretic secure aggregation problem with uncoded groupwise keys, when $1=\Ssf \leq  \Ksf-\Usf$, secure aggregation is not possible; when  $2\leq \Ssf \leq  \Ksf-\Usf$,  the communication rate of the first round must satisfy
  that 
\begin{align}
\Rsf_1\geq  1+\frac{1}{  \binom{\Ksf-1}{\Ssf-1}-1} .\label{eq:converse of R1}
\end{align} 
\hfill $\square$ 
\end{thm}
The proof of Theorem~\ref{thm:subregion} can be found in Appendix~\ref{sec:proof of subregion}. From  Theorem~\ref{thm:subregion}, when  $2\leq \Ssf \leq  \Ksf-\Usf$,  it is not enough for each user  to transmit one (normalized) linear combination of the input vector and keys. Intuitively, this is  because  the total number of  dropped users after the second round could be  
larger than or equal to $\Ssf$, which is the number of users sharing each key; 
  thus some key(s) appearing in the transmission of the first round, may not be received in the received packets of the second round due to the user dropouts. Hence, we need to transmit more than one   (normalized) linear combination in the first round. 
  It is one of our on-going works to design tight achievable schemes and converse bounds for the case  $2\leq \Ssf \leq \Ksf-\Usf$. 

\section{Proof of Theorem~\ref{thm:main result}: New Secure Aggregation Scheme}
\label{sec:novel secure aggregation scheme}
To present the proposed scheme,   we only need to focus on the case where $\Ssf=\Ksf-\Usf+1$.  
  As we explained at the end of Section~\ref{sec:system}, this is because 
if $\Ssf>\Ksf-\Usf+1$, 
we can generate any key $Z_{\Vc}$ where $\Vc\in \binom{[\Ksf]}{\Ksf-\Usf+1}$ by extracting some symbols from   $Z_{\Vc_1}$ where $\Vc_1 \in\binom{[\Ksf]}{\Ssf}$ and $\Vc \subseteq \Vc_1 $, while the users in $\Vc_1\setminus \Vc$ will not use $Z_{\Vc }$ even they know it. Thus
    a  secure aggregation scheme for the case $\Ssf=\Ksf-\Usf+1$ could also work for the case $\Ssf >\Ksf-\Usf+1$. 
    
The construction structure of the achievable scheme  is as follows. 
\begin{itemize}
\item  Since the length of each input vector  $W_i$ where $i\in [\Ksf]$ is large enough,  as explained in~\cite{ITsecureaggre2021}, 
 we can consider blocks of symbols of $W_i$ as an element of a suitably large field extension  and consider operations such as element wise sum as operations over the field extension. 
 Hence, without loss of generality, in the scheme proposed in this paper we can assume    that $\qsf$ is large enough.
We then divide each input vector $W_i$ where $i\in [\Ksf]$ into $\Usf$ non-overlapping and equal-length pieces, where the $j^{\text{th}}$ piece denoted by $W_{i,j}$ contains $\Lsf/\Usf$ symbols on $\mathbb{F}_{\qsf}$. 
In addition, for each $\Vc\in \binom{[\Ksf]}{\Ssf}$ and each $k\in\Vc$,\footnote{\label{foot:recall binom sets}Recall that $\binom{\Xc}{y}=\{\Sc \subseteq \Xc: |\Sc|=y \}$ where $|\Xc|\geq y>0$.}
 we let $Z_{\Vc,k}$ denote a vector of $\Lsf/\Usf$ uniform i.i.d. symbols on $\mathbb{F}_{\qsf}$. Then, we define a key  $Z_{\Vc}= (Z_{\Vc,k}: k\in \Vc)$ and let $Z_{\Vc}$  be shared by all users in $\Vc$. 
\item In the first  round,  each user $k\in [\Ksf]$ sends 
\begin{align}
X_{k,j} = W_{k,j} + \sum_{\Vc\in\binom{[\Ksf]}{\Ssf}: k \in \Vc} a_{\Vc, j} Z_{\Vc,k}  , \ \forall j \in [\Usf],
\label{eq:transmission of X_k,j}
\end{align}
where $a_{\Vc, j} \in \mathbb{F}_{\qsf}$ is a coefficient to be designed.\footnote{\label{foot:scalar vector product}In this paper, the product $a \bv$  where $a$ is a scalar and $\bv$ is a vector or a matrix, represents multiplying each element in $\bv$ by $a$.}  Note that each $X_{k,j}$ contains $\Lsf/\Usf$ symbols, and thus 
$X_{k}=(X_{k,1},\ldots,X_{k,\Usf})$ contains $\Lsf$ symbols, which leads to $\Rsf_1=1$.

We let   $\av_{\Vc} :=[a_{\Vc,1},\ldots,a_{\Vc,\Usf}]^{\text{\rm T}}$. 
By the security constraint, 
$W_{k}$ should be perfectly protected by the keys in $X_{k}=(X_{k,1},\ldots,X_{k,\Usf})$. Thus, by denoting  the sets $\Vc\in \binom{[\Ksf]}{\Ssf}$ where   $k \in \Vc$  by  $\Sc_{k}(1),\ldots,\Sc_{k}\left(\binom{\Ksf-1}{\Ssf-1}\right)$, we   aim to have that    the coefficients matrix  (whose dimension is   $\Usf \times \binom{\Ksf-1}{\Ssf-1}$)
\begin{align}
\left[\av_{\Sc_{k}(1)},\ldots, \av_{\Sc_{k}\left(\binom{\Ksf-1}{\Ssf-1}\right)} \right] \ \text{ has rank equal to $\Usf$}, \ \forall k\in[\Ksf]. \label{eq:full rank constraint}
\end{align}  
 If the constraints  in~\eqref{eq:full rank constraint} are satisfied, with the fact that   the keys are independent of the input vectors,    the server cannot 
get any information about $W_1,\ldots,W_{\Ksf}$ even if the server receives all $X_1,\ldots,X_{\Ksf}$ (the formal proof is given in~\eqref{eq:formal proof of Xk security} in Appendix~\ref{sec:proof of security}).

Since the set of surviving users after the first round is $\Uc_1$, the server receives $X_k$ where $k\in \Uc_1$, and thus can recover 
\begin{align}
\sum_{k\in \Uc_1} X_{k,j} &= \sum_{k\in \Uc_1} W_{k ,j} +   \sum_{\Vc\in \binom{[\Ksf]}{\Ssf}:  \Vc \cap \Uc_1 \neq \emptyset} \left( a_{\Vc, j} \sum_{k_1\in \Vc \cap \Uc_1}  Z_{\Vc,k_1} \right) \\
&= \sum_{k\in \Uc_1} W_{k ,j} +   \sum_{\Vc\in \binom{[\Ksf]}{\Ssf} } \left( a_{\Vc, j} \sum_{k_1\in \Vc \cap \Uc_1}  Z_{\Vc,k_1} \right),
 \ \forall j \in [\Usf], 
\label{eq:received packets after 1st round}
\end{align}
where~\eqref{eq:received packets after 1st round} follows since $\Ssf=\Ksf-\Usf+1>\Ksf-|\Uc_1|$.
Hence, the server still needs to recover $\sum_{\Vc\in \binom{[\Ksf]}{\Ssf}  } \left( a_{\Vc, j} \sum_{k_1\in \Vc \cap \Uc_1}  Z_{\Vc,k_1} \right)$ for each $j \in [\Usf]$ in the next   round.  We can treat 
\begin{align}
Z^{\Uc_1}_{\Vc}:= \sum_{k_1\in \Vc \cap \Uc_1}  Z_{\Vc,k_1} , \ \forall \Vc\in \binom{[\Ksf]}{\Ssf}, \label{eq:def of coded key}
\end{align}
  as one coded key, which can be encoded by all users in $ \Vc \cap \Uc_1$ and  contains $\Lsf/\Usf$ uniform and i.i.d. symbols. 
   {\bf Thus  by the construction of the first round transmission, we only need to transmit linear combinations of coded keys in the second round, such that the server can recover $\sum_{\Vc\in \binom{[\Ksf]}{\Ssf}}   a_{\Vc, j}   Z^{\Uc_1}_{\Vc} $ for each $j \in [\Usf]$.}
\item 
In the second  round, we denote    the sets in $\binom{[\Ksf]}{\Ssf}$   by $\Sc(1),\ldots,\Sc\left(\binom{\Ksf}{\Ssf}\right)$, and for each $k\in[\Ksf]$ denote   the sets  in $\binom{[\Ksf]\setminus\{k\}}{\Ssf}$   by  
$\overline{\Sc}_{k}(1),\ldots,\overline{\Sc}_k\left(\binom{\Ksf-1}{\Ssf}\right)$.
Thus the server should recover 
\begin{align}
   \begin{bmatrix}
F_1  \\
\vdots\\
F_{\Usf}
\end{bmatrix} 
= \left[ \av_{\Sc(1)}, \ldots, \av_{\Sc\left(\binom{\Ksf}{\Ssf}\right)} \right]
   \begin{bmatrix}
Z^{\Uc_1}_{\Sc(1)} \\
\vdots\\
Z^{\Uc_1}_{\Sc\left(\binom{\Ksf}{\Ssf}\right)} 
\end{bmatrix}, \label{eq:second round recover}
\end{align}
where each $F_{j}$, $j\in [\Usf]$, contains $\Lsf/\Usf$ symbols.

Note that each user $k\in \Uc_1$ cannot encode $Z^{\Uc_1}_{\Vc}$ where $\Vc\in \binom{[\Ksf]\setminus\{k\}}{\Ssf}$. 
If the $\Usf$-dimensional vectors $\av_{\Vc}$ where $\Vc\in \binom{[\Ksf]}{\Ssf}$ satisfy the constraints that 
\begin{align}
\left[ \av_{\overline{\Sc}_{k}(1)}, \ldots, \av_{\overline{\Sc}_{k}\left(\binom{\Ksf-1}{\Ssf}\right)}  \right] \ \text{ has rank equal to $\Usf-1$}, \ \forall k\in[\Ksf],
\label{eq:constraint of a for null space}
\end{align}
 then the matrix $\left[ \av_{\overline{\Sc}_{k}(1)},  \ldots, \av_{\overline{\Sc}_{k}\left(\binom{\Ksf-1}{\Ssf}\right)}  \right] $ contains exactly one linearly independent left null space vector.  {\bf To achieve~\eqref{eq:constraint of a for null space}, we will propose some interference alignment techniques to align the $\Usf$-dimensional  vectors of 
 the $\binom{\Ksf-1}{\Ssf}$  unknown keys   to a linear space spanned by    $\Usf-1$ linearly independent vectors.}

Thus 
 we can let each user $k\in \Uc_1$ transmit 
\begin{align}
Y^{\Uc_1}_{k} =    \sv_{k}    \begin{bmatrix}
F_1  \\
\vdots\\
F_{\Usf}
\end{bmatrix}   ,
\label{eq:transmission of Y_k}
\end{align}
where $\sv_{k}$  represents the left null space vector of $\left[ \av_{\overline{\Sc}_{k}(1)}, \ldots, \av_{\overline{\Sc}_{k}\left(\binom{\Ksf-1}{\Ssf}\right)}  \right]$. By construction,  in $Y^{\Uc_1}_{k} $ the coefficients of the coded keys which cannot be encoded by user $k$ are $0$.
 Note that   $Y^{\Uc_1}_{k}$ contains $\Lsf/\Usf$ symbols,  which leads to $\Rsf_2=1/\Usf$.

{\bf For the decodability, from any set of surviving users after the second round  $\Uc_2 \subseteq \Uc_1$ where $|\Uc_2| \geq \Usf$,  we should recover 
$F_1,\ldots,F_{\Usf}$
 from the second round transmission}; i.e.,
 we   aim to have 
\begin{align}
\text{any $\Usf$ vectors in $\{ \sv_{k}: k\in \Uc_1\} $ are linearly independent}. \label{eq:decodability constraint}
\end{align}
Thus from~\eqref{eq:received packets after 1st round} and~\eqref{eq:decodability constraint}, the server can recover $F_1,\ldots,F_{\Usf}$ and then recover $ \sum_{k\in \Uc_1} W_{k ,j} $ for all $j \in [\Usf]$; thus it can recover $ \sum_{k\in \Uc_1} W_{k } $. 

In addition, for the security constraint,  by construction we have  
\begin{align}
H\left(     Y^{\Uc_1}_{k}: k\in \Uc_1  \right) = \Lsf,
\label{eq:constraint on Yk}
\end{align}
which follows since each $Y^{\Uc_1}_{k}$  where $k\in \Uc_1$ is  in the linear space spanned by $F_{1},\ldots,F_{\Usf}$, where each $F_{j}$, $j\in[\Usf]$, contains $\Lsf/\Usf$ symbols.
Intuitively, from $(X_k:k\in [\Ksf])$, the server cannot get any information about $W_1,\ldots,W_{\Ksf}$. Together with $ (Y^{\Uc_1}_{k}: k\in \Uc_1 )$ whose entropy is $\Lsf$, the server can at most get $\Lsf$ symbols information about  $W_1,\ldots,W_{\Ksf}$, which are exactly the symbols in $ \sum_{k\in \Uc_1} W_{k } $.  Hence, the proposed scheme is secure. The rigorous  proof on the security constraint in~\eqref{eq:security constraint} can be found in Appendix~\ref{sec:proof of security}.

\end{itemize}
We conclude that the achieved rates are $(\Rsf_1,\Rsf_2)=( 1, 1/\Usf)$, coinciding with Theorem~\ref{thm:main result}.

{\bf For what said above, it is apparent that the key challenge  in the proposed scheme is to design  the $\Usf$-dimensional vectors $\av_{\Vc}$ where $\Vc\in \binom{[\Ksf]}{\Ssf}$, such that the constraints in~\eqref{eq:full rank constraint},~\eqref{eq:constraint of a for null space}, and~\eqref{eq:decodability constraint} are satisfied. As showed above, if such constraints are satisfied, the proposed scheme is decodable and secure.}

Another important observation is that, the   constraints in~\eqref{eq:full rank constraint},~\eqref{eq:constraint of a for null space} are not related to   $\Uc_1$; in addition, if  the constraint in~\eqref{eq:decodability constraint} is satisfied for the case $\Uc_1=[\Ksf]$, this constraint also holds for any other    $\Uc_1$. {\bf Hence, we only need to consider the case  $\Uc_1=[\Ksf]$ to design the   $\Usf$-dimensional vectors $\av_{\Vc}$ where $\Vc\in \binom{[\Ksf]}{\Ssf}$.}

In the following, 
we will further divide the considered  case $\Usf<\Ksf$ into three regimes:
  a) $\Usf\leq  \Ksf-\Usf+1$; b) $\Usf> \Ksf-\Usf+1$ and $\Usf=\Ksf-1$; c) $\Usf> \Ksf-\Usf+1$ and $\Usf<\Ksf-1$.  We  will propose our scheme  for each regime which achieves the capacity region in Theorem~\ref{thm:main result}. 
In each regime, we propose a different selection on the $\Usf$-dimensional vectors $\av_{\Vc}$ where $\Vc\in \binom{[\Ksf]}{\Ssf}$, such that the constraints in~\eqref{eq:full rank constraint},~\eqref{eq:constraint of a for null space}, and~\eqref{eq:decodability constraint} are satisfied.  


\subsection{Case $\Usf\leq \Ksf-\Usf+1$}
\label{sub:U<S}
We first illustrate the proposed scheme for this case through an example.
\begin{example}[$(\Ksf,\Usf, \Ssf))=(3,2,2)$]
\label{ex:case 1}
\rm
Consider  the  $(\Ksf,\Usf,\Ssf)=(3,2,2)$ information theoretic secure aggregation problem with uncoded groupwise keys. While illustrating the proposed
scheme through examples, we  perform a field extension  on the input vectors to a large 
enough prime field $\mathbb{F}_{\qsf}$.  In general this assumption on prime field
size is not necessary in our proposed scheme.

For each $\Vc\in \binom{[3]}{2}$,  we generate a key $Z_{\Vc}=(Z_{\Vc,k}:k\in \Vc)$ shared by users in $\Vc$, where each $Z_{\Vc,k}$ contains $\Lsf/2$  uniform and i.i.d. symbols over $\mathbb{F}_{\qsf}$.
We also divide each input vector $W_k$ where $k\in [3]$ into two pieces, $W_{k}=(W_{k,1},W_{k,2})$, where each piece contains $\Lsf/2$ uniform and i.i.d. symbols over $\mathbb{F}_{\qsf}$. 

{\it First round.}
In the first round, user $1$ transmits $X_1=(X_{1,1},X_{1,2})$, where 
\begin{align*}
&X_{1,1}=W_{1,1}+Z_{\{1,2\},1} +Z_{\{1,3\},1};\\
&X_{1,2}=W_{1,2}+Z_{\{1,2\},1}+2 Z_{\{1,3\},1}.
\end{align*}
User $2$ transmits $X_2=(X_{2,1},X_{2,2})$, where 
\begin{align*}
&X_{2,1}=W_{2,1}+Z_{\{1,2\},2}+Z_{\{2,3\},2};\\
&X_{2,2}=W_{2,2}+Z_{\{1,2\},2}+3 Z_{\{2,3\},2}.
\end{align*}
User $3$ transmits  $X_3=(X_{3,1},X_{3,2})$, where 
\begin{align*}
&X_{3,1}=W_{3,1}+Z_{\{1,3\},3}+Z_{\{2,3\},3};\\
&X_{3,2}=W_{3,2}+2Z_{\{1,3\},3}+3Z_{\{2,3\},3}.
\end{align*}
In other words, we let
\begin{align}
\av_{\{1,2\}}=[1,1]^{\text{T}}, \ \av_{\{1,3\}}=[1,2]^{\text{T}}, \ \av_{\{2,3\}}=[1,3]^{\text{T}}.
\label{eq:choice of a for example 1}
\end{align}

In $X_1$, the coefficient matrix of the keys $(Z_{\{1,2\},1},Z_{\{1,3\},1})$ is $\begin{bmatrix}
1&1  \\
1&2  
\end{bmatrix}$, which 
has rank equal to $2$ (recall that the field size is large enough), i.e., the constraint in~\eqref{eq:full rank constraint} is satisfied for user $1$. Thus  $W_1$ is perfectly protected by $(Z_{\{1,2\},1},Z_{\{1,3\},1})$ from $X_1$.
Similarly, the constraints in~\eqref{eq:full rank constraint} are satisfied for user $2,3$.

  {\it Second round.}
In the second round, we only need to consider the   case where $\Uc_1=[3]$, as explained before.
  Since $\Uc_1=[3]$, the server should recover $W_1+W_2+W_3$. 
By the definition of coded key in~\eqref{eq:def of coded key}, we define  the coded keys
\begin{align*}
&Z^{[3]}_{\{1,2\}}=Z_{\{1,2\},1}+Z_{\{1,2\},2},\\
&Z^{[3]}_{\{1,3\}}=Z_{\{1,3\},1}+Z_{\{1,3\},3},\\
&Z^{[3]}_{\{2,3\}}=Z_{\{2,3\},2}+Z_{\{2,3\},3},
\end{align*} 
each of which contains $\Lsf/2$ uniform and i.i.d. symbols.
From the transmission of the first round, the server can recover 
\begin{align*}
&X_{1,1}+X_{2,1}+X_{3,1}= W_{1,1}+W_{2,1}+W_{3,1} +Z^{[3]}_{\{1,2\}}+ Z^{[3]}_{\{1,3\}}+Z^{[3]}_{\{2,3\}};\\
&X_{1,2}+X_{2,2}+X_{3,2}=W_{1,2}+W_{2,2}+W_{3,2}+Z^{[3]}_{\{1,2\}}+2Z^{[3]}_{\{1,3\}}+3Z^{[3]}_{\{2,3\}}.
\end{align*}
 Hence, the server should further recover 
 \begin{align}
   \begin{bmatrix}
F_1  \\
F_2
\end{bmatrix}  =
  [
\av_{\{1,2\}} , \av_{\{1,3\}}, \av_{\{2,3\}}
]   
  \begin{bmatrix}
Z^{[3]}_{\{1,2\}}  \\
Z^{[3]}_{\{1,3\}}\\
Z^{[3]}_{\{2,3\}}
\end{bmatrix} =
  \begin{bmatrix}
1&1&1  \\
1&2&3
\end{bmatrix} 
   \begin{bmatrix}
Z^{[3]}_{\{1,2\}}  \\
Z^{[3]}_{\{1,3\}}\\
Z^{[3]}_{\{2,3\}}
\end{bmatrix} 
\label{eq:ex1 second round task}
 \end{align}
 totally $\Lsf$ symbols 
 in the second round. Since $|\Uc_2|\geq \Ssf=2$, the second round transmission should be designed such that from any   two of $Y^{[3]}_1, Y^{[3]}_2, Y^{[3]}_3$, we can recover~\eqref{eq:ex1 second round task}. 

For user $1$ who   cannot encode $Z^{[3]}_{\{2,3\}} $,  the   sub-matrix $[\av_{\{2,3\}}]$   has rank equal to $1$; thus the constraint in~\eqref{eq:constraint of a for null space} is 
satisfied for user $1$. The left
null space of $[\av_{\{2,3\}}]$ contains exactly one linearly independent $2$-dimensional vector, which
could be $[3,-1]$. Thus we let user $1$ transmit  
\begin{align}
Y^{[3]}_1= [3, -1]     \begin{bmatrix}
F_1  \\
F_2
\end{bmatrix} =3F_1-F_2,
\end{align}
in which the coefficient of  $Z^{[3]}_{\{2,3\}}$ is $0$.
Similarly, we  let user $2$ transmit 
\begin{align}
Y^{[3]}_2= [2, -1]     \begin{bmatrix}
F_1  \\
F_2
\end{bmatrix}=2F_1-F_2 ,
\end{align}
in which the coefficient of  $Z^{[3]}_{\{1,3\}}$ is $0$, 
and let user $3$ transmit 
 \begin{align}
Y^{[3]}_3= [1, -1]     \begin{bmatrix}
F_1  \\
F_2
\end{bmatrix} =F_1-F_2,
\end{align}
 in which the coefficient of  $Z^{[3]}_{\{1,2\}}$ is $0$.
 The constraints in~\eqref{eq:constraint of a for null space}  are also satisfied for users $2,3$.

By construction,  any two of $Y^{[3]}_1,Y^{[3]}_2,Y^{[3]}_3$ are linearly independent. Hence, for any $\Uc_2\subseteq [3]$ where $|\Uc_2|\geq 2$, the server can recover $F_1$ and $F_2$;  thus the constraint in~\eqref{eq:decodability constraint} is satisfied. 
Hence,  from the two round transmissions, the server can  recover $W_1+W_2+W_3$.

Since the constraints   in~\eqref{eq:full rank constraint},~\eqref{eq:constraint of a for null space}, and~\eqref{eq:decodability constraint} are satisfied,
by the security proof in Appendix~\ref{sec:proof of security}, the scheme is secure  for the case $\Uc_1=[3]$.

In conclusion, in the first round, each user transmits $\Lsf$ symbols. In the second round, each user in $\Uc_1$ transmits $\Lsf/2$ symbols. Hence,
the achieved rates are $(\Rsf_1,\Rsf_2)=( 1, 1/2)$, coinciding with Theorem~\ref{thm:main result}.
\hfill $\square$
\end{example}

We are now ready to generalize the proposed scheme in Example~\ref{ex:case 1} to the case where $\Usf\leq   \Ksf-\Usf+1$.
For the sake of simplicity, we directly describe the choice of the $\Usf$-dimensional vectors  and show that such choice satisfies the constraints in~\eqref{eq:full rank constraint},~\eqref{eq:constraint of a for null space}, and~\eqref{eq:decodability constraint}.

We use a cyclic key assignment, by
   defining a collection of cyclic sets
\begin{align}
\Cc:= \big\{ \{i,<i+1>_{\Ksf}, \ldots, <i+\Ksf-\Usf>_{\Ksf} \} : i\in [\Ksf]  \big\}.
\label{eq:cyclic key set}
\end{align}  
For the ease of notation, we sort the sets in $\Cc$   in an order  where the $i^{\text{th}}$ set denoted by $\Cc(i)$ is $\{i,<i+1>_{\Ksf}, \ldots, <i+\Ksf-\Usf>_{\Ksf} \}$, for each $i\in [\Ksf]$.\footnote{\label{foot:Ci example}For example, when $\Ksf=4$ and $\Usf=2$, we have $\Cc(1)=\{1,2,3\}$, $\Cc(2)=\{2,3,4\}$, $\Cc(3)=\{1,3,4\}$, and $\Cc(4)=\{1,2,4\}$.}
It can be seen that each of the sets $\Cc(k), \Cc(<k-1>_{\Ksf}), \ldots, \Cc(<k-\Ksf+\Usf>_{\Ksf})$ contains  $k$, for each $k\in [\Ksf]$.

We select the $\Usf$-dimensional vectors $\av_{\Vc}$ where $\Vc \in \binom{[\Ksf]}{\Ssf}$ as follows:
\begin{itemize}
\item if $\Vc \in \Cc$, we let $\av_{\Vc}$ be uniform and i.i.d. over $\mathbb{F}^{\Usf}_{\qsf}$;
\item  otherwise, we    let each element in $\av_{\Vc}$ be $0$.  
\end{itemize}
Next we will show that the above choice of these   $\Usf$-dimensional vectors  satisfies the constraints in~\eqref{eq:full rank constraint},~\eqref{eq:constraint of a for null space}, and~\eqref{eq:decodability constraint}, with high probability.

\paragraph*{Constraints in~\eqref{eq:full rank constraint}}
Since $\qsf$ is large enough and $\Usf\leq  \Ksf-\Usf+1$, for each $k\in [\Ksf]$ the   matrix 
$$\left[\av_{\Cc(k)},\av_{\Cc(<k-1>_{\Ksf})}, \ldots, \av_{\Cc(<k-\Ksf+\Usf>_{\Ksf})} \right]$$ whose dimension is   $\Usf \times (\Ksf-\Usf+1)$,
has rank equal to $\Usf$ with high probability;	thus the constraints in~\eqref{eq:full rank constraint} are satisfied with high probability.

\paragraph*{Constraints in~\eqref{eq:constraint of a for null space}}
Among the sets in $\Cc$, 
 each of the sets $ \Cc(<k+1>_{\Ksf}) , \Cc(<k+2>_{\Ksf}) ,  \ldots,  \Cc(<k+\Usf-1>_{\Ksf}) $ does not contain  $k$, where $k\in [\Ksf]$.
It can be seen that
 $ [\av_{\Cc(<k+1>_{\Ksf})},\av_{\Cc(<k+2>_{\Ksf})},  \ldots, \av_{\Cc(<k+\Usf-1>_{\Ksf})}]$ has    dimension equal to $\Usf \times (\Usf-1)$, and that its elements are  uniformly and i.i.d. over $\mathbb{F}_{\qsf}$. So the left null space contains 
    $\Usf-(\Usf-1)=1$ linearly independent  $\Usf$-dimensional vector with high probability, and we let $\sv_k$ be this vector. 
Hence, the constraints in~\eqref{eq:constraint of a for null space} are satisfied with high probability.

\paragraph*{Constraint in~\eqref{eq:decodability constraint}}
Recall that we only need to consider the case where  $\Uc_1=[\Ksf]$.
In the  second round transmission, the server should  recover  $\Usf$ linear combinations of coded keys, 
$$
   \begin{bmatrix}
F_1  \\
\vdots\\
F_{\Usf}
\end{bmatrix} 
= \left[ \av_{\Cc(1)}, \ldots, \av_{\Cc(\Ksf)} \right]
   \begin{bmatrix}
Z^{[\Ksf]}_{\Cc(1)} \\
\vdots\\
Z^{[\Ksf]}_{\Cc(\Ksf)} 
\end{bmatrix},
$$
from the answers of any $\Usf$ of the $\Ksf$ users, each of whom knows $\Ksf-\Usf+1$ coded keys in a cyclic way.   
This problem is   equivalent to    the distributed linearly separable computation problem in~\cite{linearcomput2020wan}, where we aim to compute $\Usf$ linear combinations of $\Ksf$ messages (whose coefficients are uniformly and i.i.d. over $\mathbb{F}_{\qsf}$) through $\Ksf$ distributed computing nodes, each of which can stores $\Ksf-\Usf+1$ messages, such that from the answers of any $\Usf$ nodes we can recover the computing task.  
From~\cite[Lemma~2]{linearcomput2020wan}, we have the following lemma.
\begin{lem}[\cite{linearcomput2020wan}]
\label{lem:from distributed computing}
For any set $\Ac\in \binom{[\Ksf]}{\Usf}$, the vectors 
$
\sv_{n} , n\in \Ac, 
$
  are linearly independent with high probability.
  \hfill $\square$ 
\end{lem}
  Thus by Lemma~\ref{lem:from distributed computing},  the   constraint in~\eqref{eq:decodability constraint} is satisfied with high probability.

In conclusion, all constraints in~\eqref{eq:full rank constraint},~\eqref{eq:constraint of a for null space}, and~\eqref{eq:decodability constraint} are satisfied with high probability.  Hence, there must exist a choice of $\left[\av_{\Cc(1)},  \ldots, \av_{\Cc(\Ksf)} \right]$ satisfying those constraints.
Thus 
the proposed scheme is decodable and secure. 
 In this case, we need the keys   $Z_{\Vc}$ where $\Vc \in \Cc$, totally $\Ksf$ keys each of which is shared by $\Ssf$ users.

\subsection{Case $\Usf> \Ksf-\Usf+1$ and $\Usf=\Ksf-1$}
\label{sub:U>S U=K-1}
When $\Usf>\Ssf$, the proposed secure aggregation scheme with cyclic assignment does not work. This is because, 
among $\Cc$, the number of sets containing each $k\in [\Ksf]$ is $\Ksf-\Usf+1<\Usf$, which are  $\Cc(k), \Cc(<k-1>_{\Ksf})  ,\ldots,  \Cc(<k-\Ksf+\Usf>_{\Ksf})  $. 
Hence,  the    coefficient matrix of keys in $X_k$, 
$\left[\av_{\Cc(k)},\av_{\Cc(<k-1>_{\Ksf})}, \ldots, \av_{\Cc(<k-\Ksf+\Usf>_{\Ksf})} \right]$, is with dimension $\Usf\times (\Ksf-\Usf+1)$ and with rank strictly less than $\Usf$. Thus the constraint in~\eqref{eq:full rank constraint} is not satisfied. In other words,   $W_k$ is not perfectly protected from $X_{k}$. 
 
 In this subsection, we present our proposed secure aggregation scheme for the case where $\Usf> \Ksf-\Usf+1$ and $\Usf=\Ksf-1$. We first illustrate the main idea through the following example.
 \begin{example}[$(\Ksf,\Usf, \Ssf))=(4,3,2)$]
\label{ex:case 2}
\rm
Consider  the  $(\Ksf,\Usf,\Ssf)=(4,3,2)$ information theoretic secure aggregation problem with uncoded groupwise keys. 
For each $\Vc\in \binom{[4]}{2}$,  we generate a key $Z_{\Vc}=(Z_{\Vc,k}:k\in \Vc)$ shared by users in $\Vc$, where each $Z_{\Vc,k}$ contains $\Lsf/3$ uniform and i.i.d. symbols over $\mathbb{F}_{\qsf}$.
We also divide each input vector $W_k$ where $k\in [4]$ into three pieces, $W_{k}=(W_{k,1},W_{k,2},W_{k,3})$, where each piece contains $\Lsf/3$ uniform and i.i.d.  symbols over $\mathbb{F}_{\qsf}$.

In the first round, each user $k\in [4]$ transmits 
\begin{align}
X_{k,j} = W_{k,j} + \sum_{\Vc\in \binom{[4]}{2}: k \in \Vc} a_{\Vc, j} Z_{\Vc,k}  , \ \forall j \in [3].
\end{align}
Now we select the $3$-dimensional vectors $\av_{\{1,2\}}$, $\av_{\{1,3\}}$, $\av_{\{1,4\}}$, $\av_{\{2,3\}}$, $\av_{\{2,4\}}$, and $\av_{\{3,4\}}$ as follows,
\begin{subequations}
\begin{align}
&\av_{\{1,2\}} = [1,0,0]^{\text{T}}, \ \av_{\{1,3\}} = [0,1,0]^{\text{T}},  \ \av_{\{1,4\}} = [0,0,1]^{\text{T}},\\
&\av_{\{2,3\}}=\av_{\{1,2\}}-\av_{\{1,3\}}=[1,-1,0]^{\text{T}}, \ \av_{\{2,4\}}=\av_{\{1,2\}}-\av_{\{1,4\}}=[1,0,-1]^{\text{T}}, \\ 
&\av_{\{3,4\}}=\av_{\{1,3\}}-\av_{\{1,4\}}=[0,1,-1]^{\text{T}}.
\end{align}
\label{eq:example 2 selected vectors}
\end{subequations}
We next show that by the above choice the constraints in~\eqref{eq:full rank constraint},~\eqref{eq:constraint of a for null space}, and~\eqref{eq:decodability constraint} are satisfied.

For user $1$, the   matrix $[\av_{\{1,2\}},\av_{\{1,3\}},\av_{\{1,4\}}]=\mathbf{I}_3$ has rank $3$, where we recall that $\mathbf{I}_3$ represents the identity matrix with dimension $3\times 3$. Hence, the constraint in~\eqref{eq:full rank constraint} is satisfied for user $1$. Thus $W_1$ is perfectly protected by $(Z_{\{1,2\},1},  , Z_{\{1,3\},1}, Z_{\{1,4\},1} )$ from $X_1$. 
For user $2$, the     matrix $[\av_{\{1,2\}},\av_{\{2,3\}},\av_{\{2,4\}}]=\begin{bmatrix}
1&1&1 \\
0&-1&0 \\
0&0&-1
\end{bmatrix}$ has rank $3$. Hence, the constraint in~\eqref{eq:full rank constraint} is satisfied for user $2$. Thus $W_2$ is perfectly protected by $(Z_{\{1,2\},2},  , Z_{\{2,3\},2}, Z_{\{2,4\},2} )$ from $X_2$. 
Similarly, 
the   constraints in~\eqref{eq:full rank constraint}  are also satisfied for users $3,4$.
 
In the second round,  we only need to consider the case $\Uc_1=[4]$, where the server should recover $W_{1}+\cdots+W_{4}$. 
By defining   the coded keys  as in~\eqref{eq:def of coded key}, 
the server needs to further recover 
\begin{align}
\begin{bmatrix}
F_1 \\
F_2 \\
F_3
\end{bmatrix}=[\av_{\{1,2\}}, \av_{\{1,3\}}, \av_{\{1,4\}}, \av_{\{2,3\}},\av_{\{2,4\}},\av_{\{3,4\}}] \begin{bmatrix}
Z^{[4]}_{\{1,2\}} \\
Z^{[4]}_{\{1,3\}} \\
Z^{[4]}_{\{1,4\}}\\
Z^{[4]}_{\{2,3\}} \\
Z^{[4]}_{\{2,4\}} \\
Z^{[4]}_{\{3,4\}}
\end{bmatrix}
=
\begin{bmatrix}
1&0&0&1&1&0\\
0&1&0&-1&0&1\\
0&0&1&0&-1&-1
\end{bmatrix}
\begin{bmatrix}
Z^{[4]}_{\{1,2\}} \\
Z^{[4]}_{\{1,3\}} \\
Z^{[4]}_{\{1,4\}}\\
Z^{[4]}_{\{2,3\}} \\
Z^{[4]}_{\{2,4\}} \\
Z^{[4]}_{\{3,4\}}
\end{bmatrix}.
\end{align}
For user $1$ who cannot encode $Z^{[4]}_{\{2,3\}},Z^{[4]}_{\{2,4\}},Z^{[4]}_{\{3,4\}}$, it can be seen that the sub-matrix 
$[\av_{\{2,3\}},\av_{\{2,4\}},\av_{\{3,4\}}]$ has rank $2$, equal to the rank of $ [\av_{\{2,3\}},\av_{\{2,4\}}]$,  since $\av_{\{2,3\}}-\av_{\{2,4\}}=-\av_{\{3,4\}}$;\footnote{\label{foot:IA}In other words,
we align the three vectors $\av_{\{2,3\}},\av_{\{2,4\}},\av_{\{3,4\}}$ into the linear space spanned by $\av_{\{2,3\}}$ and $\av_{\{2,4\}}$.}
 thus the constraint in~\eqref{eq:constraint of a for null space} is satisfied for user $1$.
Hence, the left null     space of $[\av_{\{2,3\}},\av_{\{2,4\}},\av_{\{3,4\}}]$ contains exactly one linearly independent $3$-dimensional vector, which could be $[1,1,1]$.
Thus we let user $1$ compute 
\begin{align}
Y^{[4]}_{1} = [1,1,1]\begin{bmatrix}
F_1 \\
F_2 \\
F_3
\end{bmatrix}=F_1+F_2+F_3.
\end{align}
For user $2$,  who cannot encode $Z^{[4]}_{\{1,3\}},Z^{[4]}_{\{1,4\}},Z^{[4]}_{\{3,4\}}$, it can be seen that the sub-matrix 
$[\av_{\{1,3\}},\av_{\{1,4\}},\av_{\{3,4\}}]$ has rank $2$, equal to the rank of $[\av_{\{1,3\}},\av_{\{1,4\}}]$, 
since 
 $\av_{\{3,4\}}=\av_{\{1,3\}}-\av_{\{1,4\}}$; thus the constraint in~\eqref{eq:constraint of a for null space} is satisfied for user $2$. 
Hence, the left null     space of $[\av_{\{1,3\}},\av_{\{1,4\}},\av_{\{3,4\}}]$ contains exactly one linearly independent $3$-dimensional vector, which could be $[1,0,0]$.
Thus we let user $2$ compute 
 \begin{align}
Y^{[4]}_{2} = [1,0,0]\begin{bmatrix}
F_1 \\
F_2 \\
F_3
\end{bmatrix}=F_1.
\end{align}
 Similarly, 
the constraints in~\eqref{eq:constraint of a for null space} are satisfied for users $3,4$; thus  
 we let user $3$ compute 
  \begin{align}
Y^{[4]}_{3} = [0,1,0]\begin{bmatrix}
F_1 \\
F_2 \\
F_3
\end{bmatrix}=F_2,
\end{align}
 and let user $4$ compute 
   \begin{align}
Y^{[4]}_{4} = [0,0,1]\begin{bmatrix}
F_1 \\
F_2 \\
F_3
\end{bmatrix}=F_3.
\end{align}

It can be seen that any $3$ of  $Y^{[4]}_{1},Y^{[4]}_{2},Y^{[4]}_{3},Y^{[4]}_{4}$ are linearly independent; thus the constraint  in~\eqref{eq:decodability constraint} is satisfied. 
 Hence, for any $\Uc_2\in \binom{[4]}{3}$, the server can recover $F_1,F_2,F_3$ from the second round. Thus from the two round transmissions, the server can recover $W_1+\cdots+W_4$.

Since the constraints   in~\eqref{eq:full rank constraint},~\eqref{eq:constraint of a for null space}, and~\eqref{eq:decodability constraint} are satisfied,
by the security proof in Appendix~\ref{sec:proof of security}, the scheme is secure  for the case $\Uc_1=[4]$. 

In conclusion, the  
achieved rates of the proposed scheme are $(\Rsf_1 ,\Rsf_2 ) = (1,1/3)$, coinciding with Theorem~\ref{thm:main result}. 

\hfill $\square$
\end{example}

We are now ready to generalize the proposed scheme in Example~\ref{ex:case 2} to the case where $\Usf> \Ksf-\Usf+1$ and $\Usf=\Ksf-1$. In this case, we have $\Ssf=2$.
As the previous case, we directly describe the choice of the $\Usf$-dimensional vectors  and show that such choice satisfies the constraints in~\eqref{eq:full rank constraint},~\eqref{eq:constraint of a for null space}, and~\eqref{eq:decodability constraint}.

Let us first consider the sets $\Vc \in \binom{[\Ksf]}{2}$ where $1\in \Vc$.  Each of such sets could be written as $\{1,j\}$, where $j\in [2:\Ksf-1]$.
We let 
\begin{align}
\av_{\{1,j\}}=  \ev_{\Usf,j-1} , \ \forall j\in [2:\Ksf], \label{eq:a1j}
\end{align}
  where   $\ev_{n,i}$ represents the vertical $n$-dimensional unit vector  whose entry in the  $i^{\text{th}} $ position is 1 and 0 elsewhere.
We then consider the sets $\Vc \in \binom{[2:\Ksf]}{2}$.  Each of such sets could be written as $\{i,j\}$, where $1<i<j\leq \Ksf$.
We let  
\begin{align}
\av_{\{i,j\}}=  \av_{\{1,i\}}-  \av_{\{1,j\}}=\ev_{\Usf,i-1} -\ev_{\Usf,j-1} , \  \forall 1<i<j\leq \Ksf. \label{eq:aij}
\end{align}

 Next we will show that the above choice of these   $\Usf$-dimensional vectors  satisfies the constraints in~\eqref{eq:full rank constraint},~\eqref{eq:constraint of a for null space}, and~\eqref{eq:decodability constraint}.

\paragraph*{Constraints in~\eqref{eq:full rank constraint}}
For user $1$, the   matrix $\left[\av_{\{1,2\}},\av_{\{1,3\}}, \ldots, \av_{\{1,\Ksf\}} \right]$ 
is the identity matrix $\mathbf{I}_{\Ksf-1}=\mathbf{I}_{\Usf}$, whose rank is $\Usf$; thus the constraint  in~\eqref{eq:full rank constraint} is  satisfied for user $1$.
For each user $k\in [2:\Ksf]$, by a simple linear transform on the      matrix 
\begin{align}
\left[\av_{\{1,k\}},\av_{\{2,k\}}, \ldots, \av_{\{k-1,k\}}, \av_{\{k,k+1\}},  \av_{\{k,k+2\}}, \ldots \av_{\{k,\Ksf\}} \right], \label{eq:coef matrix for k case 2}
\end{align}
  we obtain the matrix 
  \begin{align*}
&   [\av_{\{1,k\}}+\av_{\{2,k\}}, \av_{\{1,k\}}+\av_{\{3,k\}},  \ldots, \av_{\{1,k\}}+\av_{\{k-1,k\}}, \av_{\{1,k\}},  \av_{\{1,k\}}-\av_{\{k,k+1\}},\av_{\{1,k\}}-\av_{\{k,k+2\}}, \nonumber\\& \ldots, \av_{\{1,k\}}-  \av_{\{k,\Ksf\}}   ] \\
& =  [\ev_{\Usf,1},\ev_{\Usf,2}, \ldots, \ev_{\Usf,k-2}, \ev_{\Usf,k-1},  \ev_{\Usf,k}, \ev_{\Usf,k+1}, \ldots, \ev_{\Usf,\Ksf-1}] ,
  \end{align*}
  which 
 is the identity matrix $\mathbf{I}_{\Ksf-1}=\mathbf{I}_{\Usf}$ with rank equal to $\Usf$, which is also full rank. Hence, the   matrix in~\eqref{eq:coef matrix for k case 2} is full rank, with   rank equal to $\Usf$; thus the constraint  in~\eqref{eq:full rank constraint} is  satisfied for user $k$.

\paragraph*{Constraints in~\eqref{eq:constraint of a for null space}}
For user $1$, among the sets in $\Vc\in \binom{[\Ksf]}{2}$, the sets $\{2,3\},\{2,4\},\ldots,\{2,\Ksf\},$ $\{3,4\},\ldots,  \{\Ksf-1,\Ksf\}$ do not contain $1$. It can be seen that the following $\Ksf-2$ vectors, 
\begin{align}
\av_{\{2,3\}}=\ev_{\Usf,1} -\ev_{\Usf,2}, \ \av_{\{2,4\}}=\ev_{\Usf,1} -\ev_{\Usf,3}, \ \ldots, \  \av_{\{2,\Ksf\}}=\ev_{\Usf,1} -\ev_{\Usf,\Ksf-1}, \label{eq:case 2 user 1 U-1 vec}
\end{align}
are linearly independent. In addition, for each set $\{i,j\}$ where $2<i<j\leq \Ksf$, we have 
$
\av_{\{i,j\}} = \av_{\{2,j\}} - \av_{\{2,i\}}.
$
 Hence, the matrix  $\left[ \av_{\overline{\Sc}_{1}(1)}, \ldots, \av_{\overline{\Sc}_{1}\left(\binom{\Ksf-1}{2}\right)}  \right]$
has rank equal to $\Ksf-2=\Usf-1$,\footnote{\label{foot:recall overline}Recall that for each $k\in[\Ksf]$, the sets $\Vc\in \binom{[\Ksf]\setminus \{k\}}{\Ssf}$   are 
$\overline{\Sc}_{k}(1),\ldots,\overline{\Sc}_k\left(\binom{\Ksf-1}{\Ssf}\right)$.} satisfying the constraint in~\eqref{eq:constraint of a for null space}.

For each user $k\in [2:\Ksf]$, among the sets in $\Vc\in \binom{[\Ksf]}{2}$, the sets $\{1,2\},\{1,3\},\ldots,\{1,k-1\},\{1,k+1\},\ldots, \{1,\Ksf\}$ and the sets $\{i,j\}$ where $1<i<j\leq \Ksf$ and $i,j \neq k$,  do not contain $k$. It can be seen that the following  $\Ksf-2$ vectors, 
\begin{align}
\av_{\{1,2\}}=\ev_{\Usf,1}  ,   \av_{\{1,3\}}=\ev_{\Usf,2} ,   \ldots,   \av_{\{1,k-1\}}=\ev_{\Usf,k-2} ,   \av_{\{1,k+1\}}=\ev_{\Usf,k}, \ldots ,\av_{\{1,\Ksf\}}=\ev_{\Usf,\Ksf-1}, 
 \label{eq:case 2 user k U-1 vec}
\end{align}
are linearly independent. In addition, for each set $\{i,j\}$ where $1<i<j\leq \Ksf$ and $i,j \neq k$, we have 
$
\av_{\{i,j\}} = \av_{\{1,i\}} - \av_{\{1,j\}}.
$
Hence, the matrix  $\left[ \av_{\overline{\Sc}_{k}(1)}, \ldots, \av_{\overline{\Sc}_{k}\left(\binom{\Ksf-1}{2}\right)}  \right]$ 
 has rank equal to $\Ksf-2=\Usf-1$, satisfying the constraint in~\eqref{eq:constraint of a for null space}.

\paragraph*{Constraint in~\eqref{eq:decodability constraint}} 
For user $1$, recall that $\sv_1$ is a  left null space vector of the matrix 
 $\left[ \av_{\overline{\Sc}_{1}(1)}, \ldots, \av_{\overline{\Sc}_{1}\left(\binom{\Ksf-1}{2}\right)}  \right]$, whose    rank is $\Usf-1$. The left null space of $\left[ \av_{\overline{\Sc}_{1}(1)}, \ldots, \av_{\overline{\Sc}_{1}\left(\binom{\Ksf-1}{2}\right)}  \right]$ is the same as that of 
 its column-wise sub-matrix $\left[ \av_{\{2,3\}},\av_{\{2,4\}},\ldots, \av_{\{2,\Ksf\}}  \right]$, whose rank is also $\Usf-1$ and dimension is $\Usf \times (\Usf-1)$. 
 Since  
 $$
 \left[ \av_{\{2,3\}},\av_{\{2,4\}},\ldots, \av_{\{2,\Ksf\}}  \right]=\left[ \av_{\{2,3\}},\av_{\{2,4\}},\ldots, \av_{\{2,\Ksf\}}  \right]
 =[\ev_{\Usf,1} -\ev_{\Usf,2}, \ev_{\Usf,1} -\ev_{\Usf,3}, \ldots,\ev_{\Usf,1} -\ev_{\Usf,\Ksf-1}]
 $$ 
 contains exactly one linearly independent left null space vector, which could be (recall that  ${\bf 1}_{n}$    represents  the vertical $n$-dimensional   vector  whose elements are all $1$)
\begin{align}
\mathbf{1}_{\Usf}
=\sv_{1}.\label{eq:case 2 s1}
\end{align}  
 
For each user $k\in [2:\Ksf]$, $\sv_k$ is a  left null space vector of the matrix 
 $\left[ \av_{\overline{\Sc}_{k}(1)}, \ldots, \av_{\overline{\Sc}_{k}\left(\binom{\Ksf-1}{2}\right)}  \right]$, whose    rank is $\Usf-1$. The left null space of $\left[ \av_{\overline{\Sc}_{k}(1)}, \ldots, \av_{\overline{\Sc}_{k}\left(\binom{\Ksf-1}{2}\right)}  \right]$ is the same as that of 
 its column-wise sub-matrix $\left[ \av_{\{1,2\}}, \av_{\{1,3\}},\ldots, \av_{\{1,k-1\}} , \av_{\{1,k+1\}}, \ldots ,\av_{\{1,\Ksf\}}\right]$, whose rank is also $\Usf-1$ and dimension is $\Usf \times (\Usf-1)$. Since  
 $$
\left[ \av_{\{1,2\}}, \av_{\{1,3\}},\ldots, \av_{\{1,k-1\}} , \av_{\{1,k+1\}}, \ldots ,\av_{\{1,\Ksf\}}\right]
 =[\ev_{\Usf,1}  , \ev_{\Usf,2} , \ldots,\ev_{\Usf,k-2}, \ev_{\Usf,k},\ldots, \ev_{\Usf,\Ksf-1} ]
 $$ 
 contains exactly one linearly independent left null space vector, which could be 
\begin{align}
\ev_{\Usf,k-1} 
=\sv_{k}.\label{eq:case 2 sk}
\end{align}  

From~\eqref{eq:case 2 s1} and~\eqref{eq:case 2 sk}, it can be seen that any $\Usf$ vectors of $\sv_1,\ldots,\sv_{\Ksf}$ are linearly independent; thus the constraint in~\eqref{eq:decodability constraint} is satisfied.
 
In conclusion, all constraints in~\eqref{eq:full rank constraint},~\eqref{eq:constraint of a for null space}, and~\eqref{eq:decodability constraint} are satisfied; thus 
the proposed scheme is decodable and secure.  
 In this case, we need the keys   $Z_{\Vc}$ where $\Vc \in \binom{[\Ksf]}{2}$, totally $\Ksf(\Ksf-1)/2$ keys each of which is shared by $2$ users.
 
\subsection{Case $\Usf> \Ksf-\Usf+1$ and $\Usf<\Ksf-1$}
\label{sub:U>S U<K-1}
Finally, we focus on the most  involved case where  $\Usf> \Ksf-\Usf+1$ and $\Usf<\Ksf-1$. In this case, we have $\Ssf >2$ and $2\Usf>\Ksf+1$.  Recall that 
our objective is to  determine  the $\Usf$-dimensional vectors $\av_{\Vc}$ where $\Vc\in \binom{[\Ksf]}{\Ssf}$, such that  the constraints in~\eqref{eq:full rank constraint},~\eqref{eq:constraint of a for null space}, and~\eqref{eq:decodability constraint} are satisfied.
We start by illustrating the main idea through an example.
 \begin{example}[$(\Ksf,\Usf, \Ssf))=(6,4,3)$]
\label{ex:case 3}
\rm
Consider  the  $(\Ksf,\Usf,\Ssf)=(6,4,3)$ information theoretic secure aggregation problem with uncoded groupwise keys. 
We determine the $4$-dimensional vectors $\av_{\Vc}$ where $\Vc\in \binom{[6]}{3}$  as follows. 

We first consider each $\av_{\Vc}$ where $[2] \subseteq \Vc$ and let  $\av_{\Vc}$ be a distinct vertical unit vector; i.e.,  we  let 
\begin{align}
& \av_{[3]}= \ev_{4,1}, \  \av_{\{1,2,4\}}= \ev_{4,2}, \  \av_{\{1,2,5\}}= \ev_{4,3}, \  \av_{\{1,2,6\}}= \ev_{4,4}.\label{eq:ex 3 a[2]}
\end{align}
Define that $\Gc_1=\{[3],\{1,2,4\},\{1,2,5\},\{1,2,6\}\}$.

Then for each  $\av_{\Vc}$ where  $\Vc\in \binom{[6]}{3}\setminus \Gc_1$, we search for the   minimum   subset   of $\Gc_1$ the union of whose elements  is a super-set of $\Vc$; we denote this minimum   subset by $\Mc_{\Vc}$. For example, if $\Vc=\{1,3,4\}$, the   minimum   subset  of  $\Gc_1$ the union of whose elements  is a super-set of $\{1,3,4\}$,  is $\Mc_{\{1,3,4\}}=\{[3],\{1,2,4\}\}$, since $[3] \cup \{1,2,4\}=[4] \supseteq \{1,3,4\}$.  Then we let 
$\av_{\Vc}$ be a linear combination of $\av_{\Vc_1}$ where $\Vc_1 \in \Mc_{\Vc}$; i.e., (assume that the sets in $\Mc_{\Vc}$ are $\Mc_{\Vc}(1),\ldots,\Mc_{\Vc}(|\Mc_{\Vc}|)$) 
\begin{align}
\av_{\Vc} = b_{\Vc,1}\  \av_{\Mc_{\Vc}(1)} +\cdots + b_{\Vc,|\Mc_{\Vc}|} \ \av_{\Mc_{\Vc}(|\Mc_{\Vc}|)} ,\label{eq:ex3 form of a}
\end{align}
where $\bv_{\Vc}:=(b_{\Vc,1},\ldots,b_{\Vc,|\Mc_{\Vc}|})$ is an $|\Mc_{\Vc}|$-dimensional vector to be designed. 
 
By this rule, we   determine the composition  of each  $\av_{\Vc}$  (i.e., the base vertical unit vectors which compose $\av_{\Vc}$) where  $ \Vc\in \binom{[\Ksf]}{\Ssf}$, as illustrated in Table~\ref{tab:ex3}. 

 \begin{table*}
 \centering
\protect\caption{Choice of $4$-dimensional vectors $\av_{\Vc}$ in the  $(\Ksf,\Usf,\Ssf)=(6,4,3)$ information theoretic secure aggregation problem. }\label{tab:ex3}
\begin{tabular}{|c|c|c||c|c|c|}
\hline 
$\av_{\Vc}$  &  Composition & Value & $\av_{\Vc}$  &  Composition & Value 
 \tabularnewline  
\hline 
\hline 
  $\av_{[3]} $  &  $\ev_{4,1}$  & $\ev_{4,1}$ &   $\av_{\{2,3,4\}} $  &  $\ev_{4,1},\ev_{4,2}$  & $\ev_{4,1}+8 \ev_{4,2}$ \tabularnewline  
\hline 
  $\av_{\{1,2,4\}} $  &  $\ev_{4,2}$  & $\ev_{4,2}$  &   $\av_{\{2,3,5\}} $  &  $\ev_{4,1},\ev_{4,3}$  & $7\ev_{4,1}+8\ev_{4,3}$  \tabularnewline  
\hline 
  $\av_{\{1,2,5\}} $  &  $\ev_{4,3}$  & $\ev_{4,3}$     &   $\av_{\{2,3,6\}} $  &  $\ev_{4,1},\ev_{4,4}$  &   $3 \ev_{4,1}+4\ev_{4,4}$   \tabularnewline 
\hline 
  $\av_{\{1,2,6\}} $  &  $\ev_{4,4}$  & $\ev_{4,4}$  &   $\av_{\{2,4,5\}} $  &  $\ev_{4,2},\ev_{4,3}$  & $\mathbf{0}_{4}$ \tabularnewline  
\hline  
  $\av_{\{1,3,4\}} $  &  $\ev_{4,1},\ev_{4,2}$  & $\ev_{4,1}+4\ev_{4,2}$  &   $\av_{\{2,4,6\}} $  &  $\ev_{4,2},\ev_{4,4}$  & $\mathbf{0}_{4}$ \tabularnewline  
\hline 
  $\av_{\{1,3,5\}} $  &  $\ev_{4,1},\ev_{4,3}$  & $3\ev_{4,1} +4\ev_{4,3}$  &   $\av_{\{2,5,6\}} $  &  $\ev_{4,3},\ev_{4,4}$  & $\mathbf{0}_{4}$ \tabularnewline  
\hline 
  $\av_{\{1,3,6\}} $  & $\ev_{4,1},\ev_{4,4}$ & $ \ev_{4,1}+2\ev_{4,4}$   &   $ \av_{\{3,4,5\}} $  &  $\ev_{4,1},\ev_{4,2}, \ev_{4,3}$  &$\ev_{4,1}+\ev_{4,2}+ \ev_{4,3}$  \tabularnewline  
\hline 
  $\av_{\{1,4,5\}} $  &  $\ev_{4,2},\ev_{4,3}$ & $\mathbf{0}_{4}$  &   $\av_{\{3,4,6\}} $  &  $\ev_{4,1},\ev_{4,2}, \ev_{4,4}$  &$\ev_{4,1}+2\ev_{4,2}+ \ev_{4,4}$   \tabularnewline  
\hline 
  $\av_{\{1,4,6\}} $  & $\ev_{4,2},\ev_{4,4}$  & $\mathbf{0}_{4}$ &   $\av_{\{3,5,6\}} $  &  $\ev_{4,1},\ev_{4,3}, \ev_{4,4}$  &  $\ev_{4,1}+2\ev_{4,3}-\ev_{4,4}$ \tabularnewline  
\hline 
  $\av_{\{1,5,6\}} $  & $\ev_{4,3},\ev_{4,4}$ & $\mathbf{0}_{4}$  &   $\av_{\{4,5,6\}} $  &  $\ev_{4,2},\ev_{4,3}, \ev_{4,4}$  & $\mathbf{0}_{4}$  \tabularnewline  
\hline 
\end{tabular}
\end{table*}

Next we need to determine the coefficient vector   of the vertical base unit vectors    $\bv_{\Vc}$
  for each $\Vc \in \binom{[6]}{3} \setminus \Gc_1$. 
 
 For each set     $\av_{\Vc}$ where $\{3,4\} \subseteq \Vc$, we choose each element of $\bv_{\Vc}$ uniformly and i.i.d. over $\mathbb{F}_{\qsf}$. 
For example, by choosing $\bv_{\{1,3,4\}}=[1,4]$, we have 
\begin{align}
\av_{\{1,3,4\}}=\av_{[3]}+4\av_{\{1,2,4\}}= \ev_{4,1}+4\ev_{4,2}.\label{eq:ex3 a134}
\end{align}
Similarly, by choosing  $\bv_{\{2,3,4\}}=[1,8]$, $\bv_{\{3,4,5\}}=[1,1,1]$, and $\bv_{\{3,4,6\}}=[1,2,1]$, we have 
\begin{subequations}
\begin{align}
&\av_{\{2,3,4\}}=\av_{[3]}+8\av_{\{1,2,4\}}= \ev_{4,1}+8\ev_{4,2},\label{eq:ex3 a234} \\
& \av_{\{3,4,5\}}=\av_{[3]}+ \av_{\{1,2,4\}}+ \av_{\{1,2,5\}}= \ev_{4,1}+\ev_{4,2}+\ev_{4,3},\label{eq:ex3 a345} \\
& \av_{\{3,4,6\}}=\av_{[3]}+ 2\av_{\{1,2,4\}}+ \av_{\{1,2,6\}}= \ev_{4,1}+2\ev_{4,2}+\ev_{4,4}.\label{eq:ex3 a346} 
\end{align} 
\label{eq:ex3 34}
 \end{subequations} 
Define that $\Gc_2=\{\{1,3,4\},\{2,3,4\},\{3,4,5\},\{3,4,6\}\}$. 

 For each set $\Vc \in \binom{[6]}{3} \setminus (\Gc_1\cup \Gc_2)$ where $3\in \Vc$, we search for the   minimum   subset   of $\Gc_2$ the union of whose elements  is a super-set of $\Vc$; we denote this minimum   subset by $\Mc^{\prime}_{\Vc}$. 
 We let $\av_{\Vc}$ be a linear combination of $\av_{\Vc_2}$ where $\Vc_2 \in \Mc^{\prime}_{\Vc}$.
For example, if $\Vc=\{1,3,5\}$, the   minimum   subset  of  $\Gc_2$ the union of whose elements  is a super-set of $\{1,3,5\}$, is $\Mc^{\prime}_{\{1,3,5\}}=\{\{1,3,4\},\{3,4,5\}\}$.   We let $\av_{\{1,3,5\}}$ be a linear combination of $\av_{\{1,3,4\}}=\ev_{4,1}+4\ev_{4,2}$ and $\av_{\{3,4,5\}}=\ev_{4,1}+\ev_{4,2}+\ev_{4,3}$. Recall  from~\eqref{eq:ex3 form of a} that,  the base vertical unit vectors of $\av_{\{1,3,5\}}$ are $\ev_{4,1}$ and $\ev_{4,3}$, which do not contain $\ev_{4,2}$.  Hence, we let 
\begin{align}
\av_{\{1,3,5\}}= 4\av_{\{3,4,5\}} - \av_{\{1,3,4\}}=3\ev_{4,1} +4\ev_{4,3},\label{eq:ex3 a135}
\end{align}
to `zero-force' the term $\ev_{4,2}$. 
 Similarly, we let 
 \begin{subequations}
 \begin{align}
 &\av_{\{1,3,6\}} =  2\av_{\{3,4,6\}}-  \av_{\{1,3,4\}}=       \ev_{4,1}+2\ev_{4,4}, \label{eq:ex3 a136}\\
 &\av_{\{2,3,5\}} = 8 \av_{\{3,4,5\}}- \av_{\{2,3,4\}}= 7\ev_{4,1}+8\ev_{4,3}, \label{eq:ex3 235}\\
 &\av_{\{2,3,6\}}=  4 \av_{\{3,4,6\}}-\av_{\{2,3,4\}}=3 \ev_{4,1}+4\ev_{4,4}, \label{eq:ex3 236}\\
 &\av_{\{3,5,6\}}= 2\av_{\{3,4,5\}}- \av_{\{3,4,6\}}=  \ev_{4,1}+2\ev_{4,3}-\ev_{4,4},\label{eq:ex3 356}
 \end{align}
  \end{subequations} 
to `zero-force' the term $\ev_{4,2}$. 

Finally, each set $\Vc \in \binom{[6]}{3} \setminus (\Gc_1\cup \Gc_2)$ where $3\notin \Vc$, we let $\av_{\Vc}=\mathbf{0}_4$, where   $\mathbf{0}_{n}$ represents  the vertical $n$-dimensional   vector  whose elements are  all $0$.

As a result, we have determined $\av_{\Vc}$ for each $\Vc \in \binom{[6]}{3}$ as illustrated in Table~\ref{tab:ex3}. We then show the such choice satisfies  
 the constraints   in~\eqref{eq:full rank constraint},~\eqref{eq:constraint of a for null space}, and~\eqref{eq:decodability constraint}.

\paragraph*{Constraints in~\eqref{eq:full rank constraint}}
For  users  $1,2$, the   matrix $\left[\av_{[3]},\av_{\{1,2,4\}},  \av_{\{1,2,5\}} ,\av_{\{1,2,6\}} \right]$ 
is the identity matrix $\mathbf{I}_{4}$ whose rank is $4$.
For users  $3,4$,  the   matrix $\left[\av_{\{1,3,4\}},\av_{\{2,3,4\}},  \av_{\{3,4,5\}} ,\av_{\{3,4,6\}} \right]$ has rank equal to $4$.
For  user $5$,  the   matrix $\left[\av_{\{1,3,5\}},\av_{\{2,3,5\}},  \av_{\{3,4,5\}} ,\av_{\{3,5,6\}} \right]$ has rank equal to $4$.
For user $6$,  the   matrix $\left[\av_{\{1,3,6\}},\av_{\{2,3,6\}},  \av_{\{3,4,6\}} ,\av_{\{3,5,6\}} \right]$ has rank equal to $4$.
 Hence, the constraints  in~\eqref{eq:full rank constraint} are  satisfied.

 \paragraph*{Constraints in~\eqref{eq:constraint of a for null space}}
For user $1$, we first remove the columns of $0$'s from the   matrix  $\left[ \av_{\overline{\Sc}_{1}(1)}, \ldots, \av_{\overline{\Sc}_{1}\left(\binom{\Ksf-1}{3}\right)}  \right]$, to obtain 
\begin{align}
\left[ \av_{\{2,3,4\}}, \av_{\{2,3,5\}}, \av_{\{2,3,6\}}, \av_{\{3,4,5\}},\av_{\{3,4,6\}}, \av_{\{3,5,6\}}  \right].\label{eq:ex3 user 1 constraint 2}
\end{align}
By construction, we have $\av_{\{2,3,5\}},\av_{\{2,3,6\}},\av_{\{3,5,6\}}$ are linear combinations of $ \av_{\{2,3,4\}},\av_{\{3,4,5\}},\av_{\{3,4,6\}}$. In addition, 
 $ \av_{\{2,3,4\}},\av_{\{3,4,5\}},\av_{\{3,4,6\}}$ are linearly independent. Hence, the rank of the matrix in~\eqref{eq:ex3 user 1 constraint 2} is $3$, equal to the rank of $[\av_{\{2,3,4\}},\av_{\{3,4,5\}},\av_{\{3,4,6\}}]$.
 Hence, the constraint  in~\eqref{eq:constraint of a for null space} is satisfied for user $1$. Similarly, this constraint is also satisfied for user $2$.
 
 For user $3$, by construction, in each $\av_{\Vc}$ where $\Vc\in \binom{[6]\setminus \{3\}}{3}$, the coefficient of $\ev_{4,1}$ is $0$.  In addition, 
 $   \av_{\{1,2,4\}},\av_{\{1,2,5\}},\av_{\{1,2,6\}}$ are linearly independent.  Thus the   matrix
  $\left[ \av_{\overline{\Sc}_{3}(1)}, \ldots, \av_{\overline{\Sc}_{3}\left(\binom{\Ksf-1}{3}\right)}  \right]$  has rank equal to $3$, equal to the rank of $[\av_{\{1,2,4\}},\av_{\{1,2,5\}},\av_{\{1,2,6\}}]$. 
    Hence, the constraint  in~\eqref{eq:constraint of a for null space} is satisfied for user $3$. 
  Similarly, this constraint is also satisfied for each user in $\{4,5,6\}$.

 \paragraph*{Constraint in~\eqref{eq:decodability constraint}} 
For user $1$, recall that $\sv_1$ is a  left null space vector of the matrix 
 $\left[ \av_{\overline{\Sc}_{1}(1)}, \ldots, \av_{\overline{\Sc}_{1}\left(\binom{\Ksf-1}{3}\right)}  \right]$, whose    rank is $3$. 
As explained before, its column-wise submatrix\\  $[  \av_{\{2,3,4\}},\av_{\{3,4,5\}},\av_{\{3,4,6\}}]$ has the same rank. Hence, 
the left null space  of $[ \av_{\{2,3,4\}},\av_{\{3,4,5\}},\av_{\{3,4,6\}}]$ is the same as that of $\left[ \av_{\overline{\Sc}_{1}(1)}, \ldots, \av_{\overline{\Sc}_{1}\left(\binom{\Ksf-1}{3}\right)}  \right]$. 
So we let  $\sv_1$ be a  left null space vector of $[  \av_{\{2,3,4\}},\av_{\{3,4,5\}},\av_{\{3,4,6\}}]$, which could be $\sv_1=[-8, 1, 7, 6]^{\text{T}}$.
Similarly, we let  $\sv_2$ be a  left null space vector of $[  \av_{\{1,3,4\}},\av_{\{3,4,5\}},\av_{\{3,4,6\}}]$, which could be $\sv_2=[-4, 1, 3, 2]^{\text{T}}$;
we let  $\sv_3$ be a  left null space vector of $[  \av_{\{1,2,4\}},\av_{\{1,2,5\}},\av_{\{1,2,6\}}]$, which could be $\sv_3=\ev_{4,1}$;
 we let  $\sv_4$ be a  left null space vector of $[  \av_{\{1,2,3\}},\av_{\{1,2,5\}},\av_{\{1,2,6\}}]$, which could be $\sv_4=\ev_{4,2}$;
we let  $\sv_5$ be a  left null space vector of $[  \av_{\{1,2,3\}},\av_{\{1,2,4\}},\av_{\{1,2,6\}}]$, which could be $\sv_5=\ev_{4,3}$;
 we let  $\sv_6$ be a  left null space vector of $[  \av_{\{1,2,3\}},\av_{\{1,2,4\}},\av_{\{1,2,5\}}]$, which could be $\sv_6=\ev_{4,4}$.
 
 Since any two rows of $[\sv_1,\sv_2]$ are linearly independent and $[\sv_3,\sv_4,\sv_5,\sv_6]=\mathbf{I}_{4}$, we can see that any $4$ vectors of $\sv_1,\sv_2,\sv_3,\sv_4,\sv_5,\sv_6$ are linearly independent. Hence, the  constraint in~\eqref{eq:decodability constraint} is satisfied.
 
 In conclusion, all constraints in~\eqref{eq:full rank constraint},~\eqref{eq:constraint of a for null space}, and~\eqref{eq:decodability constraint} are satisfied; thus 
the proposed scheme is decodable and secure.  
\hfill $\square$

\end{example}

To summarize Example~\ref{ex:case 3}, our selection on the $\Usf$-dimensional vectors $\av_{\Vc}$ where $\Vc \in \binom{[\Ksf]}{\Ssf} $,  contains the following steps from a high-level viewpoint:
\begin{itemize}
\item  {\it First step.} Choose $\av_{\Vc}$ where $ [\Ksf-\Usf] \subseteq \Vc$ as  the base vertical unit vectors.
\item   {\it Second step.} Fix the composition  of each $\av_{\Vc}$ where $ [\Ksf-\Usf] \nsubseteq \Vc$.
\item  {\it Third step.} For   each $\av_{\Vc}$ where $ [\Ksf-\Usf] \nsubseteq \Vc$, determine the coefficients of the base vertical unit vectors which compose $\av_{\Vc}$.
 \end{itemize}

In the following, we describe the  three-step vector selection for the general case where  $\Usf> \Ksf-\Usf+1$ and $\Usf<\Ksf-1$ in detail.

{\it First step.}
For each $j\in [\Ksf-\Usf+1: \Ksf]$, we let 
\begin{align}
\av_{[\Ksf-\Usf]\cup\{j\}}=\ev_{\Usf,j-\Ksf+\Usf}.\label{eq:first class unit vector}
\end{align}
  In other words, we let $[\av_{[\Ksf-\Usf]\cup\{\Ksf-\Usf+1\}}, \av_{[\Ksf-\Usf]\cup\{\Ksf-\Usf+2\}},\ldots, \av_{[\Ksf-\Usf]\cup \{\Ksf\}}]$ be the identity matrix $\mathbf{I}_{\Usf}$.

For the ease of   notation, we define that\footnote{\label{foot:G1 ex}For example, when $(\Ksf,\Usf, \Ssf)=(8,5,4)$, we have $\Gc_1=\{[4],\{1,2,3,5\},\{1,2,3,6\},\{1,2,3,7\},\{1,2,3,8\}\}.$ } 
$$\Gc_1:=\{ [\Ksf-\Usf]\cup\{j\}  : j\in [\Ksf-\Usf+1: \Ksf]\}.$$
It can be seen that 
\begin{align}
|\Gc_1|= \Usf.\label{eq:length of G1}
\end{align}

{\it Second step.}
For each  $\av_{\Vc}$ where $\Vc \in \binom{[\Ksf]}{\Ssf} \setminus \Gc_1 $, we search for the   minimum   subset   of  $\Gc_1$, 
the union of whose elements  is a super-set of $\Vc$; we denote this minimum   subset by $\Mc_{\Vc}$.
  Then we determine the composition of  $\av_{\Vc}$,  by letting  
$\av_{\Vc}$ be a linear combination of $\av_{\Vc_1}$ where $\Vc_1 \in \Mc_{\Vc}$;  
 i.e.,
\begin{align}
\av_{\Vc} =  b_{\Vc,1} \  \av_{\Mc_{\Vc}(1)} +\cdots + b_{\Vc,|\Mc_{\Vc}|} \ \av_{\Mc_{\Vc}(|\Mc_{\Vc}|)}, \label{eq:general case composition} 
\end{align} 
where $\bv_{\Vc}:=(b_{\Vc,1},\ldots,b_{\Vc,|\Mc_{\Vc}|})$ is a $|\Mc_{\Vc}|$-dimensional vector to be designed. 

{\it Third step.}
 We divide the sets in $\binom{[\Ksf]}{\Ssf}\setminus \Gc_1$ into three classes, which are then considered sequentially. In short, for each set $\Vc$ in the first class (denoted by $\Gc_2$ to be clarified later), we choose $\bv_{\Vc}$ uniformly and i.i.d. over $\mathbb{F}_{\qsf}^{|\Mc_{\Vc}|}$; for each set $\Vc$ in the second class (denoted by $\Gc_3$ to be clarified later), we choose $\bv_{\Vc}$ such that $\av_{\Vc}$ is also a linear combination of some 
 vectors $\av_{\Vc_1}$ where $\Vc_1\in \Gc_2$; for each set $\Vc$ in the third class (i.e., $\binom{[\Ksf]}{\Ssf}\setminus (\Gc_1 \cup \Gc_2 \cup \Gc_3)$), we let $\bv_{\Vc}$ be a all-zero vector. More precisely,  
\begin{itemize}
\item We first consider the sets in\footnote{\label{foot:G2 ex}For example, when $(\Ksf,\Usf, \Ssf)=(8,5,4)$, we have $\Gc_2=\{\{1,4,5,6\},\{2,4,5,6\},\{3,4,5,6\},\{4,5,6,7\},\{4,5,6,8\}\}.$ }  
$$\Gc_2:=\left\{    [\Ksf-\Usf+1:2\Ksf-2\Usf] \cup \{j\} : j\in ([\Ksf-\Usf] \cup [2\Ksf-2\Usf+1: \Ksf])  \right\}.
$$
Recall that $2\Usf>\Ksf+1$, thus  $\Ksf>2\Ksf-2\Usf+1$ and $[2\Ksf-2\Usf+1: \Ksf]$ is not empty.
Since $\Usf<\Ksf-1$, we have $\Ksf-\Usf \geq 2$ and thus $\Gc_1 \cap \Gc_2 =\emptyset$. 
It can be seen that 
\begin{align}
|\Gc_2|= \Ksf-\Usf+ (\Ksf-2\Ksf+2\Usf)=\Usf .\label{eq:length of G2}
\end{align}
For each $\Vc\in \Gc_2$, we choose $\bv_{\Vc}$   uniformly and i.i.d. over $\mathbb{F}^{|\Mc_{\Vc}|}_{\qsf}$. 
More precisely,   
\begin{itemize}
\item  for each $j\in [\Ksf-\Usf]$, by assuming $\Vc=[\Ksf-\Usf+1:2\Ksf-2\Usf]\cup\{j\}$,  
it can be seen that 
$$
\Mc_{\Vc}=\big\{[\Ksf-\Usf]\cup \{\Ksf-\Usf+1\}, [\Ksf-\Usf]\cup \{\Ksf-\Usf+2\},\ldots, [\Ksf-\Usf]\cup \{2\Ksf-2\Usf\}  \big\},
$$
 and thus from~\eqref{eq:general case composition},  $\av_{\Vc}$ is with the form 
\begin{align}
\av_{\Vc}= b_{\Vc,1}  \ \ev_{\Usf,1}+ \cdots+ b_{\Vc, \Ksf-\Usf} \  \ev_{\Usf,\Ksf-\Usf}. \label{eq:class 1,1}
\end{align} 
We let each $ b_{\Vc,i}$, $i\in [\Ksf-\Usf]$, be chosen uniformly and i.i.d. over $\mathbb{F}_{\qsf}$;
\item  for each $j\in [2\Ksf-2\Usf+1: \Ksf]$, by assuming $\Vc=[\Ksf-\Usf+1:2\Ksf-2\Usf]\cup\{j\}$,  
it can be seen that 
\begin{align*}
\Mc_{\Vc}&=\big\{[\Ksf-\Usf]\cup \{\Ksf-\Usf+1\}, [\Ksf-\Usf]\cup \{\Ksf-\Usf+2\},\ldots, [\Ksf-\Usf]\cup \{2\Ksf-2\Usf\} , \\& [\Ksf-\Usf]\cup \{j\}   \big\},  
\end{align*}
 and thus from~\eqref{eq:general case composition},  $\av_{\Vc}$ is with the form 
\begin{align}
\av_{\Vc}= b_{\Vc,1} \  \ev_{\Usf,1}+ \cdots+ b_{\Vc, \Ksf-\Usf} \  \ev_{\Usf,\Ksf-\Usf}+  b_{\Vc, \Ksf-\Usf+1} \  \ev_{\Usf,j-\Ksf+\Usf}. \label{eq:class 1,2}
\end{align} 
We let each $ b_{\Vc,i}$, $i\in [\Ksf-\Usf+1]$, be chosen uniformly and i.i.d. over $\mathbb{F}_{\qsf}$.
\end{itemize}

\item We then consider the sets  in\footnote{\label{foot:G3 ex}For example, when $(\Ksf,\Usf, \Ssf)=(8,5,4)$, we have $\Gc_3=\{\{1,4,5,7\},\{1,4,5,8\},\{2,4,5,7\},\{2,4,5,8\},\{3,4,5,7\},$ $\{3,4,5,8\}, \{4,5,7,8\}\}.$ } 
\begin{align*}
\Gc_3 &:= \Big\{\Tc \cup [\Ksf-\Usf+1:2\Ksf-2\Usf-1]: \Tc \in \binom{[\Ksf-\Usf]\cup [2\Ksf-2\Usf+1:\Ksf]}{2}, \\& \Tc \cap [2\Ksf-2\Usf+1:\Ksf]   \neq \emptyset \Big\}.
\end{align*}
Since $\Ksf-\Usf\geq 2$, we have $\Gc_3 \cap \Gc_1 =\emptyset$; since the integer $ 2\Ksf - 2\Usf$   appears in each set  in $\Gc_2$ and does not appear in any set in $\Gc_3$, we have $\Gc_3 \cap \Gc_2=\emptyset$. It can be seen that 
\begin{align}
|\Gc_3|= \binom{\Ksf-(\Ksf-\Usf)}{2}-\binom{\Ksf-\Usf}{2}=\frac{\Ksf(2\Usf-\Ksf+1)}{2}-\Usf.\label{eq:length of G3}
\end{align}

For each $\Vc \in \Gc_3$, we search for the   minimum   subset   of $\Gc_2$ the union of whose elements  is a super-set of $\Vc$; we denote this minimum   subset by $\Mc^{\prime}_{\Vc}$.  We let $\av_{\Vc}$ be a linear combination of $\av_{\Vc_2}$ where $\Vc_2 \in \Mc^{\prime}_{\Vc}$.

More precisely, for each $ \Tc \in \binom{[\Ksf-\Usf]\cup [2\Ksf-2\Usf+1:\Ksf]}{2}$ where  $\Tc \cap [2\Ksf-2\Usf+1:\Ksf]   \neq \emptyset$,
\begin{itemize}
\item if $\Tc=\{i,j\}$ where $i\in [\Ksf-\Usf]$ and $j\in [2\Ksf-2\Usf+1:\Ksf]$, by assuming $\Vc=[\Ksf-\Usf+1:2\Ksf-2\Usf-1]\cup\{i,j\}$,  
we have 
 \begin{align*}
&   \Mc^{\prime}_{\Vc} =
\big\{[\Ksf-\Usf+1:2\Ksf-2\Usf] \cup \{i\}, [\Ksf-\Usf+1:2\Ksf-2\Usf] \cup \{j\}   \big\}.
\end{align*}
Define $ \Mc^{\prime}_{\Vc}(1)=[\Ksf-\Usf+1:2\Ksf-2\Usf] \cup \{i\}$ and  $ \Mc^{\prime}_{\Vc}(2)=[\Ksf-\Usf+1:2\Ksf-2\Usf] \cup \{j\}$. 
 Hence, we aim to let $\av_{\Vc}$   be a linear combination of 
  \begin{subequations}
\begin{align}
&\av_{\Mc^{\prime}_{\Vc}(1)}=   b_{\Mc^{\prime}_{\Vc}(1),1} \  \ev_{\Usf,1}+ \cdots+ b_{\Mc^{\prime}_{\Vc}(1), \Ksf-\Usf}  \ \ev_{\Usf,\Ksf-\Usf} ,\label{eq:from of ai first case}\\
& \text{and } \av_{\Mc^{\prime}_{\Vc}(2)}= b_{\Mc^{\prime}_{\Vc}(2),1} \  \ev_{\Usf,1}+ \cdots+ b_{\Mc^{\prime}_{\Vc}(2), \Ksf-\Usf} \  \ev_{\Usf,\Ksf-\Usf}+  b_{\Mc^{\prime}_{\Vc}(2), \Ksf-\Usf+1} \  \ev_{\Usf,j-\Ksf+\Usf},\label{eq:from of aj first case}
\end{align} 
\label{eq:from aij first case}
 \end{subequations}
where~\eqref{eq:from of ai first case} and~\eqref{eq:from of aj first case} come from~\eqref{eq:class 1,1} and~\eqref{eq:class 1,2}, respectively. Recall that each element in 
 $\bv_{\Mc^{\prime}_{\Vc}(1)}$ and  $\bv_{\Mc^{\prime}_{\Vc}(2)}$ is chosen uniformly and i.i.d.   over $\mathbb{F}_{\qsf}$.

In addition, we have 
\begin{align*}
 \Mc_{\Vc} =&\big\{[\Ksf-\Usf]\cup \{\Ksf-\Usf+1\}, [\Ksf-\Usf]\cup \{\Ksf-\Usf+2\},\ldots, [\Ksf-\Usf]\cup \{2\Ksf-2\Usf-1\},  \\& [\Ksf-\Usf]\cup \{j\} \big\}.
\end{align*}
 Hence, from~\eqref{eq:general case composition}, $\av_{\Vc}$ is   with the form 
 \begin{subequations}
\begin{align}
 \av_{\Vc} &=b_{\Vc,1} \  \av_{[\Ksf-\Usf]\cup \{\Ksf-\Usf+1\}} + \cdots+ b_{\Vc,\Ksf-\Usf-1} \  \av_{[\Ksf-\Usf]\cup \{2\Ksf-2\Usf-1\}} + b_{\Vc,\Ksf-\Usf} \  \av_{[\Ksf-\Usf]\cup \{j\}} \\
 &= b_{\Vc,1}  \ \ev_{\Usf,1} + \cdots+ b_{\Vc,\Ksf-\Usf-1} \  \ev_{\Usf,\Ksf-\Usf-1}   + b_{\Vc,\Ksf-\Usf}  \ \ev_{\Usf,j-\Ksf+\Usf}.\label{eq:form of Tij first case}
\end{align}
 \end{subequations}

 By comparing~\eqref{eq:from aij first case} with the form of $\av_{\Vc}$ in~\eqref{eq:form of Tij first case}, we need to `zero-force'  $\ev_{\Usf,\Ksf-\Usf}$, which could be done by letting 
 \begin{align}
\av_{\Vc} = b_{\Mc^{\prime}_{\Vc}(2), \Ksf-\Usf}  \  \av_{\Mc^{\prime}_{\Vc}(1)} - b_{\Mc^{\prime}_{\Vc}(1), \Ksf-\Usf}  \ \av_{\Mc^{\prime}_{\Vc}(2)}.\label{eq:value of aij first case}
 \end{align}
 \item if $\Tc=\{i,j\}$ where $2\Ksf-2\Usf+1 \leq i<j \leq \Ksf$, by assuming $\Vc=[\Ksf-\Usf+1:2\Ksf-2\Usf-1]\cup\{i,j\}$,  
it can be seen that 
\begin{align*}
&   \Mc^{\prime}_{\Vc} =
\big\{[\Ksf-\Usf+1:2\Ksf-2\Usf] \cup \{i\}, [\Ksf-\Usf+1:2\Ksf-2\Usf] \cup \{j\}   \big\}.
\end{align*}
 Hence, we aim to let $\av_{\Vc}$   be a linear combination of 
  \begin{subequations}
\begin{align}
&\av_{\Mc^{\prime}_{\Vc}(1)}=   b_{\Mc^{\prime}_{\Vc}(1),1} \  \ev_{\Usf,1}+ \cdots+ b_{\Mc^{\prime}_{\Vc}(1), \Ksf-\Usf}  \ \ev_{\Usf,\Ksf-\Usf} +  b_{\Mc^{\prime}_{\Vc}(1), \Ksf-\Usf+1} \  \ev_{\Usf,i-\Ksf+\Usf} ,\label{eq:from of ai second case}\\
& \text{and } \av_{\Mc^{\prime}_{\Vc}(2)}= b_{\Mc^{\prime}_{\Vc}(2),1} \  \ev_{\Usf,1}+ \cdots+ b_{\Mc^{\prime}_{\Vc}(2), \Ksf-\Usf}  \ \ev_{\Usf,\Ksf-\Usf}+  b_{\Mc^{\prime}_{\Vc}(2), \Ksf-\Usf+1}  \ \ev_{\Usf,j-\Ksf+\Usf},\label{eq:from of aj second case}
\end{align} 
\label{eq:from aij second case}
 \end{subequations}
where~\eqref{eq:from of ai second case} and~\eqref{eq:from of aj second case} come from~\eqref{eq:class 1,2}.

In addition, we have
\begin{align*}
 \Mc_{\Vc} =&\big\{[\Ksf-\Usf]\cup \{\Ksf-\Usf+1\}, [\Ksf-\Usf]\cup \{\Ksf-\Usf+2\},\ldots, [\Ksf-\Usf]\cup \{2\Ksf-2\Usf-1\},    \\& [\Ksf-\Usf]\cup \{i\}, [\Ksf-\Usf]\cup \{j\} \big\}.
\end{align*}
  Hence, from~\eqref{eq:general case composition}, $\av_{\Vc}$ is with the form 
 \begin{subequations}
\begin{align}
 \av_{\Vc}&= b_{\Vc,1}  \ \av_{[\Ksf-\Usf]\cup \{\Ksf-\Usf+1\}} + \cdots+ b_{\Vc,\Ksf-\Usf-1}  \ \av_{[\Ksf-\Usf]\cup \{2\Ksf-2\Usf-1\}} + b_{\Vc,\Ksf-\Usf}  \ \av_{[\Ksf-\Usf]\cup \{i\}} \nonumber \\& + b_{\Vc,\Ksf-\Usf+1}  \ \av_{[\Ksf-\Usf]\cup \{j\}} \\
 &= b_{\Vc,1}  \ \ev_{\Usf,1} + \cdots+ b_{\Vc,\Ksf-\Usf-1}  \ \ev_{\Usf,\Ksf-\Usf-1}  +b_{\Vc,\Ksf-\Usf}  \ \ev_{\Usf,i-\Ksf+\Usf}  + b_{\Vc,\Ksf-\Usf+1} \  \ev_{\Usf,j-\Ksf+\Usf}.\label{eq:form of Tij second case}
\end{align}
 \end{subequations}

 By comparing~\eqref{eq:from aij second case} with the form of $\av_{\Vc}$ in~\eqref{eq:form of Tij second case}, we need to `zero-force'  $\ev_{\Usf,\Ksf-\Usf}$, which could be done by letting 
 \begin{align}
\av_{\Vc} = b_{\Mc^{\prime}_{\Vc}(2), \Ksf-\Usf}  \  \av_{\Mc^{\prime}_{\Vc}(1)} - b_{\Mc^{\prime}_{\Vc}(1), \Ksf-\Usf}  \ \av_{\Mc^{\prime}_{\Vc}(2)}.\label{eq:value of aij second case}
 \end{align}
\item Finally, for each $\Vc\in \binom{[\Ksf]}{\Ssf} \setminus (\Gc_1\cup \Gc_2 \cup \Gc_3)$, we let 
\begin{align}
\av_{\Vc}=\mathbf{0}_{\Usf}. \label{eq:last class all zero}
\end{align}
\end{itemize} 
\end{itemize}

This concludes our selection  on $\av_{\Vc}$ where $\Vc \in \binom{[\Ksf]}{\Ssf}$.
 Next we will show that the above choice of these   $\Usf$-dimensional vectors  satisfies the constraints in~\eqref{eq:full rank constraint},~\eqref{eq:constraint of a for null space}, and~\eqref{eq:decodability constraint}, with high probability.

\paragraph*{Constraints in~\eqref{eq:full rank constraint}}
 For each user $k\in [\Ksf-\Usf]$, the   matrix  $$[\av_{[\Ksf-\Usf]\cup\{\Ksf-\Usf+1\}}, \av_{[\Ksf-\Usf]\cup\{\Ksf-\Usf+2\}},\ldots, \av_{[\Ksf-\Usf]\cup \{\Ksf\}}]$$ is the identity matrix  $\mathbf{I}_{\Usf}$, whose rank is $\Usf$.
 
 For each user $k\in [\Ksf-\Usf+1:2\Ksf-2\Usf]$,  let us focus on the matrix   
 \begin{align}
  [\av_{[\Ksf-\Usf+1:2\Ksf-2\Usf] \cup \{1\}},  \ldots, \av_{[\Ksf-\Usf+1:2\Ksf-2\Usf] \cup \{\Ksf-\Usf\}}, \av_{[\Ksf-\Usf+1:2\Ksf-2\Usf] \cup \{2\Ksf-2\Usf+1\}},\ldots, \av_{[\Ksf-\Usf+1:2\Ksf-2\Usf] \cup \{\Ksf\}}],\label{eq:user k-u+1 full rank matrix}
 \end{align}
 whose dimension is $\Usf \times \Usf$.
 By our construction, for each $j\in [\Ksf-\Usf]$, by~\eqref{eq:class 1,1} we have (assume $\Vc=[\Ksf-\Usf+1:2\Ksf-2\Usf] \cup \{j\}$)
\begin{align}
 \av_{\Vc}= b_{\Vc,1} \  \ev_{\Usf,1}+ \cdots+ b_{\Vc, \Ksf-\Usf}  \ \ev_{\Usf,\Ksf-\Usf},
 \label{eq:user k-u+1 full first case}
\end{align}
where  $ b_{[\Ksf-\Usf+1:2\Ksf-2\Usf] \cup \{j\},i}$, $i\in [\Ksf-\Usf]$, is chosen uniformly and i.i.d. over $\mathbb{F}_{\qsf}$.
In addition, 
 for each $j\in [2\Ksf-2\Usf+1:\Ksf]$, by~\eqref{eq:class 1,2} we have  (assume $\Vc=[\Ksf-\Usf+1:2\Ksf-2\Usf] \cup \{j\}$)
  \begin{align}
\av_{\Vc}& = b_{\Vc,1}  \ \ev_{\Usf,1}+ \cdots+ b_{\Vc, \Ksf-\Usf}  \ \ev_{\Usf,\Ksf-\Usf}  +  b_{\Vc, \Ksf-\Usf+1} \  \ev_{\Usf,j-\Ksf+\Usf}, \label{eq:user k-u+1 full sec case}
\end{align} 
 where each $ b_{[\Ksf-\Usf+1:2\Ksf-2\Usf] \cup \{j\},i}$, $i\in [\Ksf-\Usf+1]$, is chosen uniformly and i.i.d. over $\mathbb{F}_{\qsf}$.
 Since $\qsf$ is large enough, from~\eqref{eq:user k-u+1 full first case} and~\eqref{eq:user k-u+1 full sec case}, it can be seen that the matrix in~\eqref{eq:user k-u+1 full rank matrix} has rank equal to $\Usf$ with high probability.

  For each user $k\in [2\Ksf-2\Usf+1:\Ksf]$,  let us focus on the matrix 
  \begin{align}
  &\big[\av_{\{1\}\cup [\Ksf-\Usf+1:2\Ksf-2\Usf-1]\cup \{k\}}, \av_{\{2\}\cup [\Ksf-\Usf+1:2\Ksf-2\Usf-1]\cup \{k\}},\ldots, \av_{\{\Ksf-\Usf\}\cup [\Ksf-\Usf+1:2\Ksf-2\Usf-1]\cup \{k\}},\nonumber\\
  & \av_{[\Ksf-\Usf+1:2\Ksf-2\Usf-1]\cup\{2\Ksf-2\Usf,k\} }, \av_{[\Ksf-\Usf+1:2\Ksf-2\Usf-1]\cup\{2\Ksf-2\Usf+1,k\}},\ldots, \av_{[\Ksf-\Usf+1:2\Ksf-2\Usf-1]\cup\{ k,\Ksf\} }
   \big], \label{eq:user 2k-2u+1 full rank matrix}
  \end{align}
    whose dimension is $\Usf \times \Usf$.
For each $j\in [\Ksf-\Usf]$, by~\eqref{eq:value of aij first case}, we have 
\begin{align}
\av_{\{j\}\cup [\Ksf-\Usf+1:2\Ksf-2\Usf-1]\cup \{k\}} &= b_{[\Ksf-\Usf+1:2\Ksf-2\Usf]\cup \{k\},\Ksf-\Usf}  \ \av_{\{j\}\cup [\Ksf-\Usf+1:2\Ksf-2\Usf] } \nonumber\\&- 
b_{\{j\}\cup [\Ksf-\Usf+1:2\Ksf-2\Usf] ,\Ksf-\Usf}  \ \av_{ [\Ksf-\Usf+1:2\Ksf-2\Usf] \cup \{k\}},
\label{eq:user 2k-2u+1 full rank matrix j small}
\end{align}
where $b_{[\Ksf-\Usf+1:2\Ksf-2\Usf]\cup \{k\},\Ksf-\Usf}$ and  $b_{\{j\}\cup [\Ksf-\Usf+1:2\Ksf-2\Usf] ,\Ksf-\Usf}$ are chosen uniformly and i.i.d. over $\mathbb{F}_{\qsf}$.
   For each $j\in [2\Ksf-2\Usf+1:\Ksf]\setminus \{k\}$, by~\eqref{eq:value of aij second case}, we have 
   \begin{align}
   &\av_{[\Ksf-\Usf+1:2\Ksf-2\Usf-1]\cup\{j,k\} }= \nonumber\\
&   \begin{cases} b_{[\Ksf-\Usf+1:2\Ksf-2\Usf]\cup \{k\},\Ksf-\Usf}  \ \av_{[\Ksf-\Usf+1:2\Ksf-2\Usf]\cup \{j\}}  - 
b_{[\Ksf-\Usf+1:2\Ksf-2\Usf]\cup \{j\},\Ksf-\Usf}  \ \av_{[\Ksf-\Usf+1:2\Ksf-2\Usf]\cup \{k\}}, & \text{ if }  j<k; 
   \\ b_{[\Ksf-\Usf+1:2\Ksf-2\Usf]\cup \{j\},\Ksf-\Usf}  \ \av_{[\Ksf-\Usf+1:2\Ksf-2\Usf]\cup \{k\}}  - 
b_{[\Ksf-\Usf+1:2\Ksf-2\Usf]\cup \{k\},\Ksf-\Usf}  \ \av_{[\Ksf-\Usf+1:2\Ksf-2\Usf]\cup \{j\}},  & \text{ if }  j>k , \end{cases}
\label{eq:user 2k-2u+1 full rank matrix j large}
\end{align}     
where $b_{[\Ksf-\Usf+1:2\Ksf-2\Usf]\cup \{k\},\Ksf-\Usf}$ and $b_{[\Ksf-\Usf+1:2\Ksf-2\Usf]\cup \{j\},\Ksf-\Usf}$ are chosen uniformly and i.i.d. over $\mathbb{F}_{\qsf}$. 
In addition, as we showed before,  
    $$\av_{[\Ksf-\Usf+1:2\Ksf-2\Usf] \cup \{1\}},  \ldots, \av_{[\Ksf-\Usf+1:2\Ksf-2\Usf] \cup \{\Ksf-\Usf\}}, \av_{[\Ksf-\Usf+1:2\Ksf-2\Usf] \cup \{2\Ksf-2\Usf+1\}},\ldots, \av_{[\Ksf-\Usf+1:2\Ksf-2\Usf] \cup \{\Ksf\}}$$
   which are the columns of the matrix in~\eqref{eq:user k-u+1 full rank matrix},  are linearly independent with high probability. 
    Hence, by~\eqref{eq:user 2k-2u+1 full rank matrix j small},~\eqref{eq:user 2k-2u+1 full rank matrix j large}, and the fact  
     that $ \av_{[\Ksf-\Usf+1:2\Ksf-2\Usf-1]\cup\{2\Ksf-2\Usf,k\} }=\av_{[\Ksf-\Usf+1:2\Ksf-2\Usf]\cup\{k\} }$  is in the matrix in~\eqref{eq:user 2k-2u+1 full rank matrix}, we can see that    the matrix in~\eqref{eq:user 2k-2u+1 full rank matrix} is full rank with high probability.

 Hence, the constraints in~\eqref{eq:full rank constraint} are satisfied with high probability.

\paragraph*{Constraints in~\eqref{eq:constraint of a for null space}}
For each user $k\in [\Ksf-\Usf]$,   the sets in $\Vc\in \binom{[\Ksf]\setminus \{k\}}{\Ssf}$ do not contain $k$. 
By our construction, it can be seen that 
\begin{subequations}
\begin{align}
\binom{[\Ksf]\setminus \{k\}}{\Ssf} \cap \Gc_1 &=\emptyset,\\
  \binom{[\Ksf]\setminus \{k\}}{\Ssf} \cap \Gc_2&=\{\{j\}\cup [\Ksf-\Usf+1:2\Ksf-2\Usf]: j\in [\Ksf] \setminus (\{k\} \cup [\Ksf-\Usf+1:2\Ksf-2\Usf] ) \}, \label{eq:unknown G2}\\
 \binom{[\Ksf]\setminus \{k\}}{\Ssf} \cap \Gc_3&=\Big\{\Tc \cup [\Ksf-\Usf+1:2\Ksf-2\Usf-1]: \Tc \in \binom{([\Ksf-\Usf]\cup [2\Ksf-2\Usf+1:\Ksf]) \setminus \{k\}}{2}, \nonumber\\& \Tc \cap [2\Ksf-2\Usf+1:\Ksf]   \neq \emptyset \Big\}.\label{eq:unknown G3}
\end{align}
\end{subequations}

Focus on the sets in~\eqref{eq:unknown G2}.
Since the matrix in~\eqref{eq:user k-u+1 full rank matrix} is full rank with high probability,  
  the $\Usf-1$  vectors in  
  \begin{align}
  \{\av_{\{j\}\cup [\Ksf-\Usf+1:2\Ksf-2\Usf] } : j\in [\Ksf] \setminus (\{k\} \cup [\Ksf-\Usf+1:2\Ksf-2\Usf] ) \} \label{eq:unknown G2 set}
  \end{align} 
are linearly independent with high probability.  

Focus on the sets in~\eqref{eq:unknown G3}.
For each $\Tc \in \binom{([\Ksf-\Usf]\cup [2\Ksf-2\Usf+1:\Ksf]) \setminus \{k\}}{2}$ where $\Tc \cap [2\Ksf-2\Usf+1:\Ksf]   \neq \emptyset$, 
by assuming that $\Vc=\Tc \cup [\Ksf-\Usf+1:2\Ksf-2\Usf-1]$ and 
 $\Tc=\{i,j\}$ where $i<j$, 
it can be seen from~\eqref{eq:value of aij first case} and~\eqref{eq:value of aij second case} that
 \begin{align}
 \av_{\Vc}= b_{[\Ksf-\Usf+1:2\Ksf-2\Usf]\cup \{j\},\Ksf-\Usf} \ \av_{[\Ksf-\Usf+1:2\Ksf-2\Usf]\cup \{i\}}- b_{[\Ksf-\Usf+1:2\Ksf-2\Usf]\cup \{i\},\Ksf-\Usf} \ \av_{[\Ksf-\Usf+1:2\Ksf-2\Usf]\cup \{j\}},\label{eq:unknown G3 set}
 \end{align}
where both $\av_{[\Ksf-\Usf+1:2\Ksf-2\Usf]\cup \{i\}}$ and $\av_{[\Ksf-\Usf+1:2\Ksf-2\Usf]\cup \{j\}}$ are in~\eqref{eq:unknown G2 set}.
 
 Recall that for   each  set   $\Vc\in \binom{[\Ksf]}{\Ssf} \setminus (\Gc_1\cup\Gc_2\cup\Gc_3)$,  from~\eqref{eq:last class all zero} we have $\av_{\Vc}=\mathbf{0}_{\Usf}$. 
As a result, the matrix  $\left[ \av_{\overline{\Sc}_{k}(1)}, \ldots, \av_{\overline{\Sc}_{k}\left(\binom{\Ksf-1}{\Ssf}\right)}  \right]$ has rank equal to $\Usf-1$ with high probability, which is the same as its column-wise sub-matrix (whose dimension is $\Usf\times (\Usf-1)$)
\begin{align}
 &[ \av_{\{1\}\cup [\Ksf-\Usf+1:2\Ksf-2\Usf] }, \ldots, \av_{\{k-1\}\cup [\Ksf-\Usf+1:2\Ksf-2\Usf] },\av_{\{k+1\}\cup [\Ksf-\Usf+1:2\Ksf-2\Usf] },\ldots, \nonumber\\&
 \av_{\{\Ksf-\Usf\}\cup [\Ksf-\Usf+1:2\Ksf-2\Usf] }, \av_{[\Ksf-\Usf+1:2\Ksf-2\Usf]\cup\{\Ksf-\Usf+1\}},\ldots, \av_{[\Ksf-\Usf+1:2\Ksf-2\Usf]\cup\{\Ksf\}}]
 , \label{eq:unknown G2 matrix}
\end{align}
  where  $\av_{\{j_1\}\cup [\Ksf-\Usf+1:2\Ksf-2\Usf] }$, $j_1\in [\Ksf-\Usf]\setminus \{k\}$ is given in~\eqref{eq:user k-u+1 full first case} and 
$\av_{\{j_2\}\cup [\Ksf-\Usf+1:2\Ksf-2\Usf] }$, $j_2\in [2\Ksf-2\Usf+1:\Ksf] $ is given in~\eqref{eq:user k-u+1 full sec case}.

 For each user $k\in [\Ksf-\Usf+1:\Ksf]$, among the sets in   $\Vc\in \binom{[\Ksf]\setminus \{k\}}{\Ssf}$ which do not contain $k$, we can see that 
 in $\av_{\Vc}$ the coefficient of  $\ev_{\Usf, k-\Ksf+\Usf}$ is $0$. This could be directly checked from  the second step to select the $\Usf$-dimensional vectors,    where we fix the composition of $\av_{\Vc}$ in~\eqref{eq:general case composition}.
 Thus the rank of $\left[ \av_{\overline{\Sc}_{k}(1)}, \ldots, \av_{\overline{\Sc}_{k}\left(\binom{\Ksf-1}{\Ssf}\right)}  \right]$  is no more than $\Usf-1$. In addition, its column-wise sub-matrix  
 \begin{align}
 &[ \av_{[\Ksf-\Usf]\cup\{\Ksf-\Usf+1\} }, \ldots, \av_{[\Ksf-\Usf]\cup\{k-1\} },\av_{[\Ksf-\Usf]\cup\{k+1\} },\ldots, \av_{[\Ksf-\Usf]\cup\{\Ksf\} }]\nonumber\\
 &=[\ev_{\Usf,1},\ldots,\ev_{\Usf,k-\Ksf+\Usf-1},\ev_{\Usf,k-\Ksf+\Usf+1},\ldots,\ev_{\Usf,\Usf}  ]
 , \label{eq:unknown G1 matrix}
\end{align}
 has rank equal to $\Usf-1$. Hence, the rank of $\left[ \av_{\overline{\Sc}_{k}(1)}, \ldots, \av_{\overline{\Sc}_{k}\left(\binom{\Ksf-1}{\Ssf}\right)}  \right]$  is $\Usf-1$.

 Hence, the constraints in~\eqref{eq:constraint of a for null space} are satisfied with high probability.
 
\paragraph*{Constraint in~\eqref{eq:decodability constraint}} 
For each user $k\in [\Ksf-\Usf]$, as we showed before,  the matrix  $\left[ \av_{\overline{\Sc}_{k}(1)}, \ldots, \av_{\overline{\Sc}_{k}\left(\binom{\Ksf-1}{\Ssf}\right)}  \right]$ has the same rank equal to $\Usf-1$, as its column-wise sub-matrix  in~\eqref{eq:unknown G2 matrix}.
Hence, the left null space of the matrix  $\left[ \av_{\overline{\Sc}_{k}(1)}, \ldots, \av_{\overline{\Sc}_{k}\left(\binom{\Ksf-1}{\Ssf}\right)}  \right]$ is the same as that of its column-wise sub-matrix  in~\eqref{eq:unknown G2 matrix}. Since the matrix in~\eqref{eq:unknown G2 matrix} has dimension $\Usf\times (\Usf-1)$ and rank $\Usf-1$ with high probability, its left null space contains exactly one linearly independent left null space vector (with dimension $1\times \Usf$). 
Let $\sv_k$ be one   left null space vector of  the matrix in~\eqref{eq:unknown G2 matrix}. 

For each user $k\in [\Ksf-\Usf+1:\Ksf]$,   the matrix  $\left[ \av_{\overline{\Sc}_{k}(1)}, \ldots, \av_{\overline{\Sc}_{k}\left(\binom{\Ksf-1}{\Ssf}\right)}  \right]$ has the same rank equal to $\Usf-1$, as its column-wise sub-matrix  in~\eqref{eq:unknown G1 matrix}. Hence, the left null space of the matrix  $\left[ \av_{\overline{\Sc}_{k}(1)}, \ldots, \av_{\overline{\Sc}_{k}\left(\binom{\Ksf-1}{\Ssf}\right)}  \right]$ is the same as that of its column-wise sub-matrix  in~\eqref{eq:unknown G1 matrix}, which contains exactly one linearly independent left null space vector. One possible choice of the left null space vector could be 
\begin{align}
\sv_k=\ev^{\text{T}}_{\Usf,k-\Ksf+\Usf}. \label{eq:left null vector of user k-u+1}
\end{align}
 
The most difficult part in the proof of the constraint in~\eqref{eq:decodability constraint} is    the 
  following lemma, which  will be proved in Appendix~\ref{sec:decodability rank lemma} by     the Schwartz-Zippel lemma~\cite{Schwartz,Zippel,Demillo_Lipton}.
\begin{lem}
    \label{lem:decodability rank lemma}
For any $\Ac \subseteq [\Ksf]$ where $|\Ac|=\Usf$, the $\Usf$-dimensional vectors $\sv_{k}$ where $k\in \Ac$ are linearly independent with high probability.   
    \hfill $\square$ 
\end{lem}    

Directly from Lemma~\ref{lem:decodability rank lemma}, it can be seen that the constraint in~\eqref{eq:decodability constraint} is satisfied with high probability.

In conclusion, all constraints in~\eqref{eq:full rank constraint},~\eqref{eq:constraint of a for null space}, and~\eqref{eq:decodability constraint} are satisfied with high probability.  Hence, there must exist a choice of $\bv_{\Vc}$ where $\Vc\in \Gc_2$ satisfying those constraints.
Thus the proposed scheme is decodable and secure. 
 In this case, we need the keys   $Z_{\Vc}$ where $\Vc \in (\Gc_1\cup \Gc_2\cup \Gc_3)$. It can be seen from~\eqref{eq:length of G1},~\eqref{eq:length of G2}, and~\eqref{eq:length of G3} that there are totally 
$$
\Usf+\Usf+  \frac{\Ksf(2\Usf-\Ksf+1)}{2}-\Usf=\Usf+  \frac{\Ksf(2\Usf-\Ksf+1)}{2}
$$ 
  keys each of which is shared by $\Ssf$ users.

  \section{Experimental Results}
\label{sec:experiment}
We implement our proposed secure aggregation scheme (which is referred to as \texttt{GroupSecAgg} for the sake of simplicity)  in Python2.7 by using the MPI4py library over the Amazon EC2 cloud, which is then compared to the   original  secure aggregation  scheme in~\cite{bonawitz2017practical} (referred to as \texttt{SecAgg}), and the 
  best existing information theoretic secure aggregation scheme with offline key sharing  in~\cite{lightsec2021so} (referred to as \texttt{LightSecAgg}). We compare the   key sharing    times
of \texttt{GroupSecAgg}   and \texttt{LightSecAgg}, since the communication costs in the model aggregation phase of these two schemes are the same.  
In addition, since   \texttt{SecAgg} provides computational security instead of information theoretic security, the total size of needed keys is much smaller in \texttt{SecAgg}. Thus we compare   the model aggregation times of \texttt{GroupSecAgg}   and \texttt{SecAgg}. 
Note that  in the experiments, we only record the  the communication  time as the running time in each procedure;     the detail of running   times in each procedure  of \texttt{GroupSecAgg},     \texttt{LightSecAgg}, and \texttt{SecAgg} could be found in   Appendix~\ref{sec:data tables}.

{\bf Amazon EC2 Setup.}
The Amazon EC2 \verb"t2.large" and \verb"t2.xlarge" 
instances are selected, where we take one specific \verb"t2.xlarge" 
instance as the server and all the other instances are users.
The Amazon EC2 T2 instances have a 3.0 GHz Intel Scalable Processor, and all instances which we use in this experiment have the same  capacity of computation,  memory and network resources. The transmission speed is up to 100MB/s between the server and users. By setting the field size $\qsf$ as $7$,
we  generate the  input vectors uniformly i.i.d. over $\mathbb{F}_7$, and consider the  three   sizes of  each input vector  (100KB, 200KB, 300KB) as suggested  in~\cite{bonawitz2017practical}.  
 In the offline key sharing phase, we consider that each two users have a private link to communicate as in~\cite{lightsec2021so}; thus  between each two users,  we use the \verb"MPI.send"  command. 
For each considered system with $(\Ksf,\Usf,\Ssf)$, 
we use Monte-Carlo methods with $20$ samples and take the average times over these $20$ samples.


\begin{figure} 
  \centering
  \begin{subfigure}{0.48\textwidth}
    \centering
    \includegraphics[scale=0.5]{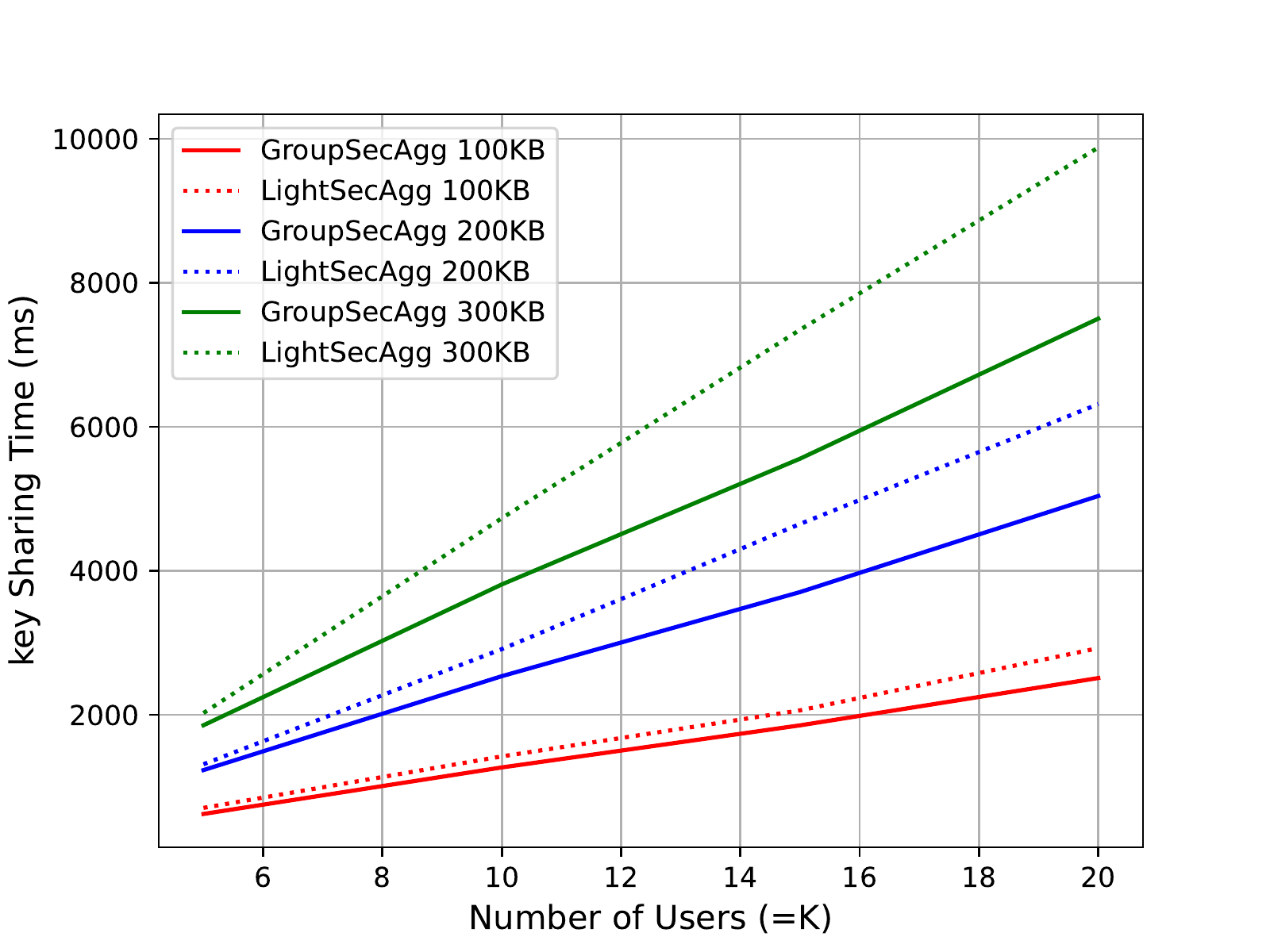}
    \caption{\texttt{GroupSecAgg} v.s. \texttt{LightSecAgg}: $\Usf=(\Ksf+1)/2$}
    \label{fig:myfiga}
  \end{subfigure} 
  \begin{subfigure}{0.48\textwidth}
    \centering
    \includegraphics[scale=0.5]{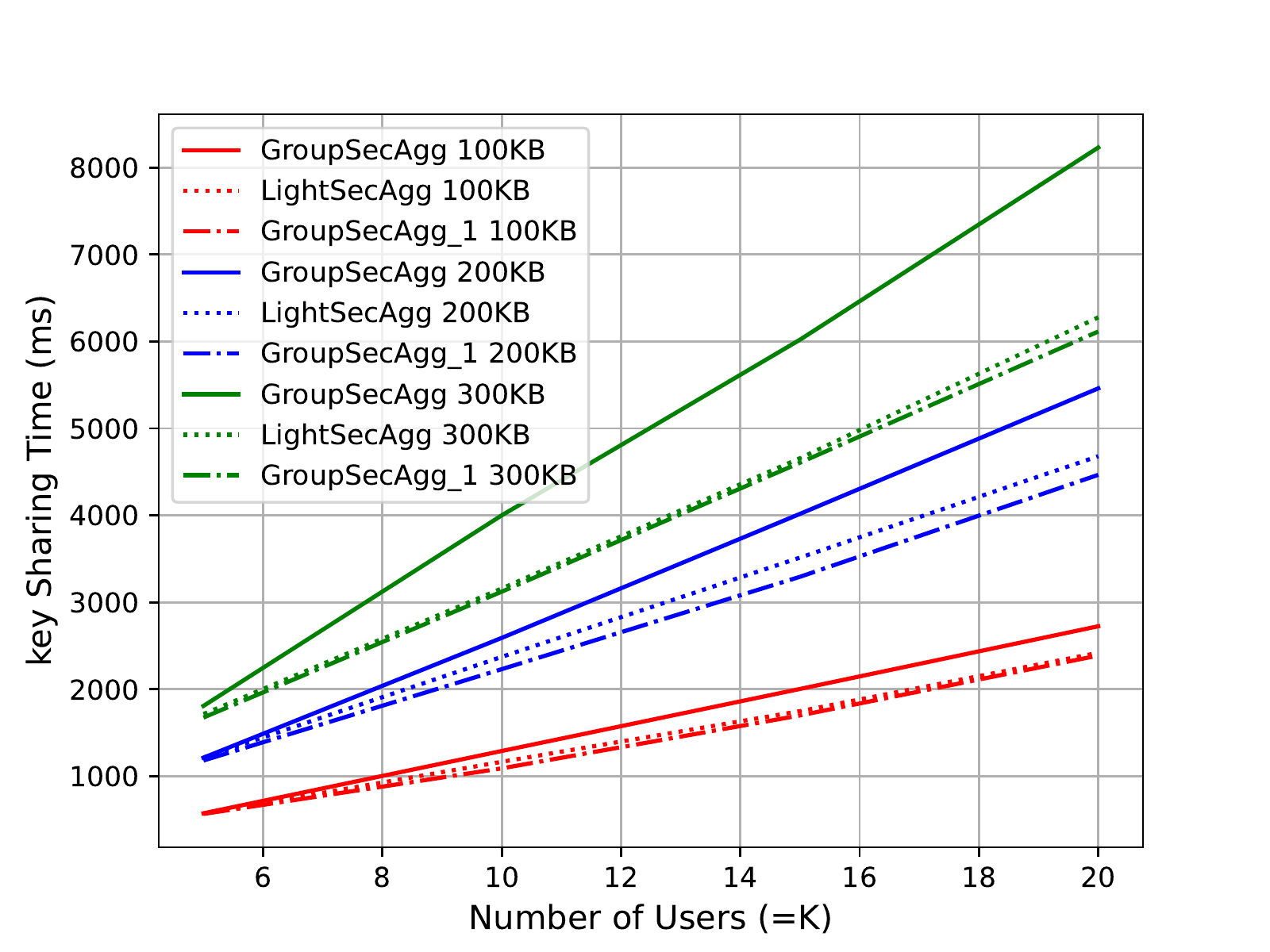}
    \caption{\texttt{GroupSecAgg} (\texttt{GroupSecAgg\_1}) v.s. \texttt{LightSecAgg}: $\Usf=\Ksf-1$}
    \label{fig:myfigb}
  \end{subfigure}
  \\
  \begin{subfigure}{0.48\textwidth}
    \centering
    \includegraphics[scale=0.5]{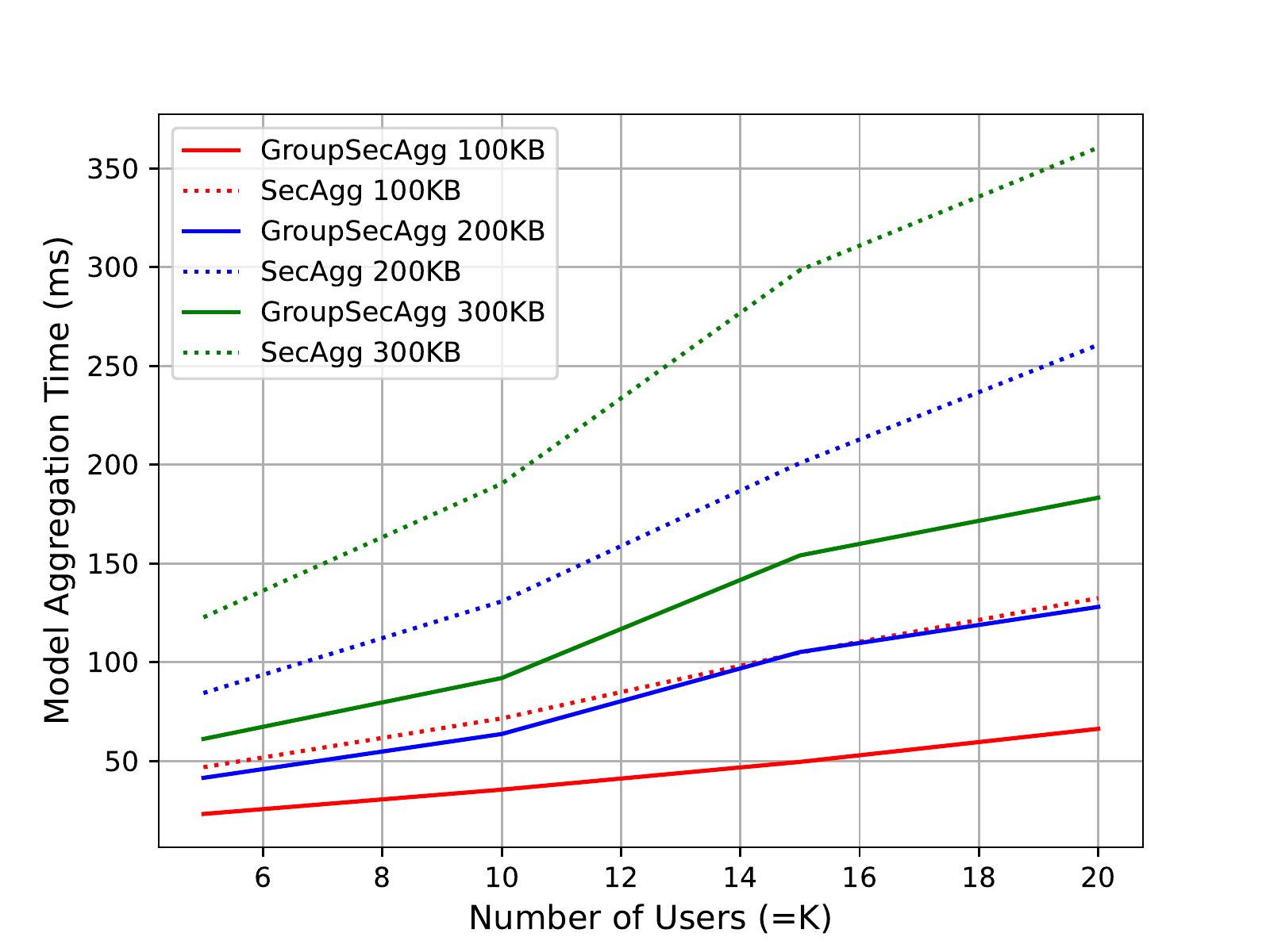}
    \caption{\texttt{GroupSecAgg} v.s. \texttt{SecAgg}: $\Usf=(\Ksf+1)/2$}
    \label{fig:myfigc}
  \end{subfigure} 
  \begin{subfigure}{0.48\textwidth}
    \centering
    \includegraphics[scale=0.5]{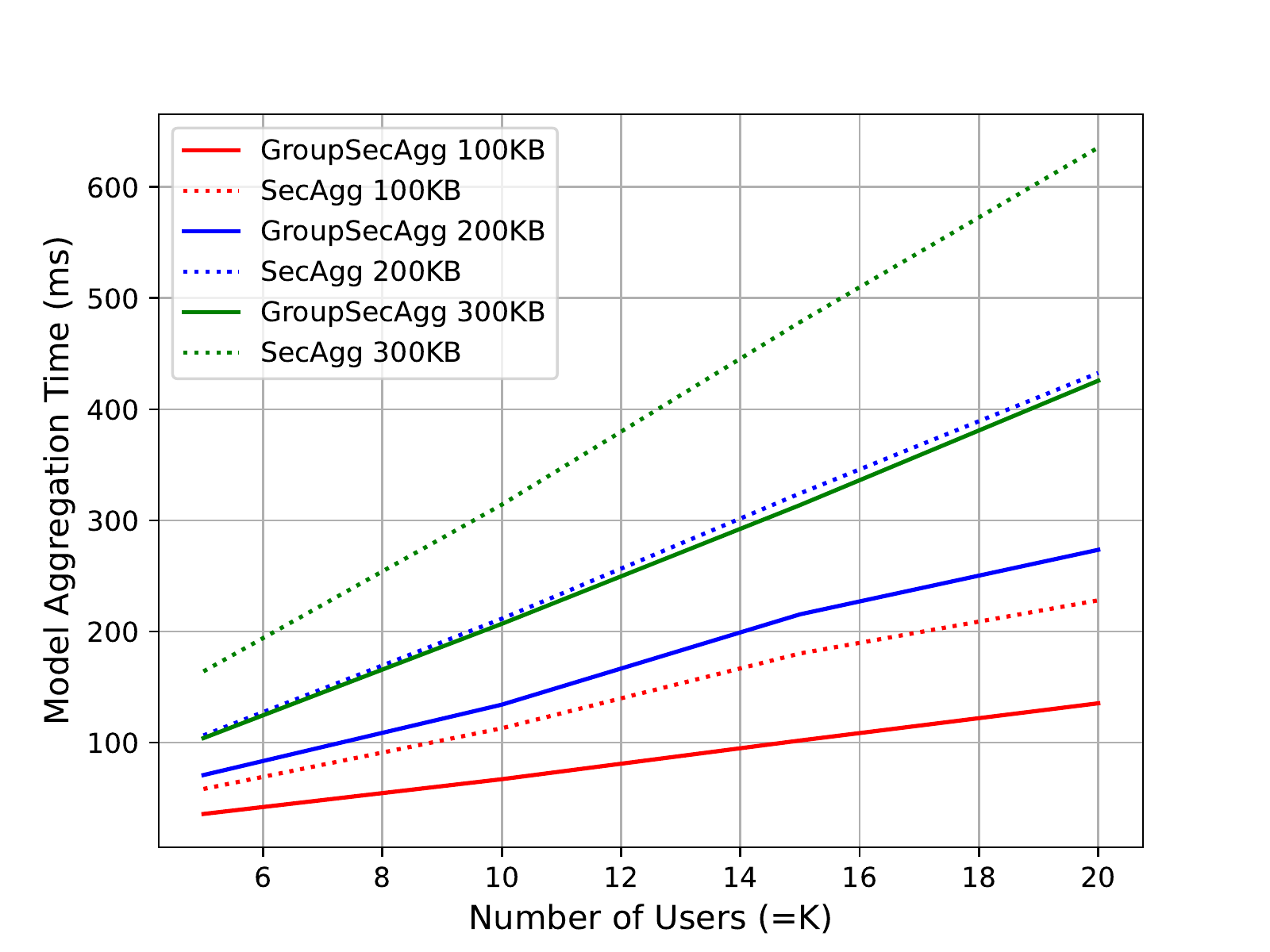}
    \caption{\texttt{GroupSecAgg} v.s. \texttt{SecAgg}: $\Usf=\Ksf-1$}
    \label{fig:myfigd}
  \end{subfigure}
  \caption{The key sharing time and the model aggregation time of \texttt{GroupSecAgg} versus \texttt{LightSecAgg} and \texttt{SecAgg}, respectively.}
  \label{fig:myfig}
\end{figure}


{\bf \texttt{GroupSecAgg} v.s. \texttt{LightSecAgg}.}
We first compare our \texttt{GroupSecAgg} with \texttt{LightSecAgg}, by  considering  the
two cases  where  $\Usf = (\Ksf + 1)/2$ illustrated in Fig.~\ref{fig:myfiga} and $\Usf =  \Ksf- 1$ illustrated in Fig.~\ref{fig:myfigb}, respectively. For each case, our \texttt{GroupSecAgg} needs $\Ssf=\Ksf-\Usf+1$.  

In Fig.~\ref{fig:myfiga}, since $\Usf = (\Ksf + 1)/2$, we have $\Usf=\Ksf-\Usf+1$ and thus our secure aggregation scheme is the one in Section~\ref{sub:U<S}. We use the cyclic key assignment; more precisely, for each $i\in [\Ksf]$, we let user $i$ randomly generate a key $Z_{\Cc(i)}$ with $(\Ksf-\Usf+1)\Lsf/\Usf=\Lsf$ symbols, and transmit   $Z_{\Cc(i)}$ to the other $\Usf-1$ users  in $\Cc(i)$, where  $\Cc$ is defined in~\eqref{eq:cyclic key set}.
Compared to \texttt{LightSecAgg}, \texttt{GroupSecAgg} reduces the key sharing time by at least $16.5\%$ and at most $31.7\%$ in Fig.~\ref{fig:myfiga}.
The improvement of \texttt{GroupSecAgg} is mainly because the number of keys is smaller than that of \texttt{LightSecAgg}, and thus  less number of connections is needed to build among users.

In Fig.~\ref{fig:myfigb}, since $\Usf =  \Ksf- 1$, our secure aggregation scheme is the one  in Section~\ref{sub:U>S U=K-1}. In this case, for each pair of users $\Vc=\{\Vc(1),\Vc(2)\}$ where $\Vc\subseteq [\Ksf]$, $|\Vc|=2$, and $\Vc(1)< \Vc(2)$, there is one key $Z_{\Vc}=\{Z_{\Vc,\Vc(1)},Z_{\Vc,\Vc(2)}\}$ with $(\Ksf-\Usf+1)\Lsf/\Usf=2\Lsf/\Usf$ symbols shared by users in $\Vc$. We consider two ways of key sharing: (i) ``\texttt{GroupSecAgg}'' in Fig.~\ref{fig:myfigb}:  user $\Vc(1)$ randomly generates $Z_{\Vc}$ and sends  $Z_{\Vc}$ to user $\Vc(2)$; (ii) ``\texttt{GroupSecAgg\_1}'' in Fig.~\ref{fig:myfigb}: user $\Vc(1)$ randomly generates $Z_{\Vc,\Vc(1)}$ and sends $Z_{\Vc,\Vc(1)}$ to user $\Vc(2)$, while user $\Vc(2)$ randomly generates $Z_{\Vc,\Vc(2)}$ and sends $Z_{\Vc,\Vc(2)}$ to user $\Vc(1)$. 
Compared to \texttt{LightSecAgg}, \texttt{GroupSecAgg} 
increase the key sharing time by at least $11.2\%$ and at most $23.7\%$ in Fig.~\ref{fig:myfigb}, while the key sharing time of \texttt{GroupSecAgg\_1} 
is close to that of \texttt{LightSecAgg}. The reason that the key sharing time of \texttt{GroupSecAgg} is more than that of \texttt{LightSecAgg} is because 
the transmissions of users  in Amazon EC2 are parallel, and in \texttt{GroupSecAgg} the users with smaller indices  transmit more keys in the key sharing phase. In \texttt{GroupSecAgg\_1}, we ``balance'' the numbers of user transmissions which reduce key sharing time.


{\bf \texttt{GroupSecAgg} v.s. \texttt{SecAgg}.}
We then compare our \texttt{GroupSecAgg} with \texttt{SecAgg}, by  considering the      two cases   where  $\Usf = (\Ksf + 1)/2$ illustrated in Fig.~\ref{fig:myfigc} and $\Usf =  \Ksf- 1$ illustrated in Fig.~\ref{fig:myfigd}, respectively.
Compared to \texttt{SecAgg}, \texttt{GroupSecAgg} reduces the model aggregation time by at least $48\%$ and at most $53\%$ in Fig.~\ref{fig:myfigc}, and reduces the model aggregation time by at least $33\%$ and at most $44\%$ in Fig.~\ref{fig:myfigd}.
From the theoretic viewpoint, this improvement is because our \texttt{GroupSecAgg} achieves the optimal communication cost in the model aggregation phase, while  \texttt{SecAgg} is sub-optimal. 

 \section{Conclusions}
 \label{sec:conclusion}
In this paper, we formulated the information theoretic secure aggregation problem with uncoded groupwise keys, where the keys are independent of each other and each of them is shared by a group of users. 	For the case   $\Ssf >\Ksf-\Usf$, we proposed a new secure aggregation scheme, which is the first scheme with uncoded   keys.  Quite surprisingly, the proposed scheme with uncoded groupwise keys  achieves the same capacity region of the communication rates in the two-round transmissions as the optimal scheme with any possible keys.  In addition, to achieve the capacity region, we showed that not all keys shared by   $\Ssf$ users are needed; instead, the number of keys used in the proposed scheme is   no more than $\Oc(\Ksf^2)$.
When $\Ssf \leq \Ksf-\Usf$, by proposing a new converse bound under the constraint of uncoded groupwise keys, we showed that uncoded groupwise keys sharing is strictly sub-optimal compared to coded keys sharing.  

 Ongoing work includes
the characterization of the capacity region for the case  $\Ssf \leq \Ksf-\Usf$ and the extension  of the proposed secure aggregation scheme to tolerate the collusion between the server and the users.  

\appendices

\section{Proof of Theorem~\ref{thm:subregion}}
\label{sec:proof of subregion}
We first consider the case $1=\Ssf\leq \Ksf-\Usf$.  In this case, it can be seen that $\Usf\leq \Ksf-1$. 
 We will show by contradiction that there does not exist any feasible secure aggregation scheme. 

Assume that there exists one feasible secure aggregation scheme.
When $ \Uc_1=[\Usf+1]$ and $\Uc_2=[2:\Usf+1]$, the server can recover $\sum_{k \in [\Usf+1]} W_{k }$; thus
\begin{subequations}
\begin{align}
0&=H\left(  W_{1}+\cdots+W_{\Usf+1}  | X_1,  (X_{k_1}, Y^{[\Usf+1]}_{k_1} :k_1\in [2:\Usf+1] ) \right) \\
&\geq  H\left(W_{1}+\cdots+W_{\Usf+1}   | X_1,    (W_{k_1}, Z_{\{k_1\}}:k_1\in [2:\Usf+1] ) \right)\label{eq:S1 impos 1}\\
&= H\left(W_1 | X_1,    (W_{k_1}, Z_{\{k_1\}}:k_1\in [2:\Usf+1] ) \right) \\
&=  H(W_1| X_1 ),
\label{eq:S1 impos 2}
\end{align}
\end{subequations}
where~\eqref{eq:S1 impos 1} follows since $(X_{k_1}, Y^{[\Usf+1]}_{k_1} :k_1\in [2:\Usf+1] ) $ is a function of $(W_{k_1}, Z_{\{k_1\}}:k_1\in [2:\Usf+1] )$ and condition does not increase entropy,~\eqref{eq:S1 impos 2} follows since $X_k$ is a function of $(W_1,Z_{\{1\}})$  and $(W_1,Z_{\{1\}})$ is independent of $(W_{2},\ldots,W_{\Usf+1},Z_{\{2\}},\ldots,Z_{\{\Usf+1\}})$.
However, by the security constraint in~\eqref{eq:security constraint}, we should have $I(X_1;W_1)=0$, which leads (recall that $W_1$ contains $\Lsf$ uniform and i.i.d. symbols over $\mathbb{F}_{\qsf}$)
\begin{align}
H(W_1|X_1)=H(W_1)-I(X_1;W_1)=\Lsf. \label{eq:contradiction}
\end{align} 
 Hence,~\eqref{eq:contradiction} contradicts to~\eqref{eq:S1 impos 2}.

In the rest of this proof, we consider the case where $2\leq \Ssf \leq \Ksf-\Usf$.
By the converse bound in Lemma~\ref{lem:converse}, we have $\Rsf_1\geq 1$. Hence, for any feasible secure aggregation scheme, 
we can assume that it achieves $\Rsf_1=1+\asf$, where $\asf \geq 0$. 
Then in the following, we focus on this scheme.

For each $k\in [\Ksf]$, when $|\Uc_1|\geq \Usf+1$, $k\in\Uc_1$,  and $\Uc_2=\Uc_1\setminus \{k\}$, the server can recover $\sum_{k_1\in \Uc_1} W_{k_1}$; thus 
we have 
\begin{subequations}
\begin{align}
0&=H\left(\sum_{k_1\in \Uc_1} W_{k_1} \Big| X_k,  (X_{k_2}, Y^{\Uc_1}_{k_2} :k_2\in \Uc_2 ) \right) \\
&\geq  H\left(\sum_{k_1\in \Uc_1} W_{k_1} \Big| X_k, Z_k,  (W_{k_2}, Z_{k_2}:k_2\in \Uc_2 ) \right)\label{eq:function and condition}\\
&=  H(W_k| X_k, Z_k, (W_{k_2}, Z_{k_2} :k_2 \in \Uc_2 ) ),
\label{eq:prove markov}
\end{align}
\end{subequations}
where~\eqref{eq:function and condition} follows since   $(X_{k_2}, Y^{\Uc_1}_{k_2} :k_2\in \Uc_2 )$ is a function of $(W_{k_2}, Z_{k_2}:k_2\in \Uc_2 )$,   and condition does not  increase entropy.
From~\eqref{eq:prove markov}, we have 
\begin{subequations}
\begin{align}
H(X_k|Z_k) &\geq H(X_k| Z_k, (W_{k_2}, Z_{k_2} :k_2 \in \Uc_2 ) ) \\
&= I(W_k;X_k| Z_k, (W_{k_2}, Z_{k_2} :k_2 \in \Uc_2 )) + H(X_k| W_k,Z_k, (W_{k_2}, Z_{k_2} :k_2 \in \Uc_2 ))\\
&=I(W_k;X_k| Z_k, (W_{k_2}, Z_{k_2} :k_2 \in \Uc_2 )) \label{eq:function of Zk Wk}\\
&= H(W_k|Z_k, (W_{k_2}, Z_{k_2} :k_2 \in \Uc_2 ) ) -H(W_k| X_k,Z_k, (W_{k_2}, Z_{k_2} :k_2 \in \Uc_2 ))\\
&\stackrel{\eqref{eq:prove markov}}{\geq}  
  H(W_k|Z_k, (W_{k_2}, Z_{k_2} :k_2 \in \Uc_2 ) )\\
&=H(W_k)=\Lsf,
\label{eq:HXk given Zk}
\end{align}
\end{subequations}
where~\eqref{eq:function of Zk Wk} follows since $X_k$ is a function of $(W_k,Z_k)$. 
From~\eqref{eq:HXk given Zk}, we have 
\begin{subequations}
\begin{align}
I(W_k; X_k | Z_k)&= H(X_k|Z_k)-H(X_k|Z_k,W_k)\\
&=H(X_k|Z_k)\\
&\stackrel{\eqref{eq:HXk given Zk}}{\geq}   \Lsf. \label{eq:I independent}
\end{align}
\end{subequations} 
From~\eqref{eq:I independent}, we have 
\begin{align}
H(X_k|Z_k)=I(W_k; X_k | Z_k) + H(X_k|Z_k, W_k)\geq  \Lsf.
\label{eq:H(X_k|Z_k)}
\end{align}
In addition, from~\eqref{eq:I independent} we also have 
\begin{align}
H(W_k|Z_k,X_k)=H(W_k|Z_k) -I(W_k;X_k|Z_k)\stackrel{\eqref{eq:I independent}}{\leq} H(W_k|Z_k)-\Lsf=0.
\label{eq:new Marcov}
\end{align}

We   define that  $\Sc^{\prime}_k:=\left\{\Vc \in \binom{[\Ksf]}{\Ssf}  :   k\in \Vc \right\}$, and sort the sets in  $\Sc^{\prime}_k$ in a lexicographic order. $\Sc^{\prime}_k(j) $ represents the $j^{\text{th}}$ set in $\Sc^{\prime}_k$, where $j\in \left[\binom{\Ksf-1}{\Ssf-1} \right]$. Since 
$2\leq \Ssf \leq \Ksf-\Usf$,  we can see that $\binom{\Ksf-1}{\Ssf-1} \geq 2$.
For any set $\Sc\subseteq \Sc^{\prime}_k$,  from~\eqref{eq:H(X_k|Z_k)} we have  
$$
\Lsf \stackrel{\eqref{eq:H(X_k|Z_k)}}{\leq}    H(X_k|Z_k) \leq H(X_k| (Z_{\Vc}:\Vc\in \Sc) ) \leq H(X_k)  \leq \Rsf_1 = \Lsf(1+\asf).
$$
Hence, we have 
\begin{align}
\Lsf \leq H(X_k| (Z_{\Vc}:\Vc\in \Sc) ) \leq \Lsf(1+\asf). \label{eq:HXk given Zv}
\end{align}

For any collections of sets $\Sc,\Sc^{\prime} \subseteq \Sc^{\prime}_k$ we have (which will be proved in Appendix~\ref{sec:prove of sub}) 
\begin{align}
&H(W_k|X_k, (Z_{\Vc_1}:\Vc_1\in  \Sc ) ) + H(W_k|X_k, (Z_{\Vc_2}: \Vc_2 \in \Sc^{\prime})) \nonumber\\& \geq  H(W_k|X_k, (Z_{\Vc_0}:\Vc_0 \in \Sc\cup \Sc^{\prime}) ) +H(W_k|X_k, (Z_{\Vc_5}: \Vc_5 \in \Sc\cap \Sc^{\prime}) ) 
\nonumber\\& - I\left(  (Z_{\Vc_4}: \Vc_4 \in \Sc \setminus \Sc^{\prime}) ; (Z_{\Vc_3}: \Vc_3 \in \Sc^{\prime}  \setminus \Sc)      | X_k,(Z_{\Vc_5}: \Vc_5 \in \Sc\cap \Sc^{\prime}) \right)
.\label{eq:from submodularity}
\end{align}  

In addition,  we have
\begin{subequations}
\begin{align}
&I\left(  (Z_{\Vc_4}: \Vc_4 \in \Sc \setminus \Sc^{\prime}) ; (Z_{\Vc_3}: \Vc_3 \in \Sc^{\prime}  \setminus \Sc)      | X_k,(Z_{\Vc_5}: \Vc_5 \in \Sc\cap \Sc^{\prime}) \right) \\& \leq I\left(  (Z_{\Vc_1}: \Vc_1 \in \Sc  ) ; (Z_{\Vc_3}: \Vc_3 \in \Sc^{\prime}  \setminus \Sc)      | X_k  \right) \\
&= H( (Z_{\Vc_1}: \Vc_1 \in \Sc  )| X_k)+ H((Z_{\Vc_3}: \Vc_3 \in \Sc^{\prime}  \setminus \Sc) |X_k) -H( (Z_{\Vc_0}: \Vc_0 \in \Sc\cup \Sc^{\prime}  )|X_k)\\
&\leq H(  Z_{\Vc_1}: \Vc_1 \in \Sc    )+ H( Z_{\Vc_3}: \Vc_3 \in \Sc^{\prime}  \setminus \Sc ) -H( (Z_{\Vc_0}: \Vc_0 \in \Sc\cup \Sc^{\prime}  )|X_k)\\
&= H(  Z_{\Vc_1}: \Vc_1 \in \Sc    )+ H( Z_{\Vc_3}: \Vc_3 \in \Sc^{\prime}  \setminus \Sc ) -   H(Z_{\Vc_0}: \Vc_0 \in \Sc\cup \Sc^{\prime} ) +I( (Z_{\Vc_0}: \Vc_0 \in \Sc\cup \Sc^{\prime}  ); X_k)\\
&=I( (Z_{\Vc_0}: \Vc_0 \in \Sc\cup \Sc^{\prime}  ); X_k)\\
&=H( X_k)-H(X_k|(Z_{\Vc_0}: \Vc_0 \in \Sc\cup \Sc^{\prime}  ))\\
&\stackrel{\eqref{eq:HXk given Zv}}{\leq}  \Lsf(1+\asf)-\Lsf=\asf\Lsf.\label{eq:derive conditional sub}
\end{align}
\end{subequations}  
By taking~\eqref{eq:derive conditional sub} into~\eqref{eq:from submodularity}, we have
\begin{align}
&H(W_k|X_k, (Z_{\Vc_1}:\Vc_1\in  \Sc ) ) + H(W_k|X_k, (Z_{\Vc_2}: \Vc_2 \in \Sc^{\prime})) \nonumber\\& \geq  H(W_k|X_k, (Z_{\Vc_0}:\Vc_0 \in \Sc\cup \Sc^{\prime}) ) +H(W_k|X_k, (Z_{\Vc_5}: \Vc_5 \in \Sc\cap \Sc^{\prime}) ) -  \asf\Lsf.\label{eq:further sub}
\end{align}

Hence, by using~\eqref{eq:further sub} iteratively, we have
\begin{subequations}
\begin{align}
\sum_{j\in \left[\binom{\Ksf-1}{\Ssf-1}\right]} H(W_k|X_k, (Z_{\Vc_1}:\Vc_1 \in \Sc^{\prime}_k\setminus \{\Sc^{\prime}_k(j)\})) &\geq  H(W_k|X_k) -\left(\binom{\Ksf-1}{\Ssf-1}-1\right)  \asf\Lsf \label{eq:contra 1} \\
& =\left(1-\Big(\binom{\Ksf-1}{\Ssf-1}-1\Big)  \asf \right)\Lsf, \label{eq:contra}
\end{align}
\end{subequations} 
where~\eqref{eq:contra} comes from the security constraint $I(W_k;X_k)=0$ and $H(W_k)=\Lsf$.\footnote{\label{foot:repeat sub}To make the derivation of~\eqref{eq:contra 1} more clear, 
  we first consider the first two terms on the LHS of~\eqref{eq:contra 1}. 
We can see that  $\big(\Sc^{\prime}_k\setminus \{\Sc^{\prime}_k(1)\}) \big) \cup \big(\Sc^{\prime}_k\setminus \{\Sc^{\prime}_k(2))\} \big)= \Sc^{\prime}_k$, and 
   $\big(\Sc^{\prime}_k\setminus \{\Sc^{\prime}_k(1)\}) \big) \cap \big(\Sc^{\prime}_k\setminus \{\Sc^{\prime}_k(2))\} \big)= \Sc^{\prime}_k\setminus \{\Sc^{\prime}_k(1),\Sc^{\prime}_k(2)\}$.  
From~\eqref{eq:further sub}, we have $\sum_{j\in [2]} H(W_k|X_k, (Z_{\Vc_1}:\Vc_1 \in \Sc^{\prime}_k\setminus \{\Sc^{\prime}_k(j)\})) \geq H(W_k|X_k, Z_k)+ H(W_k|X_k, (Z_{\Vc_1}:\Vc_1 \in \Sc^{\prime}_k\setminus \{\Sc^{\prime}_k(1),\Sc^{\prime}_k(2)\}))-\asf\Lsf$, and we recall that $H(W_k|X_k, Z_k)=0$. 
Next, from~\eqref{eq:further sub} again,  we can lower bound the sum of $H(W_k|X_k, (Z_{\Vc_1}:\Vc_1 \in \Sc^{\prime}_k\setminus \{\Sc^{\prime}_k(1),\Sc^{\prime}_k(2)\}))$ and $H(W_k|X_k, (Z_{\Vc_1}:\Vc_1 \in \Sc^{\prime}_k\setminus \{\Sc^{\prime}_k(3)\}))$, by    $H(W_k|X_k, (Z_{\Vc_1}:\Vc_1 \in \Sc^{\prime}_k\setminus \{\Sc^{\prime}_k(1),\Sc^{\prime}_k(2),\Sc^{\prime}_k(3)\})) -\asf \Lsf$. 
We repeat this  iteratively. The last (i.e., $\left(\binom{\Ksf-1}{\Ssf-1}-1\right)^{\text{th}}$) step is to lower bound the sum of $H(W_k|X_k, (Z_{\Vc_1}:\Vc_1 \in \Sc^{\prime}_k\setminus \{\Sc^{\prime}_k(1),\ldots,\Sc^{\prime}_k(\binom{\Ksf-1}{\Ssf-1}-1)\}))$ and $H(W_k|X_k, (Z_{\Vc_1}:\Vc_1 \in \Sc^{\prime}_k\setminus \{\Sc^{\prime}_k(\binom{\Ksf-1}{\Ssf-1})\}))$, by
$H(W_k|X_k) -\asf\Lsf$. 
 In conclusion, we can obtain~\eqref{eq:contra 1}.
} 

For each set  $\Vc \in \Sc^{\prime}_k$,   we have 
\begin{subequations}
\begin{align}
&H\left(W_k\Big|X_k, (W_{k_1}:k_1\in [\Ksf]\setminus \{k\}),\Big(Z_{\Vc_1}:\Vc_1 \in \binom{[\Ksf]}{\Ssf}   , \Vc_1\neq \Vc\Big) \right) \nonumber\\
& \geq I\left(W_k;Z_{\Vc} \Big|X_k,  (W_{k_1}:k_1\in [\Ksf]\setminus \{k\}),\Big(Z_{\Vc_1}:\Vc_1 \in \binom{[\Ksf]}{\Ssf}, \Vc_1\neq \Vc\Big)\right)\\
& =I(W_k;Z_{\Vc}|X_k, (Z_{\Vc_1}:\Vc_1 \in \Sc^{\prime}_k\setminus \{\Vc\}))  \label{eq:Wk's independent} \\
& =H(W_k| X_k, (Z_{\Vc_1}:\Vc_1 \in \Sc^{\prime}_k\setminus \{\Vc\})) -H(W_k| X_k, (Z_{\Vc_1}:\Vc_1 \in \Sc^{\prime}_k))\\
&=H(W_k| X_k, (Z_{\Vc_1}:\Vc_1 \in \Sc^{\prime}_k\setminus \{\Vc\})),\label{eq:after sub}
\end{align} 
\end{subequations} 
where~\eqref{eq:Wk's independent} follows since $(W_{k_1}:k_1\in [\Ksf]\setminus \{k_1\})$ and $\left(Z_{\Vc_1}:\Vc_1 \in \binom{[\Ksf]\setminus \{k\}}{\Ssf}  \right)$ are independent of $(X_k, Z_k, W_k)$,~\eqref{eq:after sub} comes from~\eqref{eq:new Marcov}.

On the other hand, when $\Uc_1=([\Ksf]\setminus \Vc) \cup \{k\}$ and $\Uc_2=[\Ksf]\setminus \Vc$,\footnote{\label{foot:possible case}This case is possible because, $|\Vc|= \Ssf \leq \Ksf-\Usf$, and thus $|[\Ksf]\setminus \Vc| \geq \Usf$.} we have 
\begin{subequations}
\begin{align}
0&=H\left(\sum_{k_1\in \Uc_1} W_{k_1} \Big| X_k, (X_{k_2}, Y^{\Uc_1}_{k_2} :k_2\in \Uc_2) \right)\\
&\geq H\left(\sum_{k_1\in \Uc_1} W_{k_1} \Big| X_k, (W_{k_2},Z_{k_2}:k_2\in \Uc_2)\right)\label{eq:k2functions}\\
&= H(W_k| X_k, (W_{k_2},Z_{k_2}:k_2\in \Uc_2))\\
&\geq H\left(W_k \Big|X_k, (W_{k_1}:k_1\in [\Ksf]\setminus \{k\}), \Big(Z_{\Vc_1}:\Vc_1 \in \binom{[\Ksf]}{\Ssf}, \Vc_1\neq \Vc\Big) \right),\label{eq:last term}
\end{align}
\end{subequations}
where~\eqref{eq:k2functions} follows since $(X_{k_2}, Y^{\Uc_1}_{k_2})$ is a function of $(W_{k_2},Z_{k_2})$, and~\eqref{eq:last term} follows since $k\notin \Uc_2$ and   $\Vc \cap \Uc_2=\emptyset$. 

From~\eqref{eq:after sub} and~\eqref{eq:last term}, we have 
\begin{align}
H(W_k| X_k, (Z_{\Vc_1}:\Vc_1 \in \Sc^{\prime}_k\setminus \Vc)) \leq 0.\label{eq:each term<0}
\end{align}
By taking~\eqref{eq:each term<0} into~\eqref{eq:contra}, we have 
\begin{subequations}
\begin{align}
&1-\left(\binom{\Ksf-1}{\Ssf-1}-1 \right) \asf \leq 0,  \\
&\Longleftrightarrow \asf \geq \frac{1}{\binom{\Ksf-1}{\Ssf-1}-1 }  .\label{eq:lower bound of a}
\end{align}
\end{subequations}
  Hence,     Theorem~\ref{thm:subregion} can be proved from $\Rsf_1=1+\asf$ and~\eqref{eq:lower bound of a}.

\section{Proof of~\eqref{eq:from submodularity}}
\label{sec:prove of sub}
The proof of~\eqref{eq:from submodularity} follows    the proof of~\cite[Proposition 3]{Wan2021combinationnet} (which shows a generalized version of the submodularity of entropy).
More precisely,  we have 
 \begin{subequations}
\begin{align}
&H(W_k|X_k, (Z_{\Vc_1}:\Vc_1\in  \Sc ) )-H(W_k|X_k, (Z_{\Vc_0}:\Vc_0\in  \Sc \cup \Sc^{\prime}) ) + H(W_k|X_k, (Z_{\Vc_2}: \Vc_2 \in \Sc^{\prime})) \nonumber\\
&=I(W_k;(Z_{\Vc_3}: \Vc_3 \in \Sc^{\prime} \setminus \Sc) | X_k, (Z_{\Vc_1}:\Vc_1\in  \Sc)  )+  H(W_k|X_k, (Z_{\Vc_2}: \Vc_2 \in \Sc^{\prime})) \\
& =  I(W_k;(Z_{\Vc_3}: \Vc_3 \in \Sc^{\prime} \setminus \Sc) | X_k, (Z_{\Vc_1}:\Vc_1\in  \Sc)  ) + H(W_k|X_k, (Z_{\Vc_0}:\Vc_0\in  \Sc \cup \Sc^{\prime}) )    \nonumber\\
& + I(W_k;(Z_{\Vc_4}: \Vc_4 \in \Sc \setminus \Sc^{\prime}) | X_k, (Z_{\Vc_2}:\Vc_2\in  \Sc^{\prime})  ).  \label{eq:submod first step} 
\end{align}
   \end{subequations}
In addition,  we have 
 \begin{subequations}
\begin{align}
 &  I(W_k;(Z_{\Vc_3}: \Vc_3 \in \Sc^{\prime} \setminus \Sc) | X_k, (Z_{\Vc_1}:\Vc_1\in  \Sc)  )    
  +I(W_k;(Z_{\Vc_4}: \Vc_4 \in \Sc \setminus \Sc^{\prime}) | X_k, (Z_{\Vc_2}:\Vc_2\in  \Sc^{\prime})  ) \nonumber\\
&=I(W_k,(Z_{\Vc_4}: \Vc_4 \in \Sc \setminus \Sc^{\prime}) ;(Z_{\Vc_3}: \Vc_3 \in \Sc^{\prime} \setminus \Sc) | X_k, (Z_{\Vc_5}:\Vc_5\in  \Sc \cap \Sc^{\prime})  ) 
\nonumber\\& - I\left(  (Z_{\Vc_4}: \Vc_4 \in \Sc \setminus \Sc^{\prime}) ; (Z_{\Vc_3}: \Vc_3 \in \Sc^{\prime}  \setminus \Sc)      | X_k,(Z_{\Vc_5}: \Vc_5 \in \Sc\cap \Sc^{\prime}) \right)\nonumber\\&  +I(W_k;(Z_{\Vc_4}: \Vc_4 \in \Sc \setminus \Sc^{\prime}) | X_k, (Z_{\Vc_2}:\Vc_2\in  \Sc^{\prime})  ) \\
& \geq  H(W_k,(Z_{\Vc_4}: \Vc_4 \in \Sc \setminus \Sc^{\prime})| X_k, (Z_{\Vc_5}:\Vc_5\in  \Sc \cap \Sc^{\prime}) )  - 
H(W_k  | X_k, (Z_{\Vc_2}:\Vc_2\in  \Sc^{\prime}))  \nonumber\\
& - I\left(  (Z_{\Vc_4}: \Vc_4 \in \Sc \setminus \Sc^{\prime}) ; (Z_{\Vc_3}: \Vc_3 \in \Sc^{\prime}  \setminus \Sc)      | X_k,(Z_{\Vc_5}: \Vc_5 \in \Sc\cap \Sc^{\prime}) \right)\nonumber\\&  +I(W_k;(Z_{\Vc_4}: \Vc_4 \in \Sc \setminus \Sc^{\prime}) | X_k, (Z_{\Vc_2}:\Vc_2\in  \Sc^{\prime})  ) \\
&=  H(W_k,(Z_{\Vc_4}: \Vc_4 \in \Sc \setminus \Sc^{\prime})| X_k, (Z_{\Vc_5}:\Vc_5\in  \Sc \cap \Sc^{\prime}) )- H(W_k | X_k, (Z_{\Vc_0}:\Vc_0\in  \Sc \cup \Sc^{\prime} )  )\nonumber \\
& - I\left(  (Z_{\Vc_4}: \Vc_4 \in \Sc \setminus \Sc^{\prime}) ; (Z_{\Vc_3}: \Vc_3 \in \Sc^{\prime}  \setminus \Sc)      | X_k,(Z_{\Vc_5}: \Vc_5 \in \Sc\cap \Sc^{\prime}) \right)   \\
& \geq H(W_k | X_k, (Z_{\Vc_5}:\Vc_5\in  \Sc \cap \Sc^{\prime}) )- H(W_k | X_k, (Z_{\Vc_0}:\Vc_0\in  \Sc \cup \Sc^{\prime} )  )\nonumber \\
& - I\left(  (Z_{\Vc_4}: \Vc_4 \in \Sc \setminus \Sc^{\prime}) ; (Z_{\Vc_3}: \Vc_3 \in \Sc^{\prime}  \setminus \Sc)      | X_k,(Z_{\Vc_5}: \Vc_5 \in \Sc\cap \Sc^{\prime}) \right).\label{eq:comp from sub prove}
\end{align}
    \end{subequations}
    By taking~\eqref{eq:comp from sub prove} into~\eqref{eq:submod first step}, we have 
    \begin{align}
   &H(W_k|X_k, (Z_{\Vc_1}:\Vc_1\in  \Sc ) )-H(W_k|X_k, (Z_{\Vc_0}:\Vc_0\in  \Sc \cup \Sc^{\prime}) ) + H(W_k|X_k, (Z_{\Vc_2}: \Vc_2 \in \Sc^{\prime})) \nonumber\\
   &\geq  H(W_k | X_k, (Z_{\Vc_5}:\Vc_5\in  \Sc \cap \Sc^{\prime}) )- I\left(  (Z_{\Vc_4}: \Vc_4 \in \Sc \setminus \Sc^{\prime}) ; (Z_{\Vc_3}: \Vc_3 \in \Sc^{\prime}  \setminus \Sc)      | X_k,(Z_{\Vc_5}: \Vc_5 \in \Sc\cap \Sc^{\prime}) \right),
    \end{align}
    which coincides with~\eqref{eq:from submodularity}. 

\section{Proof of the Security Constraint in~\eqref{eq:security constraint} for The Proposed Secure Aggregation Scheme}
\label{sec:proof of security}
Assume that in the proposed secure aggregation scheme for Theorem~\ref{thm:main result}, the $\Usf$-dimensional vectors $\av_{\Vc}$ where $\Vc\in \binom{[\Ksf]}{\Ssf}$ are determined, such that the constraints in~\eqref{eq:full rank constraint},~\eqref{eq:constraint of a for null space}, and~\eqref{eq:decodability constraint} are satisfied.

Let us then prove that the scheme is secure.  
By our construction, since the constraint in~\eqref{eq:full rank constraint} is satisfied,   we have 
\begin{subequations}
\begin{align}
I(X_1,\ldots,X_{\Ksf};W_1, \ldots, W_{\Ksf})
 &= \sum_{k\in [\Ksf]} I(X_k; W_k) \label{eq:each Xk indep}\\
&=\sum_{k\in [\Ksf]} \left( H(X_k)-H(X_k| W_k) \right)\\
&=\sum_{k\in [\Ksf]} \left( \Lsf -H(X_k| W_k)\right)\label{eq:each Xk indep2}\\
&=\sum_{k\in [\Ksf]} \left( \Lsf -\Lsf \right) =0 , \label{eq:case 1 first round secrey}
\end{align}
\label{eq:formal proof of Xk security}
\end{subequations}
where~\eqref{eq:each Xk indep} follows since $(X_1,W_1), \ldots, (X_{\Ksf},W_{\Ksf})$ are mutually independent in our scheme (Recall~\eqref{eq:key constraint} and that  $X_1,\ldots,X_{\Ksf}$ use different keys),~\eqref{eq:each Xk indep2} follows since each $W_k$ contains $\Lsf$ uniform and i.i.d. symbols over $\mathbb{F}_{\qsf}$   and the keys are independent of $W_k$, 
and~\eqref{eq:case 1 first round secrey} follows since (recall that each $Z_{\Vc,k}$ where $\Vc\in \binom{[\Ksf] }{\Ssf}$ and $k\in \Vc$ contains $\Lsf/\Usf$ uniform and i.i.d. symbols over $\mathbb{F}_{\qsf}$)
\begin{subequations}
\begin{align}
H(X_k| W_k)&=H\left(\bigg( W_{k,j} + \sum_{\Vc\in\binom{[\Ksf]}{\Ssf}: k \in \Vc} a_{\Vc, j} Z_{\Vc,k}: j\in [\Usf] \bigg) \bigg| (W_{k,j}:j\in [\Usf])  \right)\\
&\stackrel{\eqref{eq:key constraint}}{=} H\left(   \sum_{\Vc\in\binom{[\Ksf]}{\Ssf}: k \in \Vc} a_{\Vc, j} Z_{\Vc,k}: j\in [\Usf]    \right)\\
& \stackrel{\eqref{eq:full rank constraint}}{=}  \Lsf.
\end{align}
\end{subequations}

Hence, we have 
\begin{subequations}
\begin{align}
&I\left(W_1,\ldots,W_{\Ksf}; X_{1},\ldots,X_{\Ksf}, (Y^{\Uc_1}_{k}:k\in \Uc_1) \Big| \sum_{k\in \Uc_1} W_{k} \right) \nonumber\\
&=   I\left(W_1,\ldots,W_{\Ksf}; (Y^{\Uc_1}_{k}:k\in \Uc_1) \Big| \sum_{k\in \Uc_1} W_{k}, X_{1},\ldots,X_{\Ksf}\right) \label{eq:XindpendW}\\
&\leq I\left(W_1,\ldots,W_{\Ksf}; F_1,\ldots, F_{\Usf} \Big| \sum_{k\in \Uc_1} W_{k}, X_{1},\ldots,X_{\Ksf}\right)\label{eq:YfunctionF}\\
&=0, \label{eq:FfunctionXW} 
\end{align}
\end{subequations}
where~\eqref{eq:XindpendW} comes from~\eqref{eq:case 1 first round secrey},~\eqref{eq:YfunctionF} comes from  $(Y^{\Uc_1}_{k}:k\in \Uc_1) $ are in the linear space spanned by  $F_1,\ldots,F_{\Usf}$ and thus are determined by $F_1,\ldots,F_{\Usf}$,~\eqref{eq:FfunctionXW} follows since $F_1,\ldots,F_{\Usf}$ can be recovered from $\sum_{k\in \Uc_1} W_{k}$ and $\sum_{k\in \Uc_1} X_{k}$. Hence, the security constraint in~\eqref{eq:security constraint} is satisfied.

\section{Proof of Lemma~\ref{lem:decodability rank lemma}}
\label{sec:decodability rank lemma}
Consider one set $\Ac \subseteq [\Ksf]$ where $|\Ac|=\Usf$. Assume that $\Ac=\{\Ac(1),\ldots,\Ac(\Usf)\}$  where $\Ac(1)<\cdots<\Ac(\Usf)$. 
  We also assume that the sets in  
$$\Gc_2=\left\{    [\Ksf-\Usf+1:2\Ksf-2\Usf] \cup \{j\} : j\in ([\Ksf-\Usf] \cup [2\Ksf-2\Usf+1: \Ksf])  \right\}$$
are $\Gc_{2,1},\ldots,\Gc_{2,\Ksf-\Usf}, \Gc_{2,2\Ksf-2\Usf+1},\ldots,\Gc_{2,\Ksf} $,  where $\Gc_{2,j}=[\Ksf-\Usf+1:2\Ksf-2\Usf] \cup \{j\}$ for each $j\in ([\Ksf-\Usf] \cup [2\Ksf-2\Usf+1: \Ksf])$.

Recall that by our construction, 
for each user $k\in[\Ksf-\Usf]$,  $\sv_{k}$ is a left null space vector of the matrix in~\eqref{eq:unknown G2 matrix}. 
Note that each column of the matrix in~\eqref{eq:unknown G2 matrix} is $\av_{\{j\}\cup [\Ksf-\Usf+1:2\Ksf-2\Usf]}$ where $j\in ([\Ksf-\Usf]\setminus \{k\})\cup [2\Ksf-2\Usf+1:\Ksf]$. In addition, it can be seen  that $\{j\}\cup [\Ksf-\Usf+1:2\Ksf-2\Usf]$ is in $\Gc_2$; thus  
each element  of  $\bv_{\{j\}\cup [\Ksf-\Usf+1:2\Ksf-2\Usf]}$ is chosen uniformly and i.i.d. over $\mathbb{F}_{\qsf}$.
For each user   $k\in [\Ksf-\Usf+1:\Ksf]$, from~\eqref{eq:left null vector of user k-u+1} we have that $\sv_{k}=\ev^{\text{T}}_{\Usf,k-\Ksf+\Usf}$.


Hence, the determinant of  the matrix   
\begin{align}
 \begin{bmatrix}
\sv_{\Ac(1)} \\
\cdots \\
\sv_{\Ac(\Usf)}
\end{bmatrix}
\label{eq:U vectors matrix}
\end{align} 
  could be seen as $D_{\Ac}= \frac{P_{\Ac}}{Q_{\Ac}}$, where $P_{\Ac}$ and $Q_{\Ac}$ are multivariate polynomials whose variables are the elements in  $\bv_{\Vc}$ where $\Vc \in \Gc_2$.   
 Since   each  element in   $\bv_{\Vc}$ where $\Vc \in \Gc_2$
  is uniformly and i.i.d. over $\mathbb{F}_{\qsf}$ where   $\qsf$ is large enough,
by the  Schwartz-Zippel Lemma~\cite{Schwartz,Zippel,Demillo_Lipton}, if we can further show that the multivariate polynomial  $P_{\Ac}$ is   non-zero      (i.e., a multivariate polynomial whose coefficients are not all $0$), the probability that this  multivariate polynomial is equal to $0$ over all possible realization of the elements in  $\bv_{\Vc}$ where $\Vc \in \Gc_2$ goes to $0$ when $\qsf$ goes to infinity, and thus 
  the matrix in~\eqref{eq:U vectors matrix} is full rank with high probability. 
 So in the following, we need to show that $P_{\Ac}$ is   non-zero.
For the matrix (whose dimension is $\Usf \times \Usf$)
\begin{align}
{\bf G}&= 
 \begin{bmatrix}
\av_{\Gc_{2,1}},\ldots, \av_{\Gc_{2,\Ksf-\Usf}}, \av_{\Gc_{2,2\Ksf-2\Usf+1}},\ldots, \av_{\Gc_{2,\Ksf}}
\end{bmatrix}\\
&=
\begin{blockarray}{ccccccccc}
     & c_1 & c_2 & \cdots & c_{\Ksf-\Usf} & c_{\Ksf-\Usf+1} & c_{\Ksf-\Usf+2} & \cdots & c_{\Usf} \\
\begin{block}{c[cccccccc]}
  r_1 & * & * & \cdots & * & * & * & \cdots & * \\
 r_2 &  * & * & \cdots & *& * & * & \cdots & *\\
  \vdots &  \vdots& \vdots & \ddots & \vdots & \vdots &  \vdots & \ddots &  \vdots\\
  r_{\Ksf-\Usf} & * & * & \cdots & *  &* & * & \cdots & *\\
    r_{\Ksf-\Usf+1} & 0 & 0 & \cdots & 0  &* & 0 & \cdots & 0\\
    r_{\Ksf-\Usf+2} &  0 & 0 & \cdots & 0&0 & * & \cdots & 0\\
    \vdots &  \vdots& \vdots & \ddots & \vdots & \vdots &  \vdots & \ddots &  \vdots \\
   r_{\Usf} &  0 & 0 & \cdots & 0&0 & 0 & \cdots & *\\
\end{block}
\end{blockarray}  
\end{align} 
 where $r_1,\ldots,r_{\Usf}$ denote the labels of rows, $c_1,\ldots,c_{\Usf}$ denote   the labels of columns, and each `$*$' denotes a    symbol uniformly and i.i.d. over $\mathbb{F}_{\qsf}$.
 With a slight abuse of notation, we define that ${\bf G} \setminus  \av_{\Gc_{2,j}} $ where $j\in [\Ksf-\Usf]\cup [2\Ksf-2\Usf+1:\Ksf]$ as the column-wise sub-matrix  
 of ${\bf G}$ by removing the column $\av_{\Gc_{2,j}}$. 
 For each $k\in (\Ac \cap [\Ksf-\Usf])$,  by our construction, $\sv_k$ is a  left null space vector of ${\bf G} \setminus  \av_{\Gc_{2,k}}$.
 Hence, to   show that $P_{\Ac}$ is   non-zero, we need to find one realization of the `$*$'s in ${\bf G}$ such that 
 \begin{enumerate}
 \item ${\bf G} \setminus  \av_{\Gc_{2,k}}$ has rank equal to $\Usf-1$ for each $k\in (\Ac \cap [\Ksf-\Usf])$  (such that $\sv_k$ exists   by using the    Cramer's rule and thus  $Q_{\Ac}$ is not zero);
 \item the $\Usf$ rows of the matrix in~\eqref{eq:U vectors matrix}, including $\sv_{k}$ where $k\in (\Ac \cap [\Ksf-\Usf])$ and $\ev^{\text{T}}_{\Usf,j-\Ksf+\Usf}$ where $j\in (\Ac \cap [\Ksf-\Usf+1:\Ksf])$, are linearly independent (such that $D_{\Ac}$ is not zero).
 \end{enumerate}

 We divide the set  $\Ac \cap [\Ksf-\Usf+1:\Ksf]$ into two subsets, $\Sc_1=  \Ac \cap    [\Ksf-\Usf+1:2\Ksf-2\Usf]$  where  
 \begin{align}
 x=|\Sc_1|=|\Ac \cap    [\Ksf-\Usf+1:2\Ksf-2\Usf]| \leq \Ksf-\Usf, \label{eq:def of x}
 \end{align}
  and $\Sc_2=  \Ac   \cap [ 2\Ksf-2\Usf+1:\Ksf]$ where 
   \begin{align}
 y=|\Sc_2|=|\Ac   \cap [ 2\Ksf-2\Usf+1:\Ksf]| \leq 2\Usf-\Ksf. \label{eq:def of y}
 \end{align}
    For each user $j_1 \in \Sc_1$, we have $j_1-\Ksf+\Usf \in [\Ksf-\Usf]$; for each user  $j_2 \in \Sc_2$, we have   $j_2-\Ksf+\Usf \in [\Ksf-\Usf+1:\Usf]$. 
Since $x+y= |\Ac \cap [\Ksf-\Usf+1:\Ksf]|$ and $|\Ac|=\Usf$, we have       
\begin{align}
 \Usf-(\Ksf-\Usf)\leq x+y\leq  \Usf. \label{eq:x+y range}
\end{align} 
 If $x+y=\Usf$, we can see that the matrix in~\eqref{eq:U vectors matrix} is the identity matrix $\mathbf{I}_{\Usf}$ which is full rank. Hence, in the rest of the proof, we  focus on the case where  $2\Usf-\Ksf\leq x+y<\Usf$.

  By  symmetry,  we only need to consider the case where $\Ac \cap [\Ksf-\Usf]=[\Usf-x-y]$, 
   $\Sc_1 =\Ac \cap    [\Ksf-\Usf+1:2\Ksf-2\Usf]= [\Ksf-\Usf+1:\Ksf-\Usf+x]$ and $\Sc_2= \Ac   \cap [ 2\Ksf-2\Usf+1:\Ksf]=[ 2\Ksf-2\Usf+1:  2\Ksf-2\Usf+y]$, 
and find     one realization  of  the `$*$'s in ${\bf G}$ satisfying
  the   constraints 1) and 2). 
Thus the last $|\Ac \setminus [\Ksf-\Usf]|=x+y$ rows
  of the matrix in~\eqref{eq:U vectors matrix} includes $\ev^{\text{T}}_{\Usf,i}$ where $i\in ([x] \cup [\Ksf-\Usf+1 :\Ksf-\Usf+y])$.
 To determine the first $\Usf-x-y$ rows of the matrix in~\eqref{eq:U vectors matrix},  
  we select a realization of ${\bf G}$ as follows  (recall that ${\bf 0}_{m,n}$ and ${\bf 1}_{m,n}$ represents all-zero matrix and all-one matrix  of  dimension $m \times n$, respectively)  
\begin{align}
&{\bf G} = \begin{bmatrix}
\av_{\Gc_{2,1}},\ldots, \av_{\Gc_{2,\Ksf-\Usf}}, \av_{\Gc_{2,2\Ksf-2\Usf+1}},\ldots, \av_{\Gc_{2,\Ksf}}
\end{bmatrix}= \nonumber\\ 
&\begin{blockarray}{cccccc}
& c_{[g_1]} & c_{[g_1+1: \Usf-x-y]} & c_{[\Usf-x-y+1 : \Ksf-\Usf]} & c_{[\Ksf-\Usf+1 : \Ksf-\Usf+y]} &c_{[\Ksf-\Usf+y+1 : \Usf]}\\
\begin{block}{c[ccccc]}
r_{[2\Usf-\Ksf-y]} & {\bf 0}_{2\Usf-\Ksf-y,g_1} &{\bf I}_{2\Usf-\Ksf-y}&{\bf 0}_{2\Usf-\Ksf-y, g_2}&{\bf 0}_{2\Usf-\Ksf-y, y} & -{\bf I}_{2\Usf-\Ksf-y}\\ 
r_{[2\Usf-\Ksf-y+1:x]}&{\bf 0}_{g_2,g_1}& {\bf 0}_{g_2,2\Usf-\Ksf-y}& {\bf I}_{g_2}& {\bf 0}_{g_2, y}&{\bf 0}_{g_2, 2\Usf-\Ksf-y}\\
r_{[x+1:\Ksf-\Usf]} & {\bf I}_{g_1} & -{\bf 1}_{g_1,2\Usf-\Ksf-y} &  -{\bf 1}_{g_1,g_2} & {\bf 1}_{g_1,y} & {\bf 1}_{g_1,2\Usf-\Ksf-y}\\
r_{[\Ksf-\Usf+1:\Ksf-\Usf+y]} & {\bf 0}_{y,g_1} &  {\bf 0}_{y,2\Usf-\Ksf-y} & {\bf 0}_{y,g_2} & {\bf I}_{y} & {\bf 0}_{y,2\Usf-\Ksf-y}\\
r_{[\Ksf-\Usf+y+1:\Usf]} &  {\bf 0}_{2\Usf-\Ksf-y,g_1} & {\bf 0}_{2\Usf-\Ksf-y,2\Usf-\Ksf-y} & {\bf 0}_{2\Usf-\Ksf-y,g_2}& {\bf 0}_{2\Usf-\Ksf-y,y}& {\bf I}_{2\Usf-\Ksf-y}\\
\end{block}
\end{blockarray}  
\label{eq:choice of G}
\end{align}  
where  $g_1:=\Ksf-\Usf-x$, $g_2:=x+y-2\Usf+\Ksf$,  $r_{[i:j]}$ represents $r_i,r_{i+1},\ldots,r_j$, and  $c_{[i:j]}$ represents $c_i,c_{i+1},\ldots,c_j$.
Let us then derive $\sv_k$ for each user  $k\in (\Ac \cap [\Ksf-\Usf]) =[\Usf-x-y]$. 

For each user $k\in [g_1]$, the matrix ${\bf G} \setminus \av_{\Gc_{2,k}}$  has rank equal to $\Usf-1$, since one can easily check that the columns in ${\bf G}$  
  are linearly independent.  Thus ${\bf G} \setminus \av_{\Gc_{2,k}}$ contains exactly one linearly independent left null space vector. We can check that this vector could be   (recall that  $\mathbf{1}_{n}$  and $\mathbf{0}_{n}$ represent  the vertical $n$-dimensional   vector  whose elements are all $1$ and all $0$, respectively)
\begin{align}
&\sv_{k}= [{\bf 1}^{\text{T}}_{2\Usf-\Ksf-y}, \ {\bf 1}^{\text{T}}_{g_2}, \ \ev^{\text{T}}_{g_1,k}, \ -{\bf 1}^{\text{T}}_{y}, \  {\bf 0}^{\text{T}}_{2\Usf-\Ksf-y}   ], \ \forall k\in [g_1].
\label{eq:sk of first set first group}
\end{align}

 For each user $k\in [g_1+1:\Usf-x-y]$,  since  the columns in ${\bf G}$  
  are linearly independent, the matrix ${\bf G} \setminus \av_{\Gc_{2,k}}$  has rank equal to $\Usf-1$.
  Thus ${\bf G} \setminus \av_{\Gc_{2,k}}$ contains exactly one linearly independent left null space vector. We can check that this vector could be    
\begin{align}
&\sv_{k}= [\ev^{\text{T}}_{2\Usf-\Ksf-y,k-g_1}, \ {\bf 0}^{\text{T}}_{g_2}, \ {\bf 0}^{\text{T}}_{g_1}, \  {\bf 0}^{\text{T}}_{y},  \  \ev^{\text{T}}_{2\Usf-\Ksf-y,k-g_1}], \ \forall k\in [g_1+1:\Usf-x-y].
\label{eq:sk of second set first group}
\end{align}
 
 Recall that the last $ x+y$ rows
  of the matrix in~\eqref{eq:U vectors matrix} include  $\ev^{\text{T}}_{\Usf,i}$ where $i\in ([x] \cup [\Ksf-\Usf+1 :\Ksf-\Usf+y])$. Hence, together with the first $\Usf-x-y$ rows as shown in~\eqref{eq:sk of first set first group} and~\eqref{eq:sk of second set first group}, we can see that the matrix in~\eqref{eq:U vectors matrix}  is  
\begin{align}
\begin{blockarray}{ccccc}
& c_{[x]} & c_{[x+1: \Ksf-\Usf]} & c_{[\Ksf-\Usf+1 : \Ksf-\Usf+y]} & c_{[\Ksf-\Usf+y+1 : \Usf]} \\
\begin{block}{c[cccc]}
r_{[g_1]} &  {\bf 1}_{g_1,x}  &{\bf I}_{g_1} &-{\bf 1}_{g_1,y} & {\bf 0}_{g_1,2\Usf-\Ksf-y}\\ 
r_{[g_1+1:\Usf-x-y]} & ({\bf I}_{2\Usf-\Ksf-y}, \ {\bf 0}_{2\Usf-\Ksf-y,g_2})&{\bf 0}_{2\Usf-\Ksf-y,g_1}& {\bf 0}_{2\Usf-\Ksf-y,y} & {\bf I}_{2\Usf-\Ksf-y}\\
r_{[\Usf-x-y+1:\Usf-y]} & {\bf I}_{x} & {\bf 0}_{x,\Ksf-\Usf-x} & {\bf 0}_{x,y} & {\bf 0}_{x,2\Usf-\Ksf-y}\\
r_{[\Usf-y+1:\Usf]} & {\bf 0}_{y,x} & {\bf 0}_{y,\Ksf-\Usf-x} & {\bf I}_{y} & {\bf 0}_{y,2\Usf-\Ksf-y}\\ 
\end{block}
\end{blockarray} , 
\end{align}  
which is full rank. Thus we proved that with the choice of ${\bf G}$ in~\eqref{eq:choice of G}, the constraints 1) and 2) are satisfied; thus $P_{\Ac}$ is a non-zero polynomial. This completes the proof of Lemma~\ref{lem:decodability rank lemma}.

 \section{Data Tables of the Experimental Results in Section~\ref{sec:experiment}}
 \label{sec:data tables}
In the following, we consider the cases where $\Usf=(\Ksf+1)/2$ and $\Usf=\Ksf-1$, and list the running times  of each procedure in our experiments. In the   tables provided in this Section, we use ``IPS'' to represent  the size of each input vector; use ``KST'' to represent  key sharing time; use  ``R1AT'' and ``R2AT'' to represent the running times   in the first and second rounds of model aggregation, respectively; use `R3AT'' and ``R4AT'' to represent the running times   in the   third and fourth rounds of model aggregation (only needed by  \texttt{SecAgg}), respectively; use ``TMA'' to represent the total model aggregation time.

\subsubsection{Case $\Usf=(\Ksf+1)/2$: \texttt{GroupSecAgg} vs. \texttt{LightSecAgg} vs. \texttt{SecAgg}}


\begin{center}
\begin{tabular}{ |c|c|c|c|c|c| } 
\hline
\multicolumn{6}{|c|}{Running Times (ms): $\Ksf = 5, \Usf = 3$}\\
\hline
Scheme & IPS & KST & R1AT & R2AT & TMA\\
\hline
\texttt{GroupSecAgg} & 100,000 & 622.6968 & 19.4242 & 4.4532 & 23.8774\\
\hline
\texttt{LightSecAgg} & 100,000 & 707.118 & 19.2629 & 4.156 & 23.4189\\
\hline
\texttt{GroupSecAgg} & 200,000 & 1232.1053 & 31.6589 & 9.8977 & 41.5566\\
\hline
\texttt{LightSecAgg} & 200,000 & 1313.3104& 32.7853 & 9.7137 & 42.4991\\
\hline
\texttt{GroupSecAgg} & 300,000 & 1854.6576 & 42.6247 & 19.1066 & 61.7313\\
\hline
\texttt{LightSecAgg} & 300,000 & 2024.4369 & 41.3044 & 23.4556 & 64.76\\
\hline
\end{tabular}
\end{center}

\begin{center}
\begin{tabular}{ |c|c|c|c|c|c|c| } 
\hline
\multicolumn{7}{|c|}{Running Times (ms): $\Ksf = 5, \Usf = 3$}\\
\hline
Scheme & IPS   & R1AT & R2AT & R3AT & R4AT & TMA\\
\hline
\texttt{SecAgg} & 100,000 & 7.4139 & 3.53 & 38.5121 & 0.936 & 47.752\\
\hline
\texttt{SecAgg} & 200,000 & 13.7358 & 3.1454 & 77.5049 & 0.8713 & 84.845\\
\hline
\texttt{SecAgg} & 300,000 & 27.0172 & 3.1587 & 116.5683 & 0.822 & 124.073\\
\hline
\end{tabular}
\end{center}

\begin{center}
\begin{tabular}{ |c|c|c|c|c|c| } 
\hline
\multicolumn{6}{|c|}{Running Times (ms): $\Ksf = 10, \Usf = 5$}\\
\hline
Scheme & IPS & KST & R1AT & R2AT & TMA\\
\hline
\texttt{GroupSecAgg} & 100,000 & 1268.6858 & 27.843 & 7.2585 & 35.1015\\
\hline
\texttt{LightSecAgg} & 100,000 & 1422.048& 29.9577 & 7.3137 & 37.2714\\
\hline
\texttt{GroupSecAgg} & 200,000 & 2536.0957 &  49.1541 & 12.4002 & 61.5543\\
\hline
\texttt{LightSecAgg} & 200,000 & 2911.8499& 54.2802 & 11.752 & 66.0322\\
\hline
\texttt{GroupSecAgg} & 300,000 & 3810.1522 & 74.5584 & 18.7861 & 93.3445\\
\hline
\texttt{LightSecAgg} & 300,000 & 4729.238 & 73.5346 & 18.6304 & 92.165\\
\hline
\end{tabular}
\end{center}

\begin{center}
\begin{tabular}{ |c|c|c|c|c|c|c| } 
\hline
\multicolumn{7}{|c|}{Running Times (ms): $\Ksf = 10, \Usf = 5$}\\
\hline
Scheme & IPS   & R1AT & R2AT & R3AT & R4AT & TMA\\
\hline
\texttt{SecAgg} & 100,000 & 10.6461 & 4.1311 & 59.2228 & 0.8841 & 71.911\\
\hline
\texttt{SecAgg} & 200,000 & 12.1471 & 4.5201 & 118.2981 & 1.2305 & 133.9134\\
\hline
\texttt{SecAgg} & 300,000 & 9.471 & 4.0999 & 179.6695 & 1.1038 & 191.9444\\
\hline
\end{tabular}
\end{center}

\begin{center}
\begin{tabular}{ |c|c|c|c|c|c| } 
\hline
\multicolumn{6}{|c|}{Running Times (ms): $\Ksf = 15, \Usf = 8$}\\
\hline
Scheme & IPS & KST & R1AT & R2AT & TMA\\
\hline
\texttt{GroupSecAgg} & 100,000 & 1853.2537 & 45.9662 & 10.5245 & 56.4906\\
\hline
\texttt{LightSecAgg} & 100,000 & 2061.428 & 40.4577 & 10.4448 & 50.9025\\
\hline
\texttt{GroupSecAgg} & 200,000 & 3704.3875 & 90.7609 & 18.5824 & 109.3433\\
\hline
\texttt{LightSecAgg} & 200,000 & 4650.0179 & 92.1633 & 16.3515 & 108.5148\\
\hline
\texttt{GroupSecAgg} & 300,000 & 5556.7188 & 140.9258 & 22.2047 & 163.1305\\
\hline
\texttt{LightSecAgg} & 300,000 & 7346.741 & 130.8297 & 22.7997 & 153.6293\\
\hline
\end{tabular}
\end{center}

\begin{center}
\begin{tabular}{ |c|c|c|c|c|c|c| } 
\hline
\multicolumn{7}{|c|}{Running Times (ms): $\Ksf = 15, \Usf = 8$}\\
\hline
Scheme & IPS   & R1AT & R2AT & R3AT & R4AT & TMA\\
\hline
\texttt{SecAgg} & 100,000 & 12.7749 & 6.6379 & 90.3237 & 1.4144 & 107.5878\\
\hline
\texttt{SecAgg} & 200,000 & 12.0194 & 7.2726 & 182.6603 & 1.4009 & 199.3108\\
\hline
\texttt{SecAgg} & 300,000 & 13.6192 & 6.5942 & 279.8947 & 1.3217 & 298.5508\\
\hline
\end{tabular}
\end{center}

\begin{center}
\begin{tabular}{ |c|c|c|c|c|c| } 
\hline
\multicolumn{6}{|c|}{Running Times (ms): $\Ksf = 20, \Usf = 10$}\\
\hline
Scheme & IPS & KST & R1AT & R2AT & TMA\\
\hline
\texttt{GroupSecAgg} & 100,000 & 2510.2112 & 51.5425 & 14.4271 & 65.9696\\
\hline
\texttt{LightSecAgg} & 100,000 & 2925.3894& 53.7125 & 13.9064 & 67.6189\\
\hline
\texttt{GroupSecAgg} & 200,000 & 5038.559 & 110.6973 & 20.6763 & 131.3736\\
\hline
\texttt{LightSecAgg} & 200,000 & 6314.3566 & 111.0032 & 20.7812 & 131.7844\\
\hline
\texttt{GroupSecAgg} & 300,000 & 7501.9033 & 165.4 & 29.3871 & 194.787\\
\hline
\texttt{LightSecAgg} & 300,000 & 9878.072 & 158.9753 & 28.1264 & 187.1017\\
\hline
\end{tabular}
\end{center}

\begin{center}
\begin{tabular}{ |c|c|c|c|c|c|c| } 
\hline
\multicolumn{7}{|c|}{Running Times (ms): $\Ksf = 20, \Usf = 10$}\\
\hline
Scheme & IPS   & R1AT & R2AT & R3AT & R4AT & TMA\\
\hline
\texttt{SecAgg} & 100,000 & 18.0595 & 8.7712 & 112.2543 & 1.5265 & 133.58\\
\hline
\texttt{SecAgg} & 200,000 & 20.297 & 8.6479 & 236.0381 & 1.6135 & 260.7115\\
\hline
\texttt{SecAgg} & 300,000 & 34.5755 & 8.0978 & 330.9472 & 1.8907 & 359.9537\\
\hline
\end{tabular}
\end{center}

\subsubsection{Case $\Usf=\Ksf-1$: \texttt{GroupSecAgg} (\texttt{GroupSecAgg\_1}) vs. \texttt{LightSecAgg} vs. \texttt{SecAgg}}

\begin{center}
\begin{tabular}{ |c|c|c|c|c|c| } 
\hline
\multicolumn{6}{|c|}{Running Times (ms): $\Ksf = 5, \Usf = 4$}\\
\hline
Scheme & IPS & KST & R1AT & R2AT & TMA\\
\hline
\texttt{GroupSecAgg} & 100,000 & 573.5159 & 32.0896 & 4.5524 & 36.642\\
\hline
\texttt{GroupSecAgg1} & 100,000 & 565.963 & 32.0896 & 4.5524 & 36.642\\
\hline
\texttt{LightSecAgg} & 100,000 & 571.285 & 31.4279 & 4.5067 & 35.9346\\
\hline
\texttt{GroupSecAgg} & 200,000 & 1212.9158 & 61.3236 & 9.8175 & 71.1411\\
\hline
\texttt{GroupSecAgg1} & 200,000 & 1180.7742 & 61.3236 & 9.8175 & 71.1411\\
\hline
\texttt{LightSecAgg} & 200,000 & 1215.373 & 61.6626 & 9.8922 & 71.5548\\
\hline
\texttt{GroupSecAgg} & 300,000 & 1808.522 & 92.8807 & 14.3178 & 107.1984\\
\hline
\texttt{GroupSecAgg1} & 300,000 & 1674.8969 & 92.8807 & 14.3178 & 107.1984\\
\hline
\texttt{LightSecAgg} & 300,000 & 1714.3089 & 90.6547 & 15.0228 & 105.6774\\
\hline
\end{tabular}
\end{center}

\begin{center}
\begin{tabular}{ |c|c|c|c|c|c|c| } 
\hline
\multicolumn{7}{|c|}{Running Times (ms): $\Ksf = 5, \Usf = 4$}\\
\hline
Scheme & IPS   & R1AT & R2AT & R3AT & R4AT & TMA\\
\hline
\texttt{SecAgg} & 100,000 & 8.0442 & 2.3905 & 52.2406 & 0.8506 & 63.5258\\
\hline
\texttt{SecAgg} & 200,000 & 6.0251 & 4.5155 & 99.1241 & 0.9645 & 110.629\\
\hline
\texttt{SecAgg} & 300,000 & 13.3259 & 2.4766 & 148.4597 & 0.8532 & 165.1157\\
\hline
\end{tabular}
\end{center}

\begin{center}
\begin{tabular}{ |c|c|c|c|c|c| } 
\hline
\multicolumn{6}{|c|}{Running Times (ms): $\Ksf = 10, \Usf = 9$}\\
\hline
Scheme & IPS & KST & R1AT & R2AT & TMA\\
\hline
\texttt{GroupSecAgg} & 100,000 & 1290.1098 & 60.1389 & 6.5752 & 66.7142\\
\hline
\texttt{GroupSecAgg1} & 100,000 & 1090.2839 & 60.1389 & 6.5752 & 66.7142\\
\hline
\texttt{LightSecAgg} & 100,000 & 1165.683 & 61.817 & 6.1287 & 67.9457\\
\hline
\texttt{GroupSecAgg} & 200,000 & 2590.862 &  119.5961 & 13.5527 & 133.1488\\
\hline
\texttt{GroupSecAgg1} & 200,000 & 2231.250 &  119.5961 & 13.5527 & 133.1488\\
\hline
\texttt{LightSecAgg} & 200,000 & 2372.629 & 122.3105 & 12.7634 & 135.0739\\
\hline
\texttt{GroupSecAgg} & 300,000 & 4000.3806  & 189.2828 & 16.931 & 206.2138\\
\hline
\texttt{GroupSecAgg1} & 300,000 & 3122.4385  & 189.2828 & 16.931 & 206.2138\\
\hline
\texttt{LightSecAgg} & 300,000 & 3159.0967 & 192.4548 & 17.065 & 209.5198\\
\hline
\end{tabular}
\end{center}

\begin{center}
\begin{tabular}{ |c|c|c|c|c|c|c| } 
\hline
\multicolumn{7}{|c|}{Running Times (ms): $\Ksf = 10, \Usf = 9$}\\
\hline
Scheme & IPS   & R1AT & R2AT & R3AT & R4AT & TMA\\
\hline
\texttt{SecAgg} & 100,000 & 6.501 & 5.3933 & 101.6094 & 1.2792 & 114.7829\\
\hline
\texttt{SecAgg} & 200,000 & 6.1388 & 5.6953 & 202.9011 & 1.3435 & 216.0788\\
\hline
\texttt{SecAgg} & 300,000 & 5.5022 & 4.9979 & 303.9231 & 1.1925 & 315.6156\\
\hline
\end{tabular}
\end{center}

\begin{center}
\begin{tabular}{ |c|c|c|c|c|c| } 
\hline
\multicolumn{6}{|c|}{Running Times (ms): $\Ksf = 15, \Usf = 14$}\\
\hline
Scheme & IPS & KST & R1AT & R2AT & TMA\\
\hline
\texttt{GroupSecAgg} & 100,000 & 2003.5729 & 99.9402 & 2.6843 & 102.6245\\
\hline
\texttt{GroupSecAgg1} & 100,000 & 1699.1491 & 99.9402 & 2.6843 & 102.6245\\
\hline
\texttt{LightSecAgg} & 100,000 & 1748.1338 & 98.909 & 2.8935 & 101.8025\\
\hline
\texttt{GroupSecAgg} & 200,000 & 4017.3969 & 201.3512 & 13.6942 & 215.0454\\
\hline
\texttt{GroupSecAgg1} & 200,000 & 3293.6826& 201.3512 & 13.6942 & 215.0454\\
\hline
\texttt{LightSecAgg} & 200,000 & 3514.3759 & 196.7829 & 14.7847 & 211.5676\\
\hline
\texttt{GroupSecAgg} & 300,000 & 6020.5340 & 285.3097 & 21.7039 & 307.0137\\
\hline
\texttt{GroupSecAgg} & 300,000 & 4605.5441 & 285.3097 & 21.7039 & 307.0137\\
\hline
\texttt{LightSecAgg} & 300,000 & 4657.3419 & 299.4586 & 19.5312 & 318.9898\\
\hline
\end{tabular}
\end{center}

\begin{center}
\begin{tabular}{ |c|c|c|c|c|c|c| } 
\hline
\multicolumn{7}{|c|}{Running Times (ms): $\Ksf = 15, \Usf = 14$}\\
\hline
Scheme & IPS   & R1AT & R2AT & R3AT & R4AT & TMA\\
\hline
\texttt{SecAgg} & 100,000 & 17.8189 & 8.3383 & 153.163 & 1.5803 & 180.9006\\
\hline
\texttt{SecAgg} & 200,000 & 10.6163 & 7.6086 & 303.9652 & 1.6801 & 323.87\\
\hline
\texttt{SecAgg} & 300,000 & 9.1441 & 8.3122 & 459.9635 & 1.7766 & 479.196\\
\hline
\end{tabular}
\end{center}

\begin{center}
\begin{tabular}{ |c|c|c|c|c|c| } 
\hline
\multicolumn{6}{|c|}{Running Times (ms): $\Ksf = 20, \Usf = 19$}\\
\hline
Scheme & IPS & KST & R1AT & R2AT & TMA\\
\hline
\texttt{GroupSecAgg} & 100,000 & 2724.8058  & 132.5762 & 4.0336 & 136.6098\\
\hline
\texttt{GroupSecAgg1} & 100,000 & 2387.7981 & 132.5762 & 4.0336 & 136.6098\\
\hline
\texttt{LightSecAgg} & 100,000 & 2419.7531 & 128.7957 & 3.7095 & 132.5053\\
\hline
\texttt{GroupSecAgg} & 200,000 & 5460.64 & 263.683 & 16.1745 & 279.8575\\
\hline
\texttt{GroupSecAgg1} & 200,000 & 4466.254 & 263.683 & 16.1745 & 279.8575\\
\hline
\texttt{LightSecAgg} & 200,000 & 4680.175 & 260.3788 & 14.6445 & 275.0233\\
\hline
\texttt{GroupSecAgg} & 300,000 & 8230.2930 & 404.7795 & 24.8559 & 429.6354\\
\hline
\texttt{GroupSecAgg1} & 300,000 & 6114.4559 & 404.7795 & 24.8559 & 429.6354\\
\hline
\texttt{LightSecAgg} & 300,000 & 6277.1279 & 400.4311 & 25.4461 & 425.8771\\
\hline
\end{tabular}
\end{center}

\begin{center}
\begin{tabular}{ |c|c|c|c|c|c|c| } 
\hline
\multicolumn{7}{|c|}{Running Times (ms): $\Ksf = 20, \Usf = 19$}\\
\hline
Scheme & IPS   & R1AT & R2AT & R3AT & R4AT & TMA\\
\hline
\texttt{SecAgg} & 100,000 & 10.1681 & 11.6254 & 204.3118 & 2.0399 & 228.1454\\
\hline
\texttt{SecAgg} & 200,000 & 11.1408 & 11.8139 & 406.3665 & 2.5097 & 431.831\\
\hline
\texttt{SecAgg} & 300,000 & 10.2389 & 12.2843 & 611.3178 & 2.1002 & 635.9413\\
\hline
\end{tabular}
\end{center}

\bibliographystyle{IEEEtran}
\bibliography{IEEEabrv,IEEEexample}

\begin{thebibliography}{10}
\providecommand{\url}[1]{#1}
\csname url@samestyle\endcsname
\providecommand{\newblock}{\relax}
\providecommand{\bibinfo}[2]{#2}
\providecommand{\BIBentrySTDinterwordspacing}{\spaceskip=0pt\relax}
\providecommand{\BIBentryALTinterwordstretchfactor}{4}
\providecommand{\BIBentryALTinterwordspacing}{\spaceskip=\fontdimen2\font plus
\BIBentryALTinterwordstretchfactor\fontdimen3\font minus
  \fontdimen4\font\relax}
\providecommand{\BIBforeignlanguage}[2]{{%
\expandafter\ifx\csname l@#1\endcsname\relax
\typeout{** WARNING: IEEEtran.bst: No hyphenation pattern has been}%
\typeout{** loaded for the language `#1'. Using the pattern for}%
\typeout{** the default language instead.}%
\else
\language=\csname l@#1\endcsname
\fi
#2}}
\providecommand{\BIBdecl}{\relax}
\BIBdecl

\bibitem{mcmahan2017communication}
B.~McMahan, E.~Moore, D.~Ramage, S.~Hampson, and B.~A. y~Arcas,
  ``Communication-efficient learning of deep networks from decentralized
  data,'' in \emph{Artificial intelligence and statistics}.\hskip 1em plus
  0.5em minus 0.4em\relax PMLR, 2017, pp. 1273--1282.

\bibitem{yang2019federated}
Q.~Yang, Y.~Liu, T.~Chen, and Y.~Tong, ``Federated machine learning: Concept
  and applications,'' \emph{ACM Transactions on Intelligent Systems and
  Technology (TIST)}, vol.~10, no.~2, p.~12, 2019.

\bibitem{li2020federated}
T.~Li, A.~K. Sahu, A.~Talwalkar, and V.~Smith, ``Federated learning:
  Challenges, methods, and future directions,'' \emph{IEEE Signal Processing
  Magazine}, vol.~37, no.~3, pp. 50--60.

\bibitem{mcmahan2021advances}
H.~B. McMahan \emph{et~al.}, ``Advances and open problems in federated
  learning,'' \emph{Foundations and Trends{\textregistered} in Machine
  Learning}, vol.~14, no.~1, 2021.

\bibitem{bonawitz2017practical}
K.~Bonawitz, V.~Ivanov, B.~Kreuter, A.~Marcedone, H.~B. McMahan, S.~Patel,
  D.~Ramage, A.~Segal, and K.~Seth, ``Practical secure aggregation for
  privacy-preserving machine learning,'' in \emph{2017 ACM SIGSAC Conference on
  Computer and Communications Security}, 2017, pp. 1175--1191.

\bibitem{bell2020secure}
J.~H. Bell, K.~A. Bonawitz, A.~Gasc{\'o}n, T.~Lepoint, and M.~Raykova, ``Secure
  single-server aggregation with (poly) logarithmic overhead,'' in
  \emph{Proceedings of the 2020 ACM SIGSAC Conference on Computer and
  Communications Security}, 2020, pp. 1253--1269.

\bibitem{choi2020communication}
B.~Choi, J.-y. Sohn, D.-J. Han, and J.~Moon, ``Communication-computation
  efficient secure aggregation for federated learning,''
  \emph{arXiv:2012.05433}, Dec. 2020.

\bibitem{ITsecureaggre2021}
Y.~Zhao and H.~Sun, ``Information theoretic secure aggregation with user
  dropouts,'' \emph{arXiv:2101.07750}, Jan. 2021.

\bibitem{lightsec2021so}
J.~So, C.~He, C.-S. Yang, S.~Li, Q.~Yu, R.~E. Ali, B.~Guler, and S.~Avestimehr,
  ``{LightSecAgg}: a lightweight and versatile design for secure aggregation in
  federated learning,'' \emph{arXiv:2109.14236}, Feb. 2022.

\bibitem{so2021turbo}
J.~So, B.~G{\"u}ler, and A.~S. Avestimehr, ``Turbo-aggregate: Breaking the
  quadratic aggregation barrier in secure federated learning,'' \emph{IEEE
  Journal on Selected Areas in Info. Theory}, vol.~2, no.~1, pp. 479--489,
  2021.

\bibitem{kadhe2020fastsecagg}
S.~Kadhe, N.~Rajaraman, O.~O. Koyluoglu, and K.~Ramchandran, ``{FastSecAgg}:
  Scalable secure aggregation for privacy-preserving federated learning,''
  \emph{arXiv:2009.11248}, Sep. 2020.

\bibitem{nezhad2022swiftagg}
T.~Jahani-Nezhad, M.~A. Maddah-Ali, S.~Li, and G.~Caire, ``{SwiftAgg+}:
  Achieving asymptotically optimal communication load in secure aggregation for
  federated learning,'' \emph{arXiv:2202.04169}, Mar. 2022.

\bibitem{hellman1976newdirection}
W.~Diffie and M.~Hellman, ``New directions in cryptography,'' \emph{IEEE Trans.
  Infor. Theory}, vol.~22, no.~6, pp. 644--654, Nov. 1976.

\bibitem{shannonsecurity}
C.~E. Shannon, ``Communication theory of secrecy systems,'' \emph{in The Bell
  System Technical Journal}, vol.~28, no.~4, pp. 656--715, Oct. 1949.

\bibitem{maurer1993secretkey}
U.~M. Maurer, ``Secret key agreement by public discussion from common
  information,'' \emph{IEEE Trans. Infor. Theory}, vol.~39, no.~3, pp.
  733--742, May 1993.

\bibitem{ahlswede1993commonran}
R.~Ahlswede and I.~Csiszar, ``Common randomness in information theory and
  cryptography. {I}. secret sharing,'' \emph{IEEE Trans. Infor. Theory},
  vol.~39, no.~4, pp. 1121--1132, Jul. 1993.

\bibitem{csiszar2004secrey}
I.~Csiszar and P.~Narayan, ``Secrecy capacities for multiple terminals,''
  \emph{IEEE Trans. Infor. Theory}, vol.~50, no.~12, pp. 3047--3061, Dec. 2004.

\bibitem{gohari2010itkeyaggre}
A.~A. Gohari and V.~Anantharam, ``Information-theoretic key agreement of
  multiple terminals—part {I},'' \emph{IEEE Trans. Infor. Theory}, vol.~56,
  no.~8, pp. 3973--3996, Aug. 2010.

\bibitem{sun2020securegroupcast}
H.~Sun, ``Secure groupcast with shared keys,'' \emph{arXiv:2003.11995}, Mar.
  2020.

\bibitem{sun2020compound}
------, ``Compound secure groupcast: Key assignment for selected
  broadcasting,'' \emph{arXiv:2004.14986}, Apr. 2020.

\bibitem{kuserinterference}
V.~R. Cadambe and S.~A. Jafar, ``Interference alignment and degrees of freedom
  of the k-user interference channel,'' \emph{IEEE Trans. Infor. Theory},
  vol.~54, no.~8, pp. 3425--3441, Aug. 2008.

\bibitem{Wan2022securecomp}
K.~Wan, H.~Sun, M.~Ji, and G.~Caire, ``On secure distributed linearly separable
  computation,'' \emph{IEEE Journal on Selected Areas in Communications},
  vol.~40, no.~3, pp. 912--926, Mar. 2022.

\bibitem{linearcomput2020wan}
------, ``Distributed linearly separable computation,'' \emph{IEEE Trans. Inf.
  Theory}, vol.~68, no.~2, pp. 1259--1278, Feb. 2022.

\bibitem{Schwartz}
J.~T. Schwartz, ``Fast probabilistic algorithms for verification of polynomial
  identities,'' \emph{Journal of the ACM (JACM)}, vol.~27, no.~4, pp. 701--717,
  1980.

\bibitem{Zippel}
R.~Zippel, ``Probabilistic algorithms for sparse polynomials,'' in
  \emph{International symposium on symbolic and algebraic manipulation}.\hskip
  1em plus 0.5em minus 0.4em\relax Springer, 1979, pp. 216--226.

\bibitem{Demillo_Lipton}
R.~A. Demillo and R.~J. Lipton, ``A probabilistic remark on algebraic program
  testing,'' \emph{Information Processing Letters}, vol.~7, no.~4, pp.
  193--195, 1978.

\bibitem{Wan2021combinationnet}
K.~Wan, D.~Tuninetti, M.~Ji, and P.~Piantanida, ``Combination networks with
  end-user-caches: Novel achievable and converse bounds under uncoded cache
  placement,'' \emph{IEEE Trans. Inf. Theory}, vol.~68, no.~2, pp. 806--827,
  Feb. 2022.

\end{thebibliography}

\end{document}